\documentclass[reprint, prx, aps, notitlepage, nofootinbib,longbibliography,floatfix, superscriptaddress]{revtex4-1}
\usepackage{amsmath,amstext,amssymb,graphicx,bm,dcolumn,color,times, braket,caption}
\usepackage[colorlinks=true,citecolor=blue,urlcolor=red]{hyperref}
\usepackage[table]{xcolor}
\usepackage{graphicx}
\usepackage{amssymb}
\usepackage{amsmath}
\usepackage{amsthm}
\usepackage{amsfonts}
\usepackage{braket}
\usepackage{lineno}
\usepackage{subfig}

\begin{document}

\title{Synchronizing clocks via satellites using entangled photons: \\Effect of relative velocity on precision}

\author{Stav Haldar}\email{hstav1@lsu.edu}\affiliation{Department of Physics \& Astronomy, Louisiana State University, Baton Rouge, LA 70803, USA}
\author{Ivan Agullo}\affiliation{Department of Physics \& Astronomy, Louisiana State University, Baton Rouge, LA 70803, USA}
\author{James E. Troupe}\affiliation{Xairos Systems, Inc., Denver, CO, USA}

\begin{abstract}
A satellite-based scheme to perform clock synchronization between ground stations spread across the globe using quantum resources
was proposed in \cite{paper1}, based on the quantum clock synchronization (QCS) protocol developed in  \cite{AntiaTroupeQCS}.  Such a scheme could achieve synchronization up to the picosecond level over distances of thousands of kilometers. Nonetheless, the implementation of this QCS protocol is yet to be demonstrated experimentally in situations where the satellite velocities cannot be neglected, as is the case in many realistic scenarios. In this work, we develop analytical and numerical tools to study the effect of the relative velocity between the satellite and ground stations on the success of the QCS protocol.
We conclude that the protocol can still run successfully if the data acquisition  window is chosen appropriately. As a demonstration, we simulate the synchronization outcomes for cities across the continental United States using a single satellite in a LEO orbit, low-cost entanglement sources, portable atomic clocks, and avalanche detectors. We conclude that, after including the effect of relative motion, sub-nanosecond to picosecond level precision can still be achieved over distance scales of $\approx 4000$ kms. Such high precision synchronization is currently not achievable over long distances ($\gtrsim 100 km$) with standard classical techniques including the GPS.
The simulation tools developed in this work are in principle applicable to other means of synchronizing clocks using entangled photons \cite{Jon_QCS_2018, QClock_network_sat}, which are expected to form the basis of future quantum networks like the Quantum Internet, distributed quantum sensing and Quantum GPS \cite{white_paper}.
\end{abstract}

\maketitle
\section{Introduction}
\label{section1}
A satellite based quantum clock synchronization (QCS) scheme for time distribution amongst a network of ground based clocks was introduced in \cite{paper1}. We concluded that such a QCS scheme could achieve sub-nanosecond precision at the global scale utilising modest optical sources, modestly stable clocks, and a small constellation of LEO satellites. In this scheme, establishment of elementary links between satellite and individual ground stations is achieved via the exchange of entangled photons and, subsequently, the satellites act as intermediaries for synchronizing different ground stations. This scheme is based on the QCS protocol introduced in  \cite{AntiaTroupeQCS}, which has been experimentally demonstrated to achieve  picosecond sync level precision between stationary clocks on Earth with as few as 20 detected entangled photon pairs \cite{Lee2019} (for a more recent demonstration also see Ref. \cite{spiess_exp}).

Given the large number of variables involved when moving satellites are considered, our previous  analysis in \cite{paper1} was performed under some simplifying assumptions. 
The main one was to use the rate of exchange of entangled photons between 
satellites and ground stations as a proxy for the precision at which they can synchronize ---rather than a detailed calculation of the correlation functions involved in the protocol, which ultimately determine the sync precision. This assumption was backed by numerical simulations, and allowed us to focus attention  on kinematic aspects involving losses, beam spreading, sync area coverage, etc. 
In this work, we go a step beyond and quantitatively analyze the effects of the relative instantaneous velocities on the performance of the protocol, quantifying the way relative motion limits the sync precision, and determining the optimal data acquisition time. The main assumption was that if the rate of exchange of entangled photons in the elementary link is greater than a cut off, the QCS protocol succeeds at a certain precision. This assumption although backed by static simulations, fails to hold in the dynamic picture, where the precision of the QCS protocol is effected by the relative motion between the satellite and the ground station.
The present work determines the degree at which the QCS protocol, which has successfully tested to synchronise stationary clocks, can be generalized to moving clocks, and hence to space based applications, which have shown great promise in extending the scale and efficiency of quantum networks\cite{Sidu2015, LSU_satsim}.  

The main goal of this paper is thus to assess the feasibility of a QCS network by determining the achievable precision, network scale and connectivity, primarily in terms of the performance of the elementary links.

Our analysis has applicability beyond the concrete QCS protocol analyzed in this article, in particular to the protocols based on distributing entanglement from moving satellites \cite{Sidu2015, LSU_satsim}, an application that we explore in a separate publication.

Before describing our methods and results, it is important to briefly motivate, (a) the need for a QCS network and (b) the advantages of using entangled photons and a constellation of satellites to establish such a network of synchronized clocks.

\begin{figure}
    \centering
    \includegraphics[scale = 0.5]{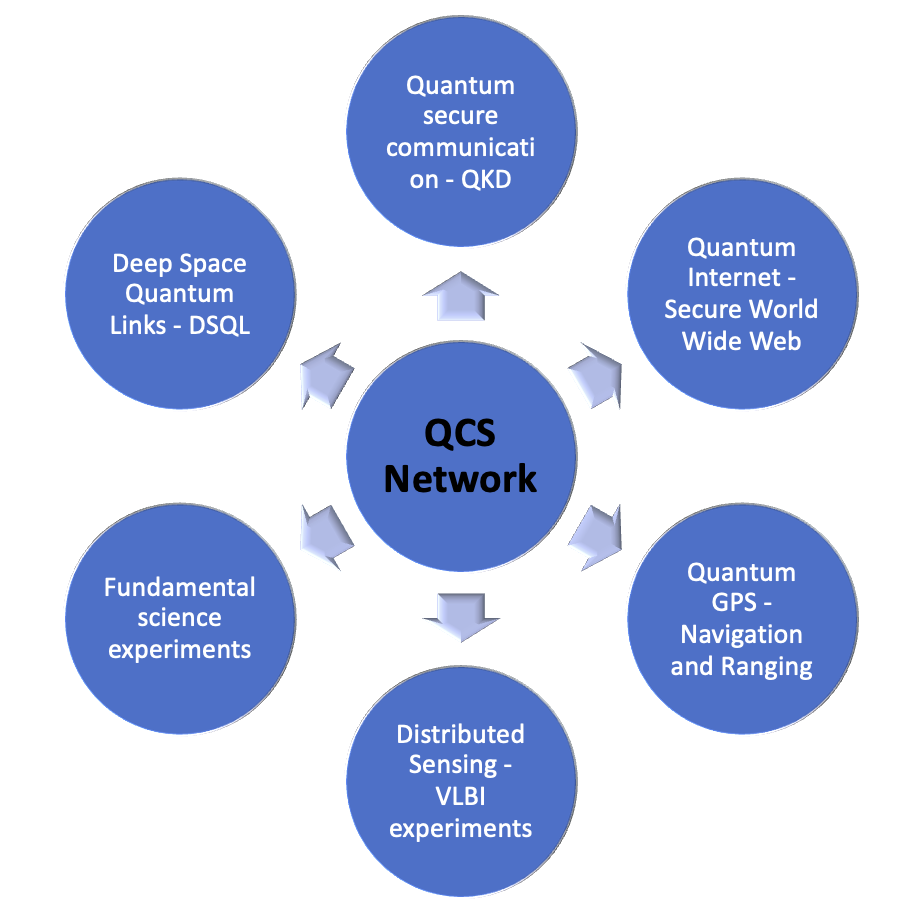}
    \caption{Application spaces for the  QCS network --- a hybrid quantum network of satellite and ground based clocks.}
    \label{fig:QCS_applications}
\end{figure}

The ability to measure, hold and distribute time at high precision determines the limits of our scientific explorations. From a technological point of view, precise time  measurement  and  synchronisation  is  an indispensable feature of communication and networking protocols, navigation and ranging, astronomical, geological and meteorological measurements, amongst others \cite{Sidu2015, opticalclocks2018, RTI, BEC_space_interference_Lachmann2021, PTP}.
It must be further stressed that most realisations of quantum protocols such as teleportation and quantum key distribution, have an inherent requirement for continuous, high accuracy clock synchronization at the sub-nanosecond level \cite{white_paper, GE_2021}. Tremendous advances have been made in classical techniques for clock synchronization \cite{Newbury2019_quadcopter}, for example using radio-frequency pulses \cite{rf_picosecond}, and optical frequency combs \cite{Newbury2016_freq_combs}. Although these techniques can provide synchronisation in the picosecond to femtosecond range, considerable challenges exist in their long distance ($>$100 km) implementations, such as large computational overhead and transfer of technology to satellite payloads \cite{Newbury2019_quadcopter, Newbury2016_freq_combs}. It is thus important to look at QCS networks from the perspective of complementing these state-of-the-art synchronisation and time distribution protocols for long distance applications. At the same time, we also point out new application spaces for QCS networks, given the inherent advantages of quantum communication over classical communication (see Figure \ref{fig:QCS_applications}) \cite{Yin2020,Sidhu2022,SAGE,Anthony_HOM_grav,Bell_test_weak_grav, Einstein_eq_Qopt, Qoptics_space_exp, Wang2022, Xiang2022, Lanning2022, FanYuan2021, FanYuan2022, caleffi2022_1, caleffi2022_2}. 
Quantum state transfer protocols such as teleportation and key distribution have been carried out over continental distances of more than 1000 km \cite{Pan_exp_crypto, Pan_exp_teleport, Pan_satellite_QCS, Nano_sat}. To contrast with the limits of fiber optic based quantum communication, we note that the quantum repeater-less fiber optic based secret key rate bound is surpassed beyond 215 km for a satellite at altitude of 530 km \cite{Pan_exp_crypto, Sidu2015}. A key role in developing feasible long distance quantum communication implementations has thus been played by the transition to hybrid space-terrestrial quantum communication network architectures \cite{Nano_sat, cubesat:2020, Oi2017}. This combines satellites and ground stations equipped with optical telescopes with metropolitan scale fiber optic networks. This is because large-scale realizations of fully terrestrial quantum networks are hindered by the exponential losses associated with ground-based communication channels (primarily fiber optic cables) \cite{LSU_satsim}. Large number of high fidelity quantum repeaters and/or quantum memories could improve the situation to some extent, but their current performance levels are below those needed for mature applications \cite{Sidu2015}, and furthermore, it would very likely be impractical to place such devices in difficult terrain, e.g. mountains or oceans.

Finally, we justify the use of incoherent optical communication (single photon detection) via entangled photons for distributing time, when the same  can be achieved classically. Sub nano-second synchronization is currently not easily achievable over long distances, as GPS only allows synchronization with a precision of $>$ 40 nanoseconds (95\% of the time). Further, for very long optical links with high loss, e.g. those between LEO satellites and ground stations, high precision synchronization will have to be achieved with a relatively small number of received photons. In this regard, it is important to point out that optical communication techniques that use single photon detection (non-coherent/quantum measurement) have a provable advantage in terms of power efficiency (bits per photon) over the coherent optical communication techniques which are generally used in high precision, classical optics based clock synchronization, e.g. classical O-TWTFT \cite{Dolinar2011}. As mentioned earlier, picosecond level QCS was demonstrated in a ground based setting to work with as few as 20 detected entangled photons. This makes it an ideal candidate for high-loss satellite based quantum communication channels.

The structure of the paper is as follows -- In Section \ref{sec:2}, we describe the QCS protocol and find analytical results for the achievable precision and network scale. We consider the effects of relative motion between clocks (range-rate change), channel loss and background noise. In Section \ref{sec:3}, we provide details about the simulation techniques used, and in sections \ref{sec:4} and \ref{sec:5}, we present the results of these simulations, developing tools to assess and describe the QCS network outcomes. More specifically, in Section \ref{sec:5} we show that for a QCS network of 4 cities ---New York, Atlanta, Los Angeles and Seattle--- sub-nanosecond to picosecond level clock sync precision can be achieved by using  modest resources. We present our conclusions and directions for future work in Section \ref{sec:6}.  
\section{Description of the protocol}
\label{sec:2}
We briefly describe a QCS protocol to remotely synchronise a network of clocks located on the Earth using satellites in Earth's orbits as intermediaries. The main resource used in the protocol is time correlated photons generated out of an spontaneous parametric down conversion (SPDC) source. As an added advantage, the entanglement/quantum correlations in the polarization degree of freedom can be utilised to enhance the security of the protocol. For more details on the security analysis and experimental implementation, we point the reader to foundational work on this protocol \cite{AntiaTroupeQCS, Lee2019}.
To begin the discussion, we describe a simpler scenario which is entirely ground-based and involves stationary clocks, to familiarize ourselves with the cogs and wheels of the protocol. This would make the subsequent discussion of the satellite-based version involving moving clocks cogent.

\subsection{Ground-based protocol}
\label{subsec:2.1}
\begin{figure}
    \centering
    \includegraphics[width = \linewidth]{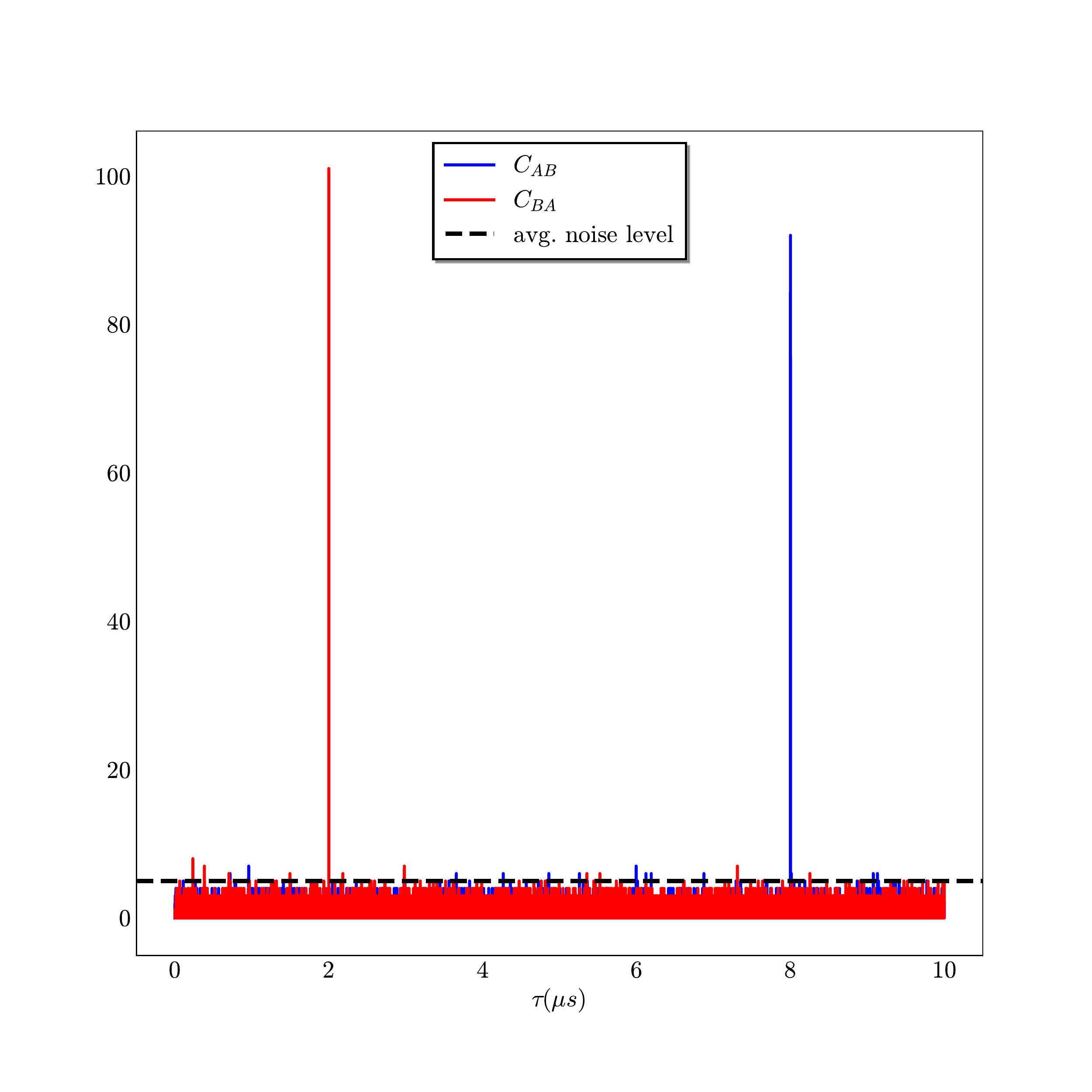}
    \caption{Here, we show the timestamp cross-correlation functions obtained via a Monte Carlo simulation of the QCS protocol. The effects of channel loss, noise and dark counts have been included. The average noise level shown is a rough guide to eye to show the height the genuine peaks must surpass in order for the protocol to be successful (signal-to-noise ratio). The background noise + dark count combined rate is assumed to be $10^7$ photons/s, the source rate is also $10^7$ ebits/s. The channel loss is 20 dB (representative of an uplink between a ground station and a polar satellite when the satellite is overhead the ground station). We have chosen $t_{\rm Acq} = 1 \rm ms$ in this plot.  The offset between the clocks has been chosen to be $6 \mu s$; this number can be correctly obtained from the figure as the gap between the two peaks. The least count of the timestamps is 0.5 ns. Since we have two unique peaks in this case, the offset can thus be determined at this precision (0.5 ns).}
    \label{fig:cc_no_motion}
\end{figure}

Consider two clocks at A and B with Alice and Bob, which are assumed to have the same frequency within the precision of the synchronization task and are stable enough to maintain this frequency during the time in which the sync protocol is executed ---we discuss below how to relax this assumption. Now, consider that they have a constant offset of $\Delta$. The task of the protocol is to find the value of $\Delta$. Say B is ahead of A.
Consider both Alice and Bob have an SPDC source to generate entangled photon pairs. Alice generates a pair at a random time $t_a$; one photon from the pair is locally detected and time stamped by Alice; the other photon travels to Bob and is detected and time stamped at time $t_b$, using a single photon detector. If $t_{ab}$ is the travel time for the photon between A and B then,
\begin{equation}
\label{eqn:1}
    t_b' = t_a + t_{ab} + \Delta \, .
\end{equation}
Similarly, for a photon pair generated at Bob's end and detected at A:
\begin{equation}
\label{eqn:2}
    t_a = t_b' + t_{ba} - \Delta \, .
\end{equation}
Consider the distributions $A(t)$ and $B(t)$ which count the photon detection events at A and B respectively.
\begin{equation}
\label{eqn:3}
    A(t) = \Sigma_{i} \delta(t - t_a^i)
\end{equation}
Similarly, for photons  detected by Bob
\begin{equation}
\label{eqn:4}
    B(t') = \Sigma_{j} \delta(t' - t_b'^j) \, .
\end{equation}
Finally consider the time-stamp cross-correlation function for these detection events, $C_{AB}(\tau)$ defined as below:
\begin{equation}
\label{eqn:5}
    C_{AB}(\tau) = \int_0^{t_{\rm Acq}} A(t)B(t+\tau)dt \, .
\end{equation}
Here, the acquisition time $t_{\rm Acq}$ is the total time for which the protocol runs, i.e., photon detection timestamps collected over a time window $t_{\rm Acq}$ are used to find the cross-correlation functions. We will later show that the choice of $t_{\rm Acq}$ plays an important role in the success of the QCS protocol.
Using Equations (\ref{eqn:1})-(\ref{eqn:4}), it is easy to see that $C_{AB}(\tau)$ has a peak at:
\begin{equation}
\label{eqn:6}
    \tau^{ab}_{\rm max} = t_{ab} + \Delta \, .
\end{equation}
A similarly defined correlation function $C_{BA}(\tau)$
\begin{equation}
    C_{BA}(\tau) = \int_0^{t_{\rm Acq}} B(t')A(t'+\tau)dt' \, ,
\end{equation}
has a peak at:
\begin{equation}
\label{eqn:8}
    \tau^{ba}_{\rm max} = t_{ba} - \Delta \, .
\end{equation}
If we assume reciprocity of the time of travel \cite{Taylor2020, AntiaTroupeQCS}, i.e., $t_{ab} = t_{ba}$, adding and subtracting Equations (\ref{eqn:6}) and (\ref{eqn:8}) we get the time of travel and the offset as:
\begin{equation}
\label{eqn:9}
    \Delta = \frac{\tau^{ab}_{\rm max} - \tau^{ba}_{\rm max}}{2} \, .
\end{equation}
\begin{equation}
\label{eqn:10}
    t_{ab} = t_{ba} =  \frac{\tau^{ab}_{\rm max} + \tau^{ba}_{\rm max}}{2}\, .
\end{equation}
So, looking at the peaks of these two-way correlation functions, both the time of travel (\ref{eqn:9}) and the offset (\ref{eqn:10}) can be evaluated (see Fig.~\ref{fig:cc_no_motion} for an example). It is noteworthy that although an entanglement source is used to produce the photons that are then timestamped for the QCS protocol, the polarization degree of freedom which has the quantum correlations is never invoked. The protocol only relies on the tight time-correlations between such photons. Therefore, the quantumness of the protocol comes from the use of single photon sources and detectors and the randomness of the time of birth of the SPDC photon pairs. The entanglement between photons is not explicitly used, although it can play an important role in adding an extra layer of security to the protocol.

Some effects which can deteriorate the success rate and precision of this protocol include channel loss, dark counts and background noise. Further, since, here we limit ourselves to determining a fixed offset between the two clocks, any relative frequency differences and/or drifts lead to spreading of the cross-correlation function peaks. This effect is very similar to relative motion between the clocks changing the time of travel between the two. This is the main challenge we seek to address in this work and we begin a detailed discussion in the next section. For effects of frequency differences and drifts  see also Refs. \cite{ho:09, Lee2019, spiess_exp}.
\subsection{Challenges for a satellite based version of the protocol: How to synchronize clocks in relative motion?}
\label{subsec:2.2}
Consider now that Alice and Bob are separated in a way that makes it inefficient to exchange photons directly between the two parties. A quantum network between different cities is an example of such a scenario. The distances are large enough ($\approx$ 1000 km) to make direct communication through standard optical fiber channels (even with repeaters) less efficient and resource consuming than communication through a network of intermediary satellites in low Earth orbits.
The satellites are to be used as intermediaries in the sense that ground station A can be synced to a satellite and then the same satellite could be synced to the ground station B. If all three clocks involved are relatively stable within the time this protocol is executed then the clocks at A and B can be successfully synced in this way. 
Consider now the elementary link/task of this protocol, which is to sync a ground station to a satellite. The roles of Alice and Bob are thus taken up by the ground station and satellite respectively.
The main difference between the task in the ground based case and this satellite based case is the relative motion between the clocks that need to be synchronized. Satellite velocities in LEOs can be of the order of a few kilometers per second with respect to the ground stations. Throughout this work we will assume circular orbits for satellites (and ground stations). Therefore, the link distance which is now a function of time, changes by $\Delta d_{ab}$ in time t, given by:
\begin{equation}
\label{eqn:11}
	\Delta d_{ab}(t) = v_{\rm rel}^{\rm rad} t \, ,
\end{equation}
where $v_{\rm rel}^{\rm rad}$ is the relative radial velocity of the satellite w.r.t. the ground station and is given in terms of the position vectors $\mathbf{r_{sat}}$, $\mathbf{r_{gs}}$ (w.r.t. the center of the Earth), and angular velocities $\mathbf{\omega_{sat}}$, $\mathbf{\omega_{gs}}$ of the satellite and ground station respectively:
\begin{equation}
\label{eqn:12}
	v_{\rm rel}^{\rm rad} = (\mathbf{r_{sat}} \times \mathbf{\omega_{sat}} -  \mathbf{r_{gs}} \times \mathbf{\omega_{gs}}) . \frac{(\mathbf{r_{sat}} - \mathbf{r_{gs}})}{\vert (\mathbf{r_{sat}} - \mathbf{r_{gs}}) \vert}  \, .
\end{equation}
Thus, the time of travel for a photon moving between A and B also becomes a function of time, given by:  
\begin{equation}
\label{eqn:13}
    t_{ab}(t) = \frac{d_{ab}(t)}{c},
\end{equation}
where $c$ is the speed of light in vacuum (we are ignoring the small variation in the speed of light when it enters the atmosphere, since the thickness of the atmosphere is $\approx 10 \rm{km} \ll d_{ab}$). Assuming a small enough acquisition time $t_{\rm Acq}$ such that the relative radial velocity does not change appreciably within the precision levels of the protocol and also any relativistic effects can be ignored, during the interval $t_{\rm Acq}$ the time of travel changes by $\Delta t_{ab}$ given by \footnote{For the GPS, the combined special and general relativistic effects are $\approx 0.5 ns/s$\cite{Zhu_GPS_relativity}. We will see in Section \ref{sec:4} that the optimal acquisition time is in the sub-millisecond range, which means relativistic effects only become important when the precision required is in the sub-picosecond range}
\begin{equation}
\label{eqn:15}
    \Delta t_{ab} = \frac{\Delta d_{ab}(t_{\rm Acq})}{c},
\end{equation}
When looked at in conjugation with Equations (\ref{eqn:6}) and (\ref{eqn:8}), this tells us that the correlation function no longer has a single sharp peak. New peaks keep on appearing adjacent to the first peaked
as the time of travel changes with time, forming a broad band. The correlation function has, therefore, multiple adjacent peaks given by the modified form of Equation (\ref{eqn:6}): 
\begin{equation}
\label{eqn:16}
    \tau_{ab}(t) = t_{ab}(t)  + \Delta \, .
\end{equation}
Now, the pertinent question is: how can a unique and accurate value of time offset $\Delta$ be evaluated, up to a certain precision, given these new features of the correlation function $C_{AB}$?

In the rest of the paper we take a two pronged approach. First of all,  we define and analytically calculate the signal to noise ratio (SNR) of the time stamp correlation functions, and define from it the achievable precision. Secondly, we perform Monte Carlo simulations of the QCS protocol and,  finally, we combine the intuition gained by the Monte Carlo simulations and analytical results to perform numerical simulations for a full QCS network to calculate  figures of merit, taking into account the motion of multiple satellites and ground stations and the dynamics of lossy channels between them. 
\begin{figure}
    \centering
    \includegraphics[width = 0.7\linewidth]{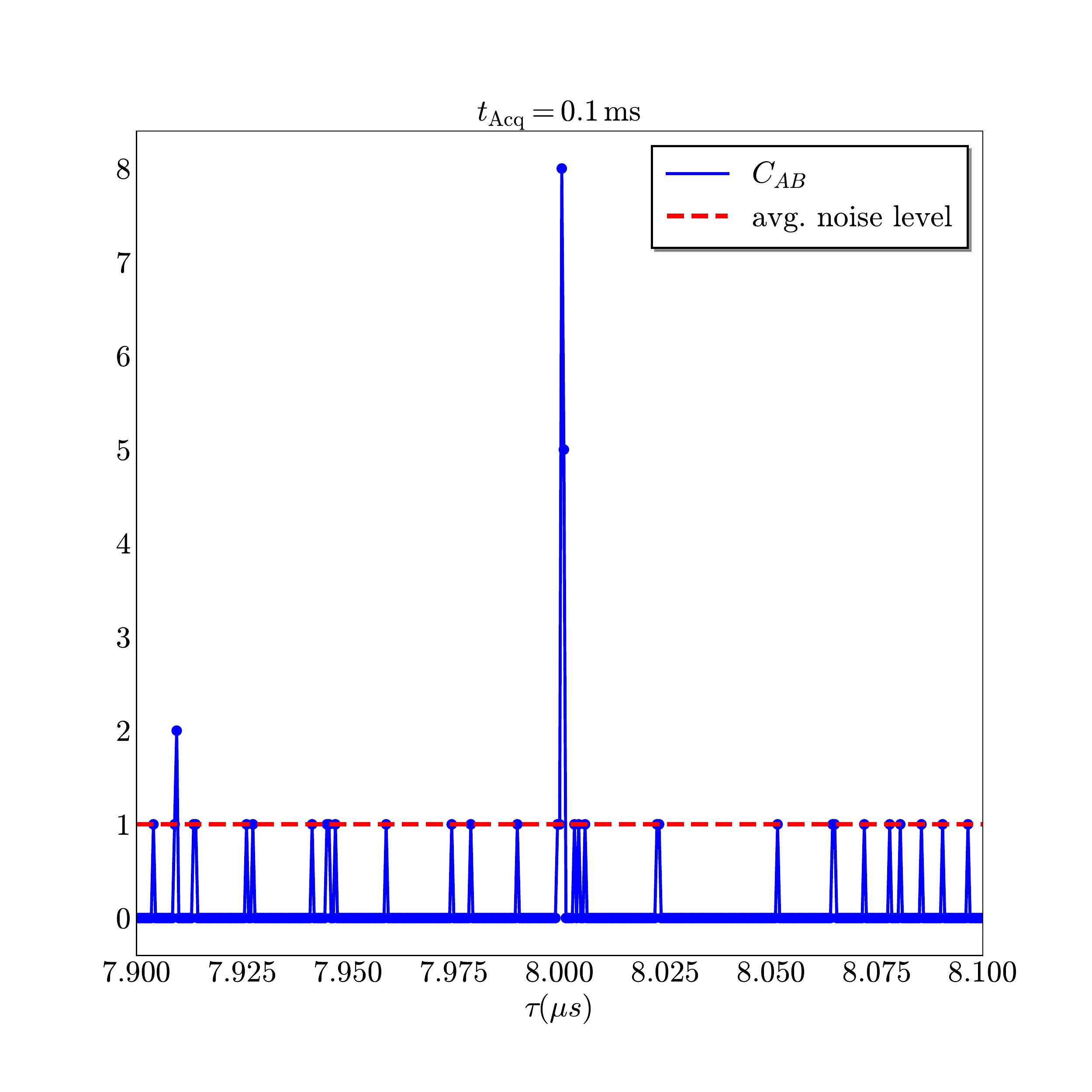}
    \includegraphics[width = 0.7\linewidth]{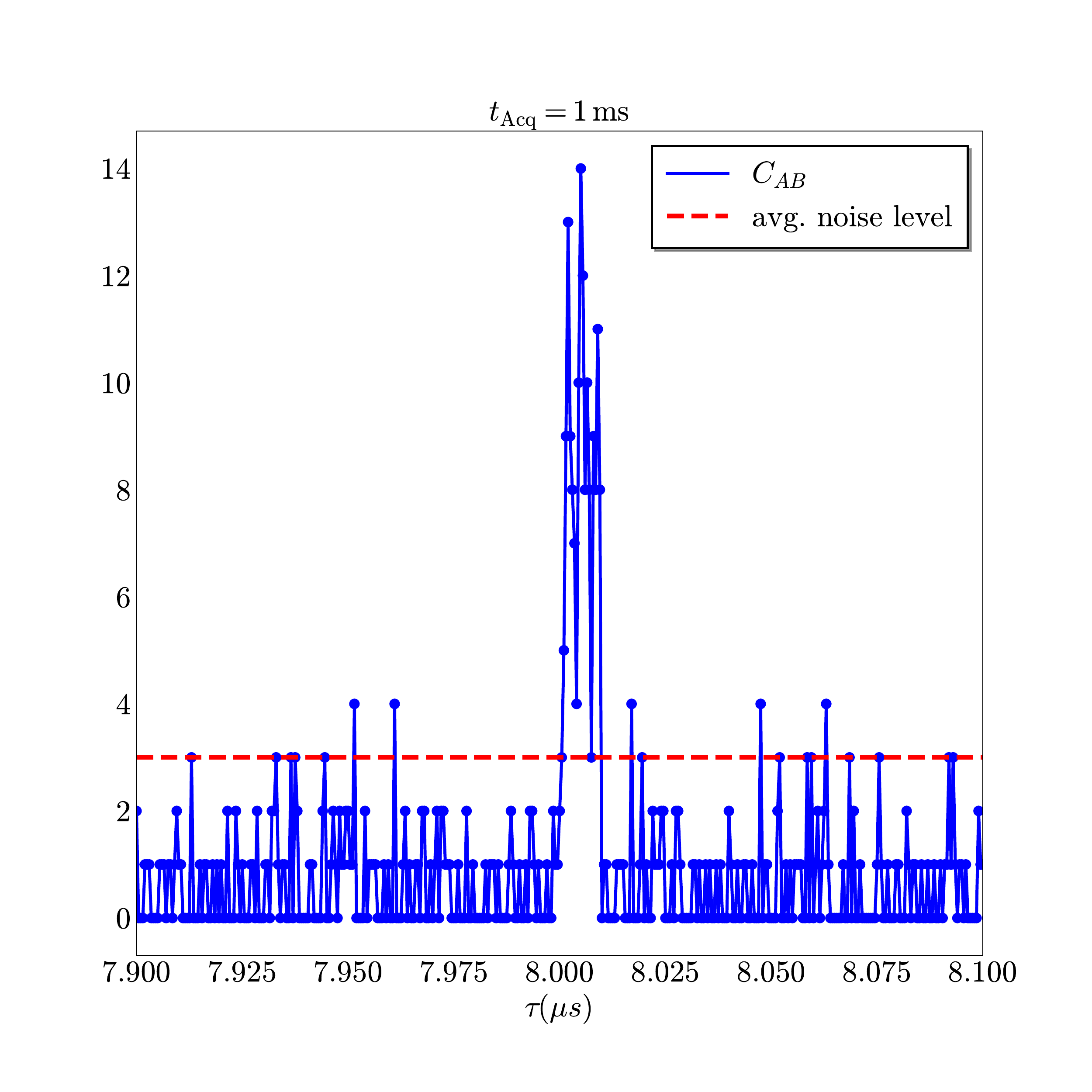}
    \includegraphics[width = 0.7\linewidth]{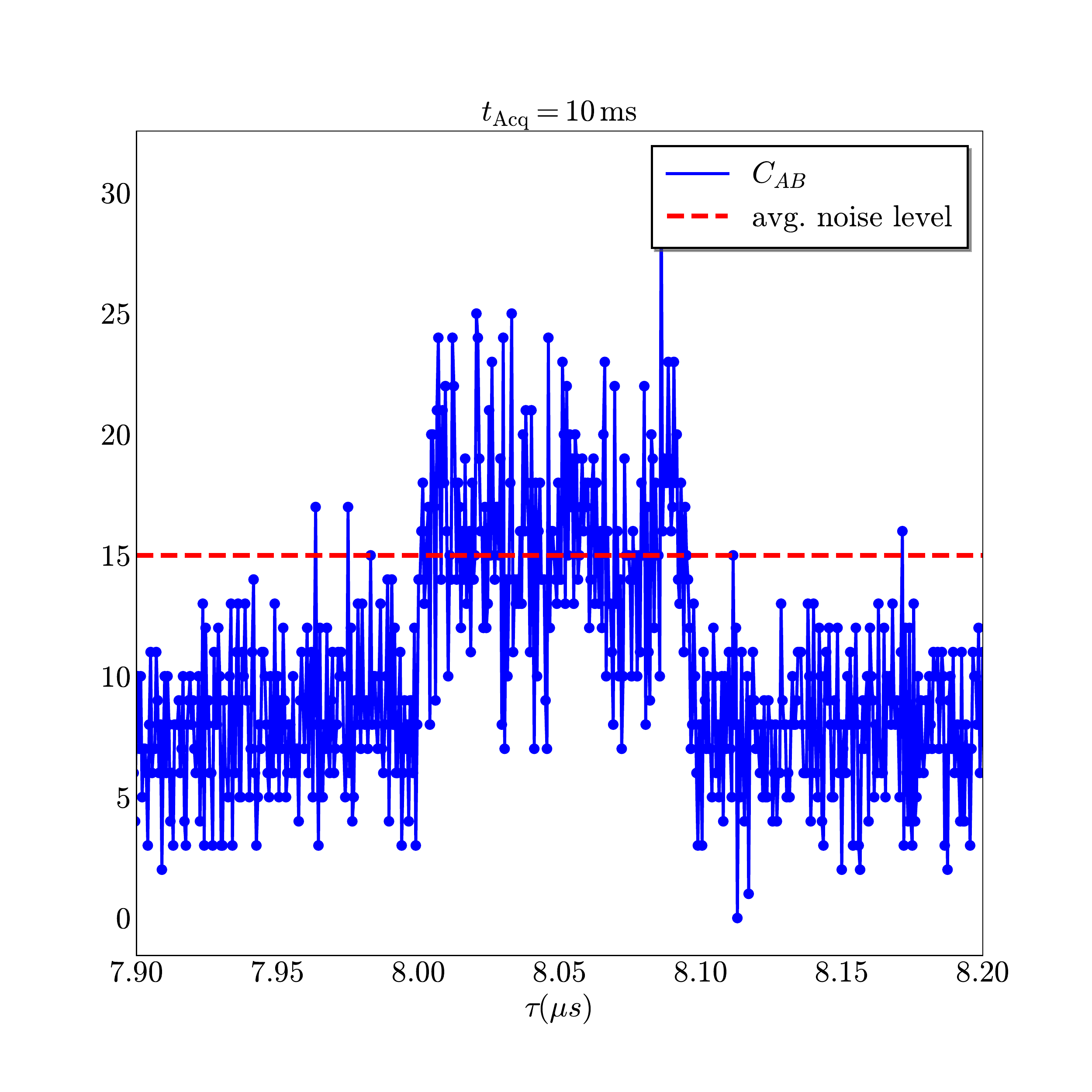}
    \caption{Here, we show the timestamp cross-correlation functions obtained via a Monte Carlo simulation of the QCS protocol between two clocks in relative motion (one on a satellite and one on ground station). As the acquisition time increases from $0.1 \mu s$ in the topmost plot to $1 \mu s$ in the center and finally $10 \mu s$ on the bottom, multiple peaks start appearing due change in time of travel, i.e, the peak broadens. Further, the gap between the these peaks and average noise level (indicated by the red dotted line as a guide to the eye) also reduces with increasing $t_{\rm Acq}$, reducing the SNR of the true peaks. The background noise rate is assumed to be $10^6$ photons/s, the source rate is $10^7$ ebits/s. The channel loss is around 20 dB (uplink loss between a ground station and a polar satellite at 500 km altitude, varies slightly as the satellite and ground station both move during the acquisition time). The least count of the timestamps is 0.5 ns.}
    \label{fig:cc_motion}
\end{figure}

\section{Behaviour of the correlation functions}
\label{3}

The success and  precision of the protocol depends on the sharpness of the correlation function peak. This sharpness can be quantified through the signal to noise ratio (SNR), defined as the quotient of the height of the peak above the average noise level (spurious peaks) and the standard deviation of the noise (see Appendix \ref{sec:appendixA}).  
For simplicity of notation, we now drop the indices A and B, since the analysis applies equally to both correlation functions. 

The height of the peak in $C$ if the two parties were stationary is given by
\begin{equation}
\label{eqn:17}
    C(\tau_{\rm max}) = R \, \eta\,  t_{\rm Acq} \, ,
\end{equation}
where $R$ is the source rate, at which the entangled photons are produced, and $\eta$ is the probability of detection at the receiver's end. $\eta$ includes the efficiency of the detectors as well as the 
effects of losses in the propagation. Hence, $\eta$ depends on the distance between the two parties. More specifically, the transmissivity $\eta$ has three factors \cite{LSU_satsim}:
\begin{itemize}
    \item Free space transmittance: This includes effects of beam broadening and finite apertures of the transmitting and receiving telescopes. The free space transmittance follows an inverse square fall off law with the link distance and is given by:
    \begin{eqnarray}
    {\eta }_{\text{fs}}(L)=1-\exp \left(-\frac{2{R}^{2}}{(w(L))^{2}}\right),
    \label{eqn:fs_loss}
    \end{eqnarray}
    where $R$ is the receiving telescope's radius and  $w(L)$ is the beam waist after traversing the link distance $L$. The latter is given by:
    \begin{eqnarray}
    w(L):={w}_{0}\sqrt{1+{\left(\frac{L}{{L}_{{\rm{R}}}}\right)}^{2}},
    \label{eqn:fs_loss2}
    \end{eqnarray}
    and the Rayleigh range ${L}_{{\rm{R}}}:=\pi {w}_{0}^{2}{\lambda }^{-1}$. $\lambda$ is the source wavelength, and $w_0$ is the initial beam-waist radius. $w_0 = 0.8r$, where $r$ is the transmitting telescope's radius and we use a 80\% fill fraction.
    The above relation holds under the approximation of the beam as a zeroth order Gaussian spatial mode.
    \item Atmospheric transmittance: Here we assume the atmosphere to be a homogeneous absorptive medium following the Beer-Lambert's Law 
    (transmittance falls exponentially with increasing distance that is travelled through the medium). Thus, the atmospheric transmittance is given by
    \begin{eqnarray}
        {\eta }_{{\rm{atm}}}(L,h)=\left\{\begin{array}{*{20}{l}}{\left({\eta }_{\rm{atm}}^{\rm{zen}}\right)}^{{{\rm{sec}}}\,\zeta}, & {\rm{if}}\,-\frac{\pi }{2}<\zeta <\frac{\pi }{2},\\ 0, & {\rm{if}}\,| \zeta | \ge \frac{\pi }{2},\end{array}\right.
        \label{eqn:atm_loss}
    \end{eqnarray}
    $\eta_{\rm{atm}}^{zen}$ is the atmospheric transmittance at zenith ($\zeta= 0$) and the zenith angle $\zeta$ for circular orbits is given by:
    \begin{eqnarray}
        \cos \zeta =\frac{h}{L}-\frac{1}{2}\frac{{L}^{2}-{h}^{2}}{{R}_{E}L},
        \label{eqn:atm_loss2}
    \end{eqnarray}
    where $R_E$ is the radius of Earth.
    \item Detector inefficiencies: $\kappa_{\rm{sat}}$ and $\kappa_{\rm{grd}}$ are efficiencies of the detectors at the satellite and ground station, respectively.
\end{itemize}

Thus, the total efficiency of the channel (uplink or downlink) is given by:
$\eta = \eta_{\rm{atm}} \eta_{\rm{fs}} \kappa_{\rm{sat}} \kappa_{\rm{grd}}$.

Finally, it is important to note that when relative motion is not considered, Equation \eqref{eqn:17} shows that the height $C(\tau_{\rm max})$ keeps on increasing with the acquisition time $t_{\rm Acq}$.

On the other hand, this is no longer true
in the case with relative motion. The height of a peak only rises till the time of travel $t_{ab}$ changes by unit precision. Let $t_{\rm bin}$ be the maximum achievable precision (least count or bin size for time stamps in the simulation). In that case, the peak in $C(\tau_{\rm max})$ due to correlated photons rises only until the acquisition time reaches the value $t_{\rm Acq}^{\rm opt}$ given by the following condition:
\begin{equation}
\label{eqn:18}
    \Delta t_{ab}(t_{\rm Acq}^{\rm opt}) = t_{\rm bin} \, .
\end{equation}
After $t_{\rm Acq}^{\rm opt}$, a new peaked starts raising, adjacent to the first peak. 
(The reason for referring to this as the optimal acquisition time will become obvious  later in this section.)
Using Equations (\ref{eqn:11}) and (\ref{eqn:15}) in (\ref{eqn:18}), we can obtain the value of  $t_{\rm Acq}^{\rm opt}$ in terms of other parameters:
\begin{equation}
\label{eqn:19}
    t_{\rm Acq}^{\rm opt} = \mathcal{K} \, t_{\rm bin} \, ,
\end{equation}
where the geometrical factor $\mathcal{K}$ is given by
$\mathcal{K} = \frac{c}{v_{\rm rel}^{\rm rad}}$;  $\mathcal{K}$ depends, via $v_{\rm rel}$, on the distance between A and B; this will be important shortly. 

Therefore, $t_{\rm Acq}^{\rm opt}$ quantifies the acquisition time at which the position of the peak in the correlation function  shifts by one unit precision $t_{\rm bin}$, from $\tau_{\rm max}$ to $\tau_{\rm max}+t_{\rm bin}$.  
In turn, this determines the maximum height of any peak generated by true correlations, which is given by:
\begin{equation}
\label{eqn:20}
    C(\tau_{\rm max}) = R\, \,  \eta \mathcal{K} \, t_{\rm bin} \, .
\end{equation} 
Clearly, the SNR will have different behaviours for the two regimes defined by $t_{\rm Acq} < t_{\rm Acq}^{\rm opt}$ and $t_{\rm Acq} > t_{\rm Acq}^{\rm opt}$. We derive the SNR for these two regimes in the Appendix \ref{sec:appendixA}. Assuming that photons from background noise appear at a rate $R_{\rm bkg}$, we obtain  the following expressions for the SNR:
\begin{itemize}
    \item for $t_{\rm Acq} < t_{\rm Acq}^{\rm opt}$,
    \begin{equation}
    \label{eqn:27}
    \rm SNR \approx \sqrt{\frac{\eta}{t_{\rm bin}(1 + R_{\rm bkg}/R\eta)}} \sqrt{t_{\rm Acq}} \, .
    \end{equation}
    Therefore, for $t_{\rm Acq} < t_{\rm Acq}^{\rm opt}$ the SNR increases with the acquisition time as  $\sqrt{t_{\rm Acq}}$. This is due to the fact that, for $t_{\rm Acq} < t_{\rm Acq}^{\rm opt}$, the height of the peak grows faster than the noise.
    
    \item for $t_{\rm Acq} > t_{\rm Acq}^{\rm opt}$,
    \begin{equation}
    \label{eqn:29}
        \rm SNR \approx \mathcal{K} \sqrt{\frac{\eta t_{\rm bin}}{\sqrt{(1 + R_{\rm bkg}/R\eta)}}} \frac{1}{\sqrt{t_{\rm Acq}}}\, .
    \end{equation}
    Therefore, for $t_{\rm Acq} > t_{\rm Acq}^{\rm opt}$, the SNR decreases with increasing $t_{\rm Acq}$. Again, this is expected because in this regime the height of the peak no longer increases, while noise keeps accumulating.
    
\end{itemize}

This analysis clearly indicates that the SNR is maximum for $t_{\rm Acq} = t_{\rm Acq}^{\rm opt}$. Its maximum value is
\begin{equation}
\label{eqn:30}
    \rm SNR_{max} \approx \sqrt{\frac{\eta\mathcal{K}}{(1 + R_{\rm bkg}/R\eta)}} \, .
\end{equation}

On the other hand, in order to get a clearly defined peak recognizable out of the noise,  the SNR must be greater than a threshold value $\rm SNR_{th}$. The higher our choise for  this threshold is, the lower is the probability of misidentifying a peak. For a choice if $\rm SNR_{th}$, the condition for a peak to be identifiable is:
\begin{eqnarray}
\label{eqn:31}
    \rm SNR_{max} \geq \rm SNR_{th}, \ \  \rm {or, \ equivalently} \\ \nonumber
    \sqrt{\frac{\eta\mathcal{K}}{(1 + R_{\rm bkg}/R\eta)}} \geq \rm SNR_{th} \, .
\end{eqnarray}
For a given level of background noise $R_{\rm bkg}$, the above condition provides a constraint on the relative position of the ground station w.r.t. the satellite, setting a scale for the service area of the satellite (region on Earth where it can provide synchronization). At the same time, it also puts bounds on the levels of tolerable noise for the sync to be successful for a given  configuration of the satellite and ground station positions.
Note that this condition does not interfere with sync precision of the protocol, since it is independent of $t_{\rm bin}$. Further, for simplicity, we have ignored here the effect of detector jitter (which leads to loss of time-stamp precision) on the sync precision. We refer the reader to Appendix \ref{sec:appendixC} for an analysis of these effects.

Apart from the SNR condition just discussed, for a peak in the correlation function to be visible, we must impose  an extra requirement on the absolute of the peak's height. This condition comes from the simple fact that one must collect at least a few photons with the correct correlations within the acquisition time; i.e., the peak must have a minimum height. Since the ebit generation through an SPDC source and the losses are both  random processes, by setting a minimum threshold for the mean number of ebits detected, we can ensure that at least a few ebits get shared between the ground station and satellite with high probability within the acquisition window.  Let $N_{\rm min}$ be that photon number threshold; if $R$ is the source rate and $\eta$ the probability of detection, we obtain the following constraint 
\begin{equation}
\label{eqn:34}
    R\, \eta \, t_{\rm Acq}^{\rm opt} \geq N_{\rm min} \, .
\end{equation}
Using Equation (\ref{eqn:19}), (\ref{eqn:34}) can be rewritten as
\begin{equation}
\label{eqn:35}
    t_{\rm bin} \geq \frac{N_{\rm min}}{R\eta \mathcal{K}} \, .
\end{equation}
Unlike Equation (\ref{eqn:31}), this constraint involves the sync precision $t_{\rm bin}$. For a required level of precision, this sets a constraint on the product of geometrical factors $\eta$ and $\mathcal{K}$, and thus it also defines a scale for the service area of the satellite, via the equation:
\begin{equation}
\label{eqn:36}
    \eta(\vert \mathbf{r_{\rm gs}} - \mathbf{r_{\rm sat}}\vert) \mathcal{K}(\mathbf{r_{\rm gs}}, \mathbf{r_{\rm sat}}) \geq \frac{N_{\rm min}}{Rt_{\rm bin}} \, .
\end{equation}
For a reasonable value of $\rm SNR_{th}$ and as long $R_{bkg}$ is not very large (compared to $R$), Equation (\ref{eqn:36}) is a stronger condition than Equation (\ref{eqn:31}), and hence it determines the serviceable region (later we will call this the precision ``shadow'' of the satellite). This is the Equation we will use later to determine the size of sync networks.
Only for very high noise levels, the $\rm SNR$ condition may become the dominant constraint on the network size. 
Finally, we stress that Equation \eqref{eqn:35} puts a limit on the achievable precision only when the acquisition time is set to its optimal values. This assumes working at the maximum achievable $\rm SNR$. The working precision of the protocol could be higher if we loosen this constraint. For more details and an operational view of the protocol see Appendix \ref{sec:appendixB}.  
\subsection{A simple example}
\label{subsec:simple_example}
\begin{figure}
    \centering
    \includegraphics[width = \linewidth]{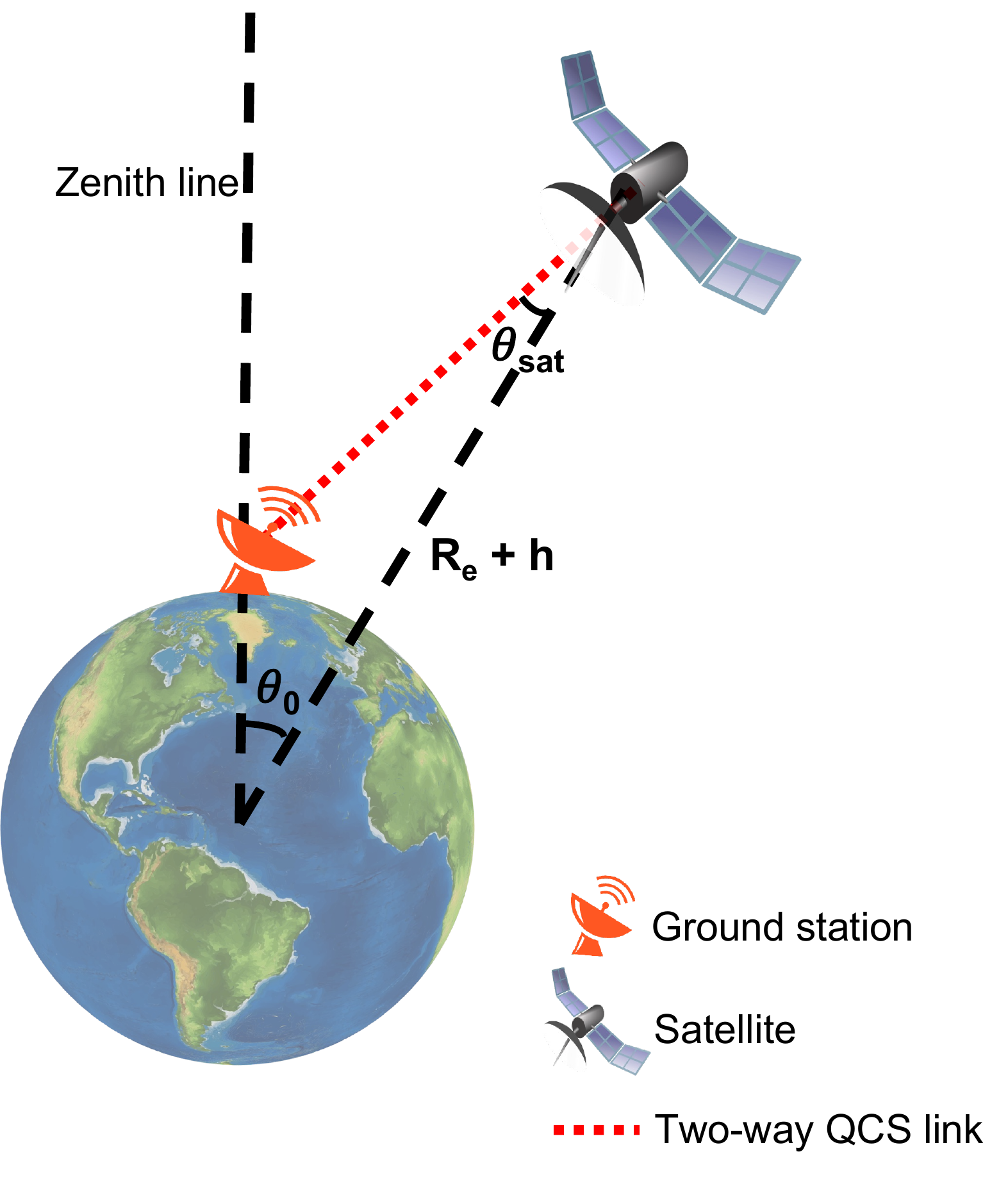}
    \caption{A simple configuration for the QCS protocol. A satellite orbiting in the plane formed by the two lines
    joining the center of the Earth to the ground station (zenith line) and to the satellite respectively.}
    \label{fig:figure_sat_orbit}
\end{figure}

In order to build some intuition, let us begin by looking at a simple example.  Consider a ground station at a random location on Earth's surface. For illustrative purposes, in this example we assume  that  the line joining the center of the Earth to the ground station (zenith line) lie in the plane defined by the satellite's orbit (Figure \ref{fig:figure_sat_orbit}). The satellite is at altitude $h$ above the Earth's surface. We assume circular orbit for the satellite and also for the ground station (ignoring any topographical features and the non-spherical geometry of the Earth). $R_e$ is the Earth's radius. Say the protocol starts when the satellite makes an angle $\theta_0$ with the zenith. Expressions for $\mathcal{K}$ and $\rm SNR$ can be calculated analytically for this case. 
\begin{equation}
\label{eqn:37}
    \mathcal{K} = \frac{2c(R_e^2 + (R_e+h)^2 -2R_e(R_e+h)\cos\theta_0)^{1/2}}{2R_e(R_e+h)\omega\sin\theta_0}
\end{equation}

In Equation (\ref{eqn:35}), on the RHS $\eta$ and $\mathcal{K}$ depend on the geometrical factors (h, $\theta_0$). Hence, Equation (\ref{eqn:35}) can be interpreted as the best precision (smallest $t_{\rm bin}$) that can be achieved at a given relative position of the ground station and the satellite. At the same time, from a different point of view, it can be considered as a limit on the relative angular separation between the satellite and ground station up to which a certain precision can be achieved. For example, given a fixed satellite altitude h, the protocol cannot be successfully conducted beyond an angle $\theta^{crit}_0$ at a required precision of $t_{\rm bin}$, where $\theta^{crit}_0$ is determined from
\begin{equation}
\label{eqn:38}
    \eta(\theta_0^{\rm crit}, h) \mathcal{K}(\theta_0^{\rm crit}, h) \geq \frac{N_{\rm min}}{Rt_{\rm bin}}.
\end{equation}

\begin{figure}
    \centering
    \includegraphics[width =  \linewidth]{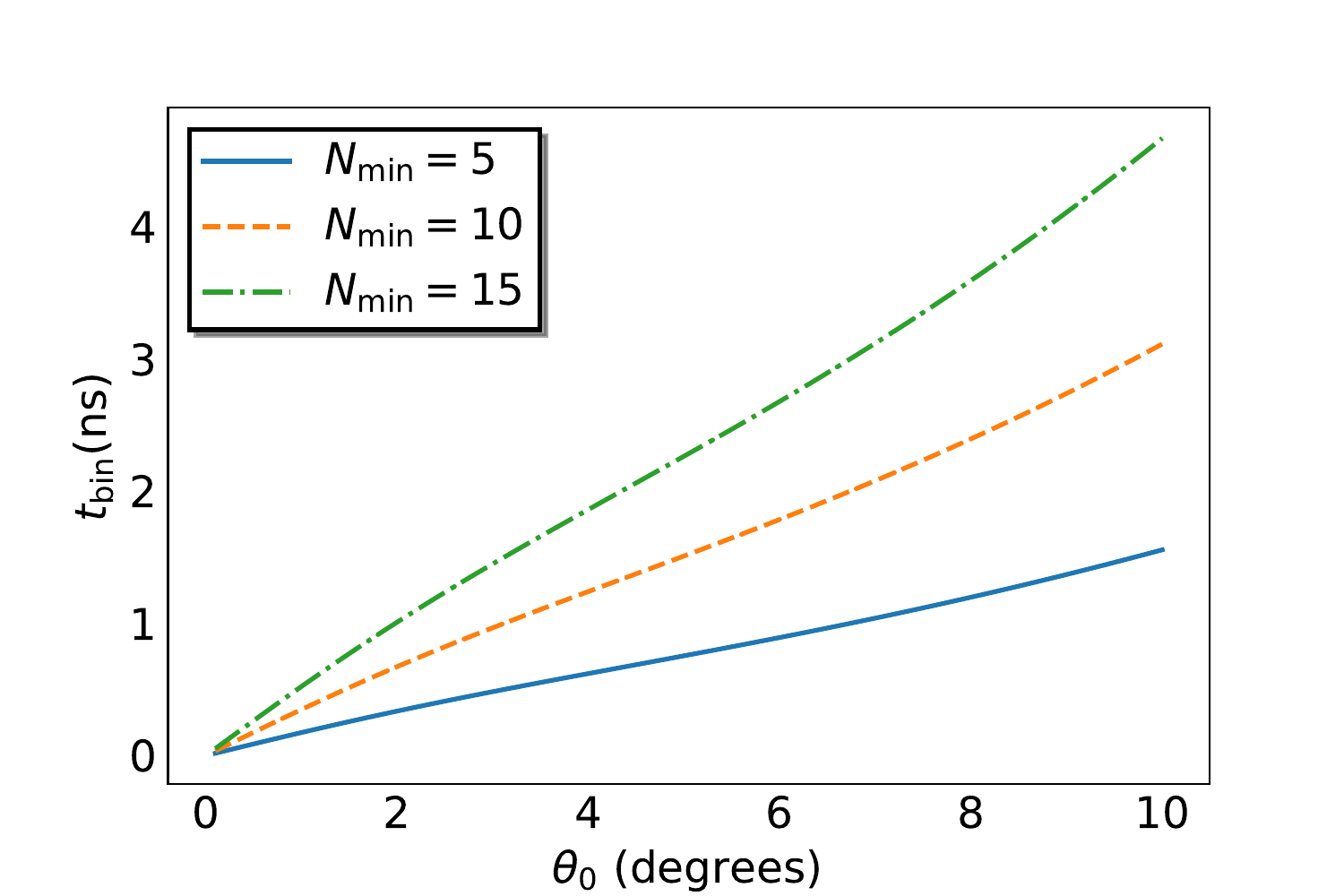}
    \includegraphics[width =  \linewidth]{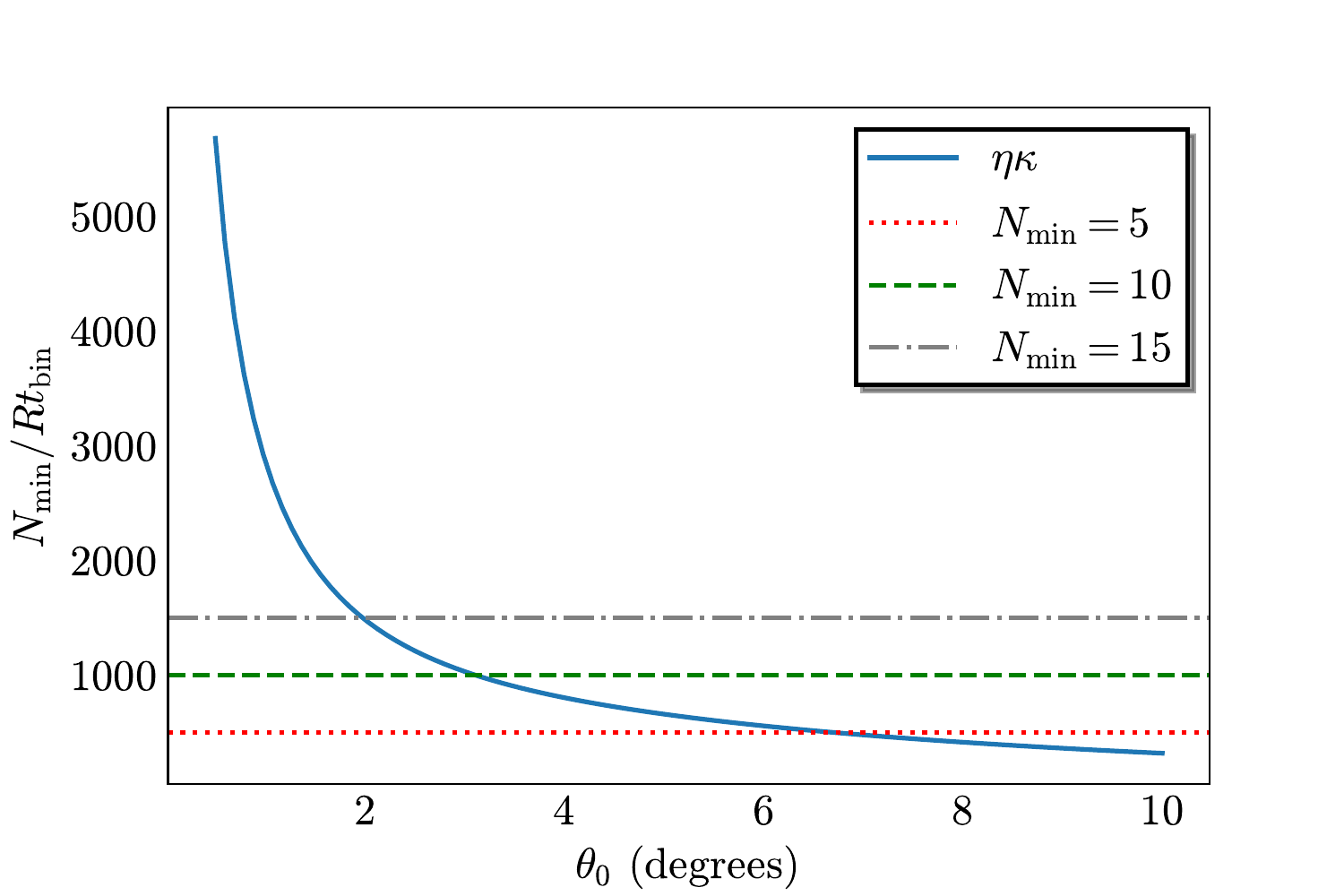}
    \caption{Two ways of looking at the achievable precision - (Top) Maximum achievable precision as a function of the gs-sat angular separation and (Bottom) maximum separation allowed (serviceable area) for a required level of precision. $\theta_0^{\rm crit}$, the critical angular separations for 1 ns precision are given by the intersection of the blue continuous curve with the horizontal lines, corresponding  to different choices of $N_{\rm min}$}
    \label{fig:figure2}
\end{figure}

\begin{figure}
    \centering
    \includegraphics[width = \linewidth]{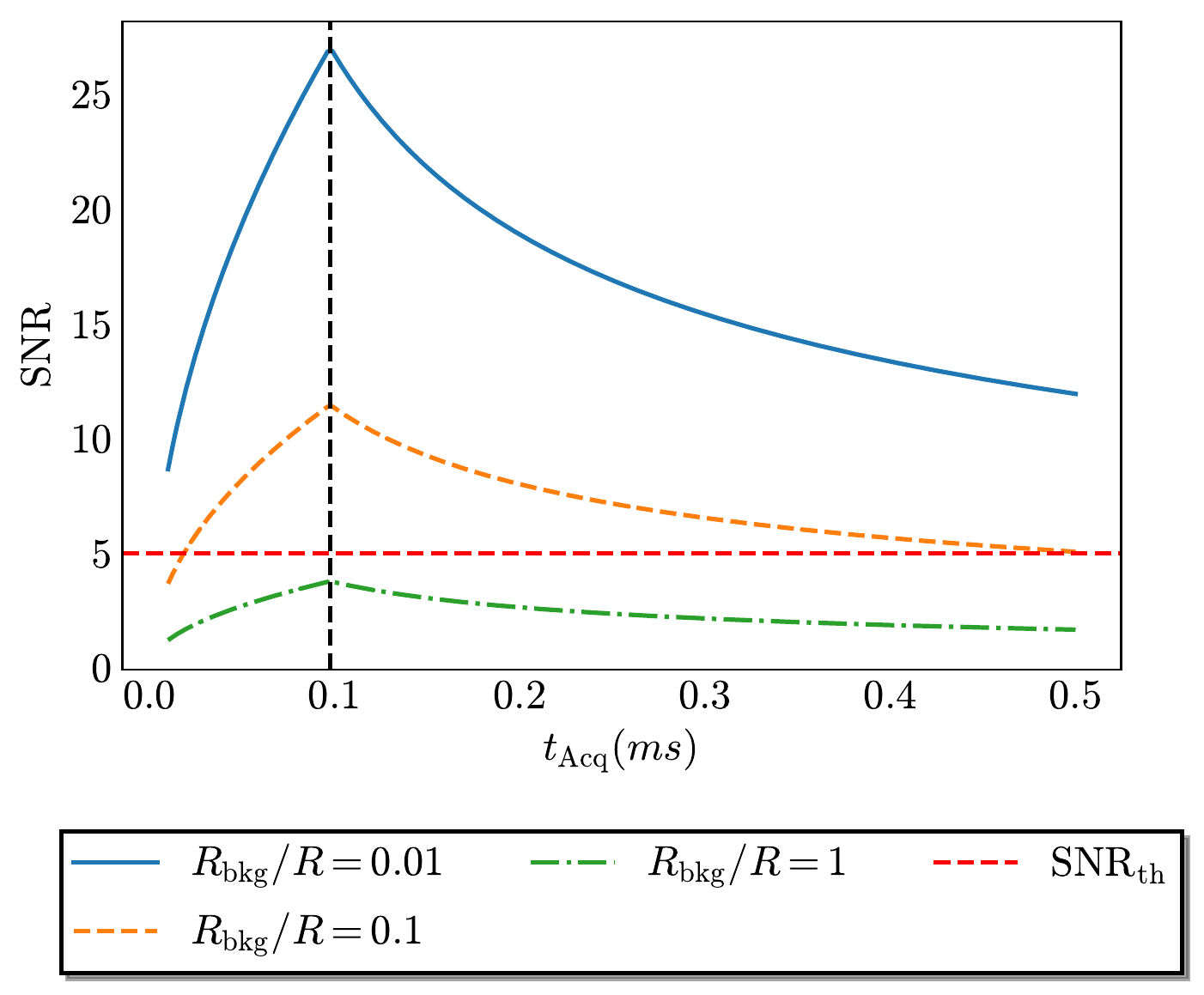}
    \includegraphics[width = \linewidth]{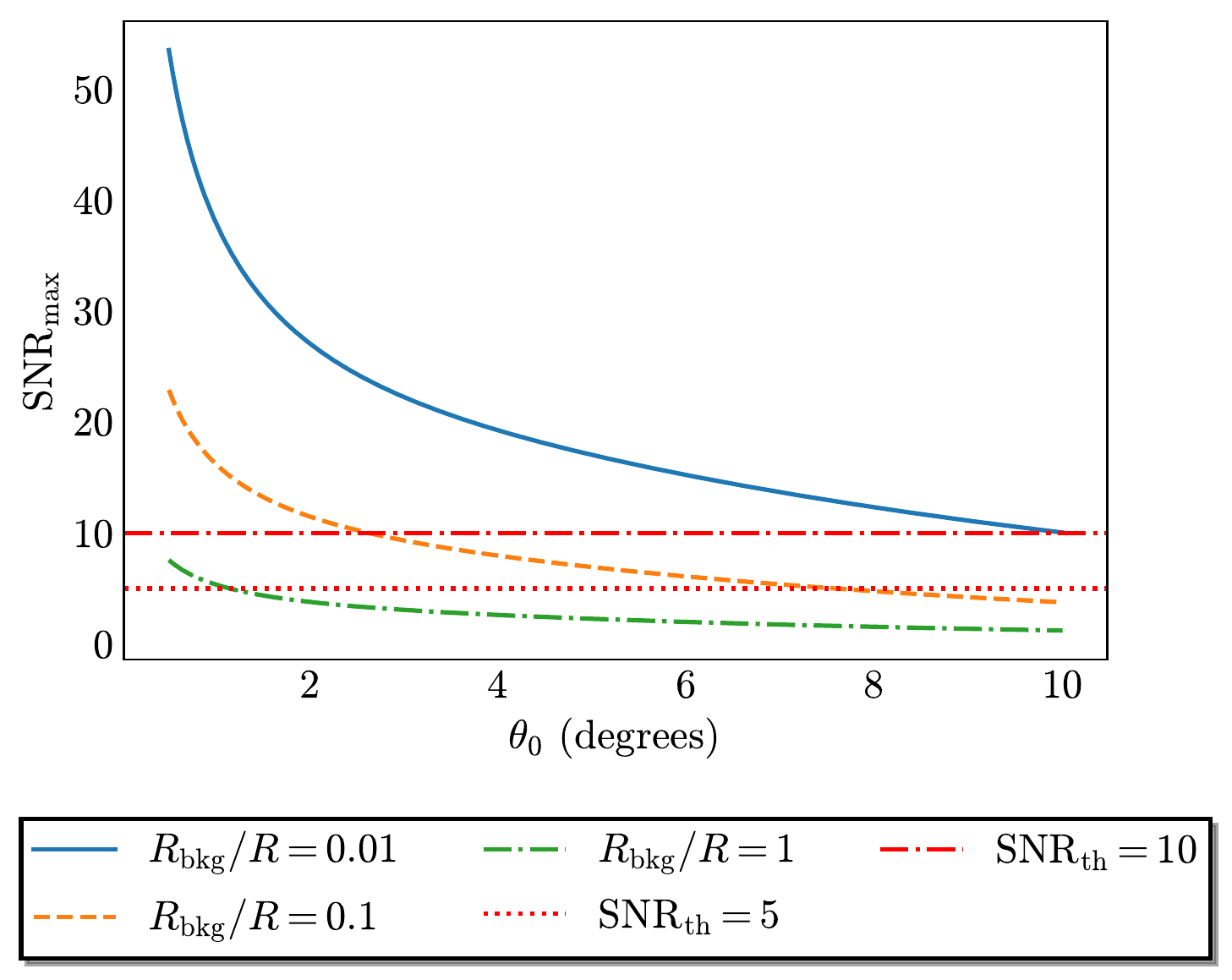}
    \caption{(Top) SNR v/s acquisition time for different noise levels (angular separation $\theta_0 = 2^\circ$), clearly maximum SNR is achieved at $t_{\rm Acq} = t_{\rm Acq}^{\rm opt}$ (black dotted line) as expected according to Equation (\ref{eqn:30}). Also, at higher noise background levels SNR drops below the threshold of $\rm SNR_{th} = 5$. (Bottom) SNR v/s angular separation for different background rates, clearly $\rm SNR>\rm SNR_{max}$ for even very large angular separations when $R_{\rm bkg}\ll R$. On the other hand when $R_{\rm bkg} \approx R$, $\rm SNR$ drops below the threshold of $\rm SNR_{th} = 5$ for a much smaller angular separation between the satellite and ground station}
    \label{fig:figure3}
\end{figure}

Figure \ref{fig:figure2} illustrates these two viewpoints of looking at the precision. Also, Figure \ref{fig:figure3} shows the dependence of SNR on background rates, acquisition time and angular separation.  
As a quantitative example, for h = 500 km and $N_{\rm min} = 5$ a precision of $t_{\rm bin}=1$ ns can be achieved up to $\theta_0 = \theta_0^{\rm crit} \approx 3^0$, with a source rate of $10^7 ebits/s$. This angle, when translated to the coverage angle of the satellite, is $\theta_{sat} \approx 34^0$. That is, all ground stations falling in this angular region will be able to sync with the satellite at 1 ns precision.

As was mentioned in the previous section, even within these limits set by $N_{\rm min}$, the background noise dictates the SNR and hence it must be low enough such that Equation (\ref{eqn:31}) is also satisfied. We show in the subsequent section, using simulation results and also from analytical results obtained in this section, that Equation (\ref{eqn:31}) is comfortably satisfied for $\rm SNR_{th} \approx 5$, whenever the background rate is around 2 to 5 times lower than the source rate. For details, see Figures \ref{fig:figure3} (Bottom) and \ref{fig:figure6}.

\section{Simulation methods: Synchronization of a satellite and a single ground station}
\label{sec:3}
The simulation of this scenario  consists mainly of three parts: (1) simulating the photon generation events; (2) simulating the dynamics of the satellite and ground station and; (3) simulating the lossy quantum communication channel between the two parties.
We use standard Monte-Carlo techniques to randomly generate photon time stamps at the source, assuming a constant rate $R$ of entangled bits (e-bit) production. We also generate background photons independently at the ground station and satellite at constant rates given by $R_{\rm bkg}^{gs}$ and $R_{\rm bkg}^{sat}$, respectively (it would be straightforward to add random variability to these rates, mimicking fluctuations in the noise).
Further, at every time step we update the positions of the ground station and the satellite to evaluate the link distance and time of travel for the photons that are produced in that time step. The link distance enters as an input into the efficiency $\eta$ of the free space communication channel. Further $\theta_0$ which is the angle between the ground station zenith and the satellite is calculated from the position vectors. 
We take into account the various losses by calculating effective efficiency factors both for the uplink (ground station to satellite) and downlink (satellite to ground station) $\eta_{(up)}$ and $\eta_{(dwn)}$ respectively. $\eta_{(up)}$ is the probability that a photon generated out of the SPDC source at the ground station will lead to a double detection event, i.e., one partner will be detected at A locally and the other at B after travelling through space and the atmosphere. Similarly for $\eta_{(down)}$. 

We model the satellite-ground quantum communication channel as follows. For concreteness, let us focus on a downlink channel. The following will hold similarly for an uplink channel. We assume clear skies and approximate the downlink channel as only lossy (background noise is accounted for at the detectors). That is, photons are either transmitted through the channel or lost in transmission. The dominant sources of loss are (1) beam spreading (free-space diffraction loss), (2) atmospheric absorption/scattering, and (3) non-ideal photodetectors on the satellite and on the ground. We characterize these loss mechanisms by their transmittance values, which is the fraction of the received optical power to the transmitted power (which is also equal to the probability to transmit/detect a single photon). Let these transmittance values be, respectively,
\begin{equation*}
  \eta^{(dwn)}_{fs}(L), \quad \eta^{(dwn)}_{atm}(L), \quad \kappa_{sat}, \quad \kappa_{gs}\, ,
\end{equation*}
where the superscripts refer to the downlink, $fs$ refers to free-space diffraction loss, $atm$ to atmospheric loss, $L$ is the link distance (physical distance) between satellite and receiver (which in turn depends on the satellite altitude, $h$, position of the satellite in its orbit, and position of the ground station on Earth's surface), $\kappa$ denotes non-ideal detection efficiencies for the onboard satellite detector array ($sat$) as well as the detector array at the ground station ($gs$), and all transmittance values are less than or equal to $1$. Simple analytic formulae are used to estimate the free-space and atmospheric transmittance values in accordance with \cite{LSU_satsim}. See Equations \ref{eqn:fs_loss}--\ref{eqn:atm_loss2}.
Therefore the overall efficiencies are given as:
\begin{eqnarray}
\label{eqn:39-40}
    \eta_{(dwn)} &=& \eta^{(dwn)}_{fs}\eta^{(dwn)}_{atm}\kappa_{sat}\kappa_{gs}\, , \\
    \eta_{(up)} &=& \eta^{(up)}_{fs}\eta^{(up)}_{atm}\kappa_{sat}\kappa_{gs}\, . 
\end{eqnarray}
Finally, following standard Monte Carlo techniques, our code works by registering a joint detection event (two photons from the same pair detected at A and B) every time two random numbers $r$, $r'$ independently generated from a uniform distributions $\in [0,1]$ satisfy the following conditions in the same time step: 
\begin{eqnarray}
    r &<& Rt_{\rm bin} \label{eqn:41} \\
    r' &<& \eta \label{eqn:42}
    \label{eqn:43}
\end{eqnarray}
For the noise photons to be detected at A or B, a condition analogous to Equation (\ref{eqn:41}), $r'' < R_{\rm bkg}t_{\rm bin}$ is used, where, $r''$ is also a random number chosen independently from a uniform distribution $\in [0,1]$.
Further, a condition that the satellite be above the ground station's horizon is already imposed via Equation (\ref{eqn:atm_loss}).

Once the photons are generated and time stamped, the correlation functions are calculated by counting the number of photons that are generated a time interval $\tau$  apart from each other. This is the value of the correlation function $C(\tau)$.

For larger configurations involving multiple satellites and ground stations, and for longer simulation periods ($\approx 1$ day), it becomes computationally expensive to run the Monte Carlo simulation at the sub-nanosecond time resolution. It is also not necessary to do so, since all the information about the success and quality of the time synchronization can be evaluated from Equations  (\ref{eqn:31}) and (\ref{eqn:35}). See Figure \ref{fig:figure7} for an illustration.

\section{Simulation results}
\label{sec:4}
Now we discuss the results obtained from the simulation described in the preceding section. First, in Section \ref{subsec:4.1}, we look at the simplest scenario of a single  ground station located within the orbital plane of the satellite (Figure \ref{fig:figure_sat_orbit}),  in order to quantitatively verify the analytical results obtained in Section \ref{3}. Next, in Section \ref{subsec:4.2}, we discuss the results for more generic orbits, but restricting ourselves still to a single satellite and ground station pair. We will then introduce the idea of the precision shadow of the satellite, which will help us optimize the satellite configuration and trajectories for the more complicated multiple ground station scenario, which we pursue in Section \ref{subsec:simple_example}.
For convenience, the operational parameters used for all simulations are listed in table \ref{tab:table1}.
\begin{table}[]
\label{tab:table1}
\centering
\begin{tabular}{|l|l|}
\hline
Altitude of satellite                              & h = 500 km         \\
\hline
Operational wavelength                            & $\lambda$ = 810 nm         \\
\hline
Radii of telescopes                  & ($r_{sat}$, $r_{gs}$) = (10 cm, 60 cm) \\
\hline
Detector efficiencies                        & ($\kappa_{sat}$,$\kappa_{gs}$) = (0.5, 0.5)     \\
\hline
Source rate                                       & $R$ = $10^7$ entangled-pairs/s   \\
\hline
Simulation time-step (max precision)    & $t_{\rm bin}$ = 0.5 ns \\
\hline
\end{tabular}
\caption{Various operational parameters for the simulation - notation and choice of values}
\centering
\label{tab:table1}
\end{table}

\subsection{Single satellite and ground station in the same plane}
\label{subsec:4.1}
First, let us study the peak spreading effect due to relative motion. We choose $\theta_0 = 2^{\circ}$ (which is below $\theta_0^{crit} \approx 3^{\circ}$ given by Equation (\ref{eqn:38})). From Equation (\ref{eqn:19}) and setting $h = 500$ km, we find $t_{\rm Acq}^{\rm opt} = 4.8 \times 10^{-5} s$. 
Figure \ref{fig:figure4} below shows the results for two different acquisition times, falling in the two regimes $t_{\rm Acq} = 5 \times 10^{-5} s \approx t_{\rm Acq}^{\rm opt}$ and $t_{\rm Acq} = 10^{-3} s \gg t_{\rm Acq}^{\rm opt}$. In the first case, it is clear that sharp peaks can be identified for both the correlation functions $C_{AB}$ and $C_{BA}$. This is quantitatively portrayed by the adjoining plot showing the $\rm SNR$ as a function of $\tau$. Clearly, only one value of $\tau$ shows an $\rm SNR$ significantly higher than ${\rm SNR}_{th} = 5$. On the other hand, in the latter case it is easy to see the peak broadening, as anticipated in the discussion from Section \ref{3}. The $\rm SNR$ values also highlight the advent of multiple peaks, and the consequent decrease in precision of synchronisation.

\begin{figure*}
    \centering
    \includegraphics[width = 0.45\linewidth]{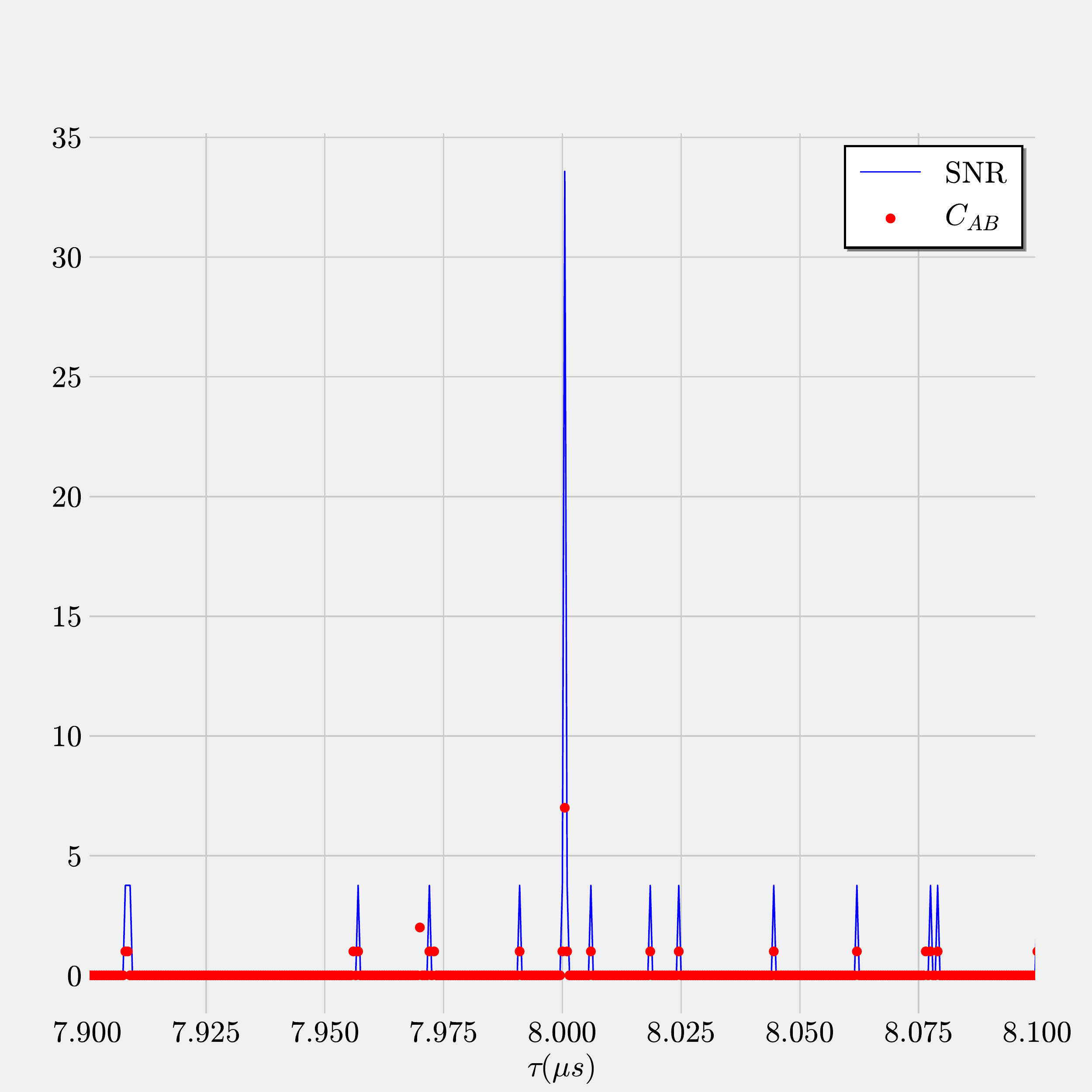}
    \includegraphics[width = 0.45\linewidth]{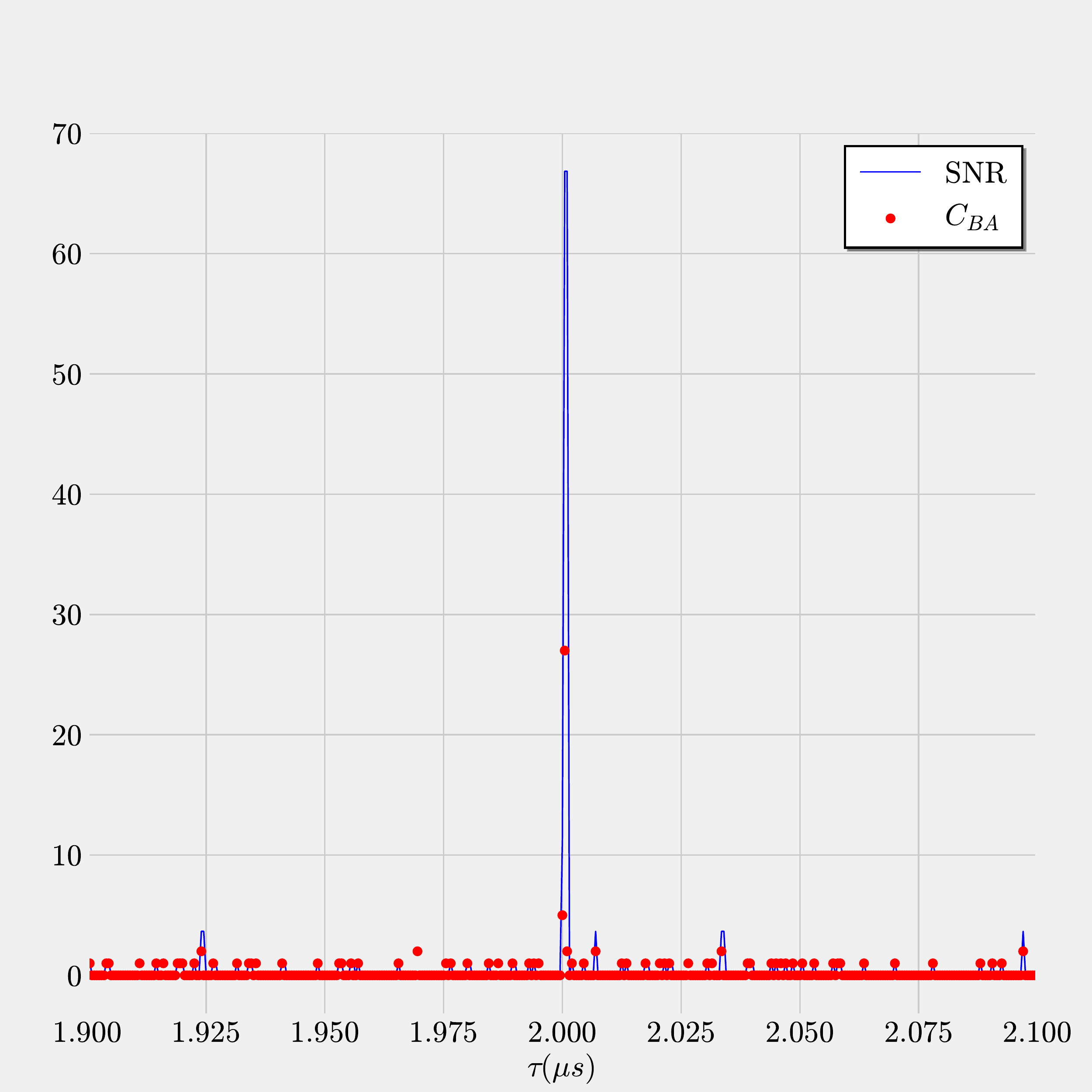}
    \includegraphics[width = 0.45\linewidth]{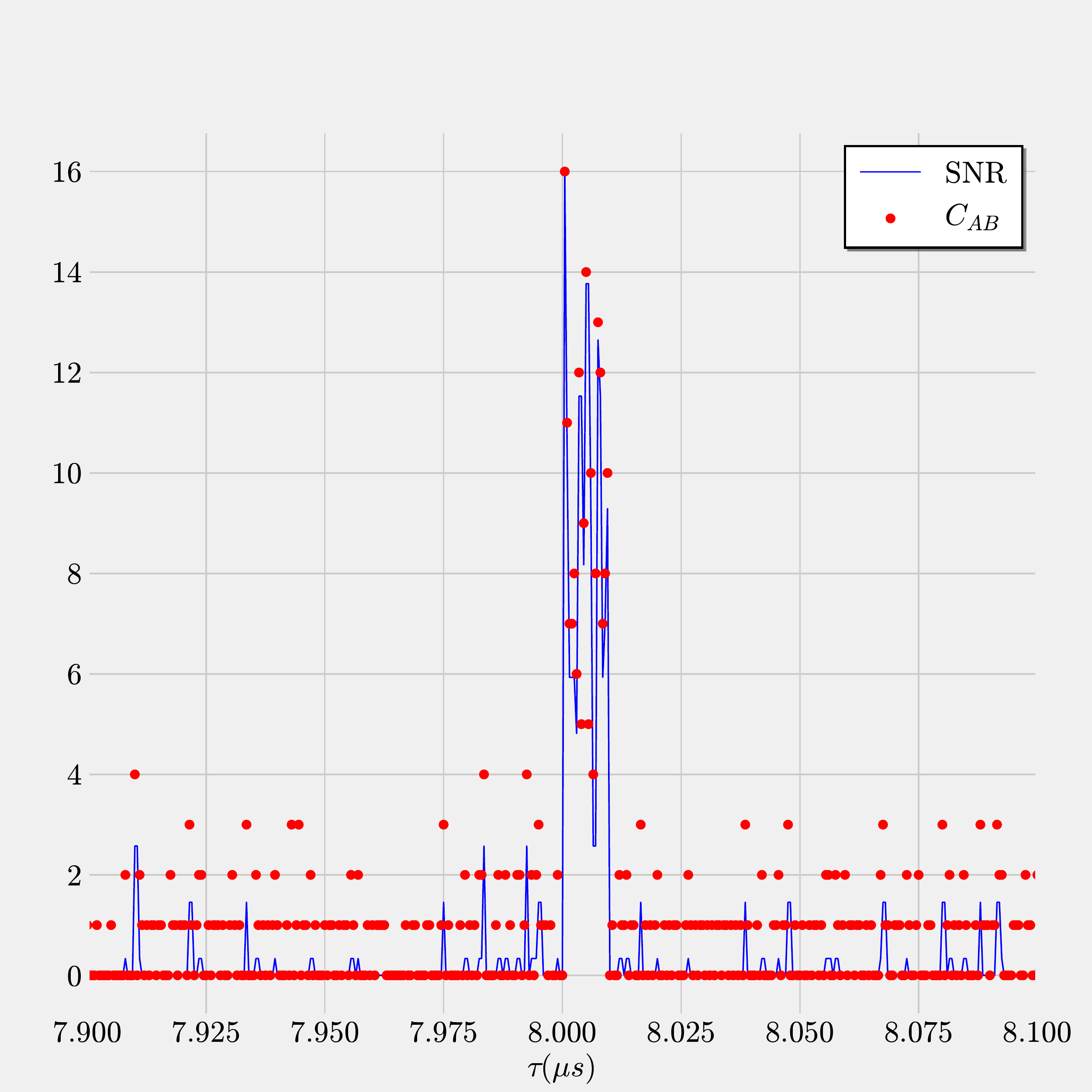}
    \includegraphics[width = 0.45\linewidth]{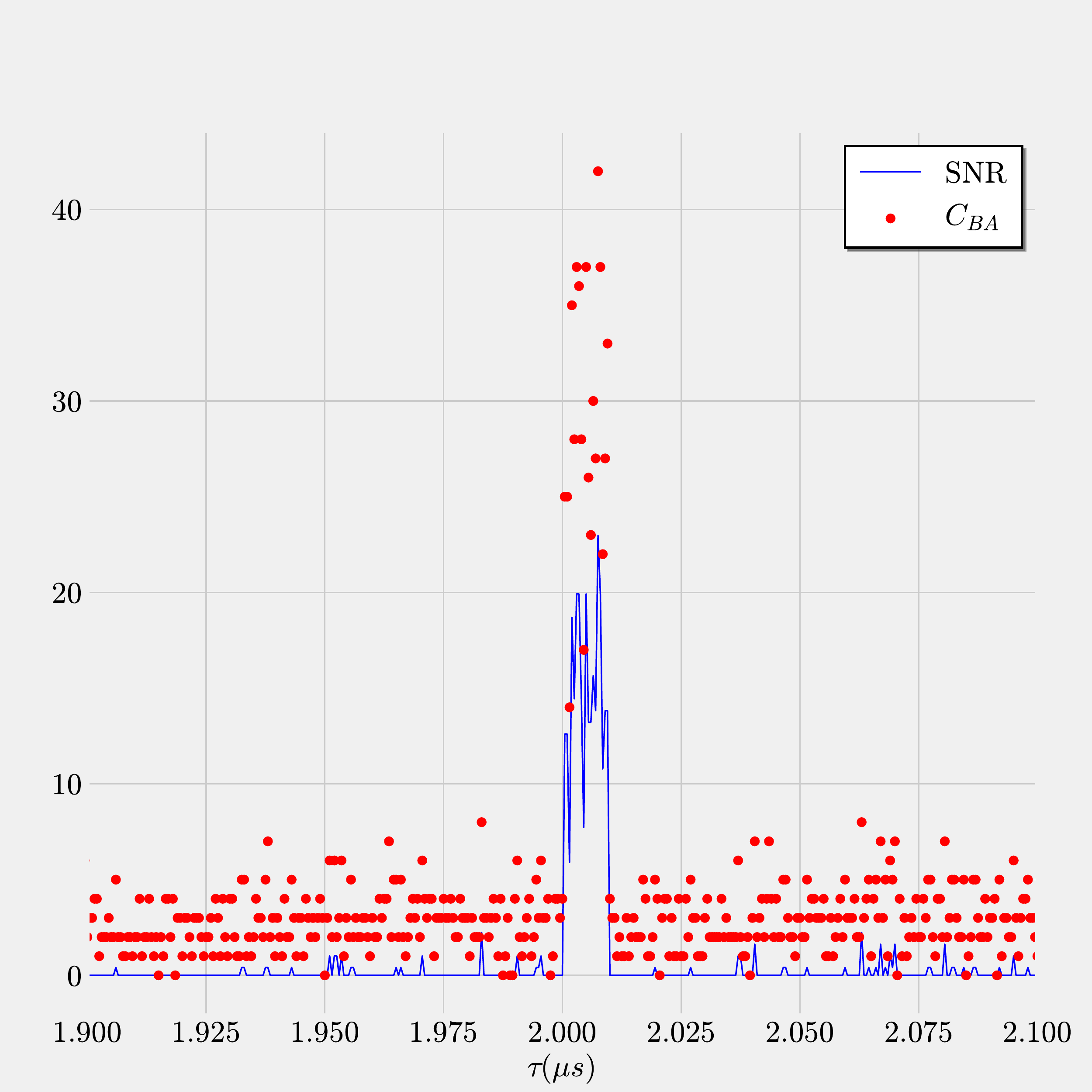}
    \caption{Uplink (left) and downlink (right) correlation functions (red dots) and respective SNRs (blue continuous line) of all peaks as a function of the time-shift $\tau$. For genuine correlations, the expected peaks are at $8\mu$s for the uplink and $2\mu$s for the downlink. Top plots represent the $t_{\rm Acq} \approx t_{\rm Acq}^{\rm opt}$ case. Notice that only one peak with substantially high SNR exists; all other peaks are also below the chosen threshold ${\rm SNR}_{th} = 5$. For the bottom plots, on the other hand, $t_{\rm Acq} \gg t_{\rm Acq}^{\rm opt}$, and several peaks with high SNR ($>5$) appear. This introduces uncertainty in the identification of the time offset, directly reduces the sync precision. 
    Peak-spreading due to motion of the satellite is responsible for the reduced precision. These simulation results  support the results in Section \ref{3} that acquisition times of the order of $t_{\rm Acq}^{\rm opt}$ must be chosen for the protocol to work at optimal level of precision. The value  $R_{\rm bkg} = 10^6 \rm /s$ has been used for background rate in these simulations. The values used for other parameters are summarized in Table \ref{tab:table1}.}
    \label{fig:figure4}
\end{figure*}

\begin{figure*}
    \centering
    \includegraphics[width=0.45\linewidth]{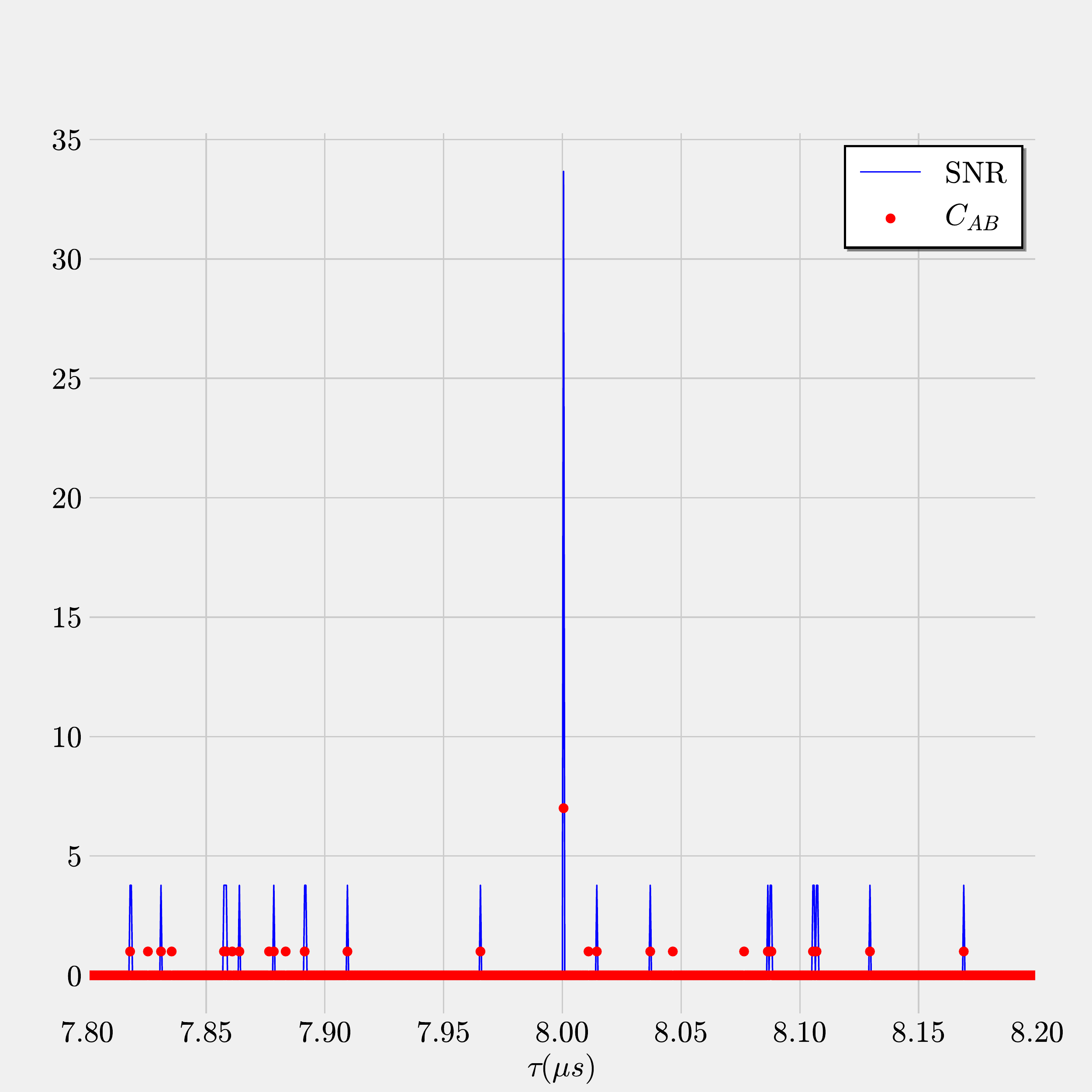}
    \includegraphics[width=0.45\linewidth]{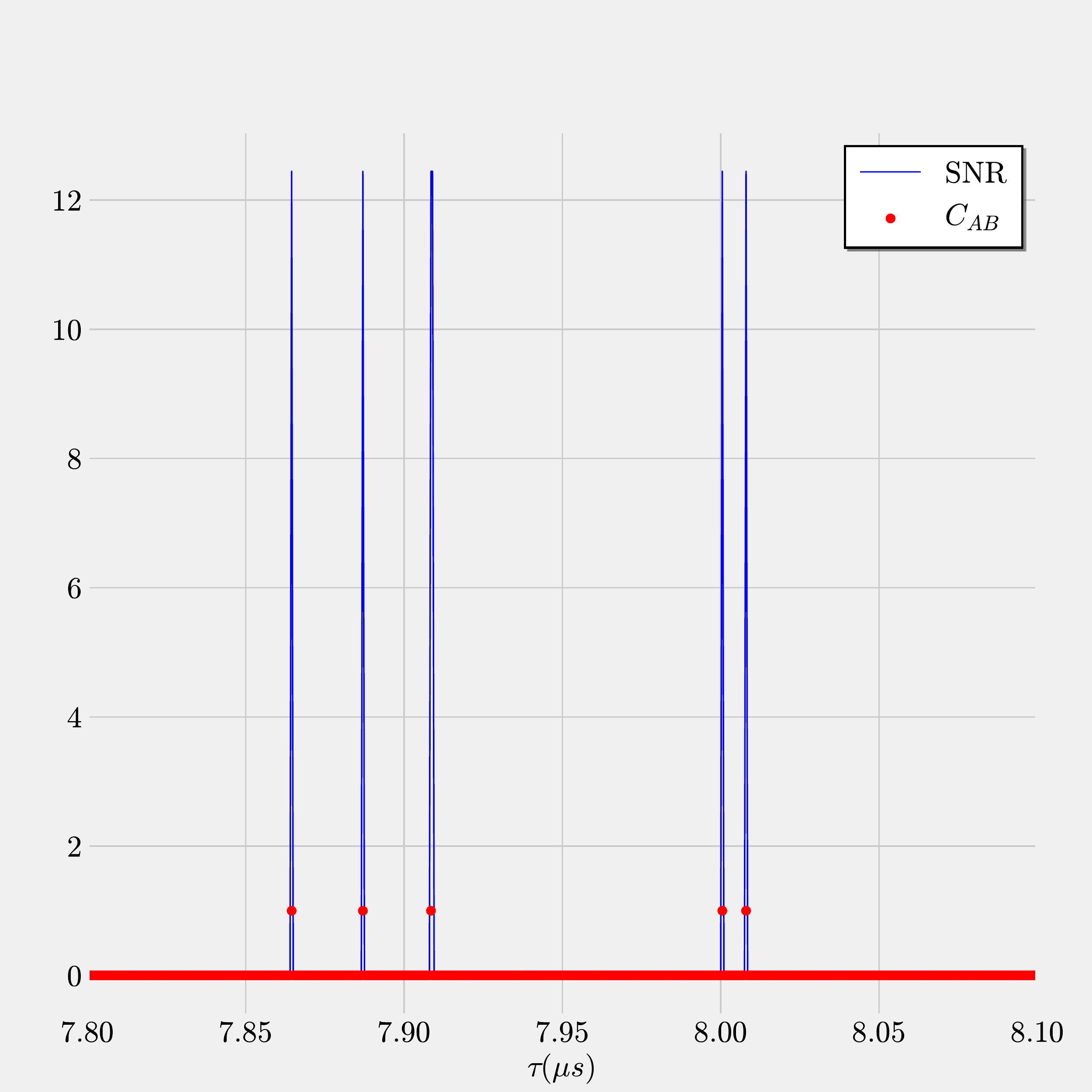}
    \caption{Uplink correlation function $C_{AB}$ for $\theta_0 = \theta^{\rm crit}_0 \approx 3^{\circ}$ (left) (the uplink is the weaker link, and determines the success or failure of the protocol). Clearly, there is a unique peak marking the correct time correlations. For comparison, the right panel shows a similar plot for $\theta_0 = 10^{\circ}$ is shown, where multiple peaks are observed. The values for background noise and acquisition time used in these plots are $R_{\rm bkg} = 10^6 \rm /s$ and $t_{\rm Acq} = 5 \times 10^{-5} s$}, respectively.
    \label{fig:figure5}
\end{figure*}

Next, we investigate the effect of increasing angular separation on the time synchronisation. Equation (\ref{eqn:38}) gives an estimate of the maximum critical angle for a given choice of $N_{\rm min}$ at a required level of precision. We choose  $t_{\rm Acq} \approx t_{\rm Acq}^{\rm opt}$  from now on, having shown that the protocol can work at optimal precision only under that condition. From Figure \ref{fig:figure5}, it is evident that, at angles greater than the critical angle, even with no background noise, enough photons cannot be collected within the acquisition time to achieve a clear peak. The SNR values also indicate this effect. Obviously, collecting  photons for a longer time will not alleviate the situation since the SNR is just going to fall for $t_{\rm Acq} > t_{\rm Acq}^{\rm opt}$ due to the peak broadening effects just discussed. 

\begin{figure*}
    \centering
    \includegraphics[width=0.3\linewidth]{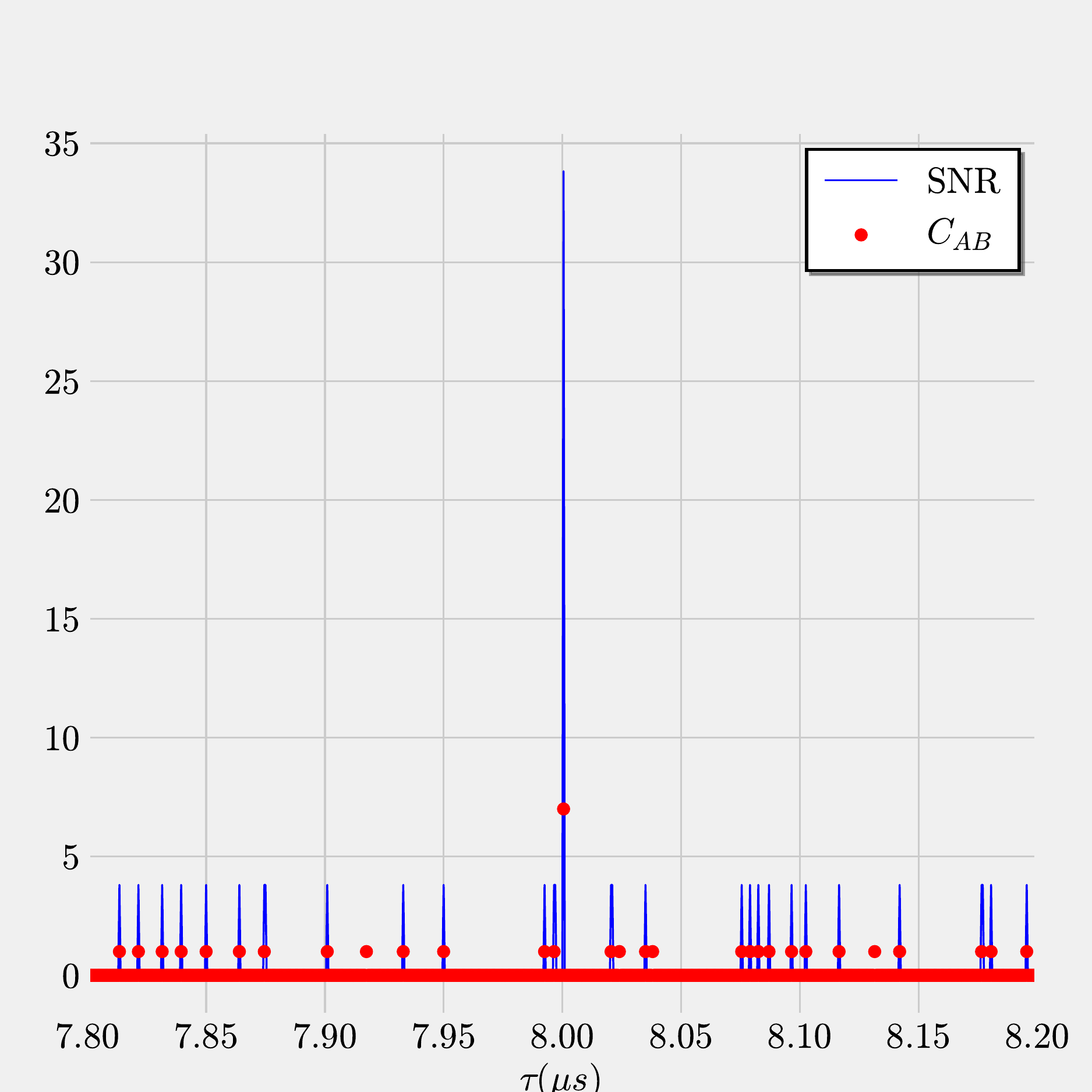}
    \includegraphics[width=0.3\linewidth]{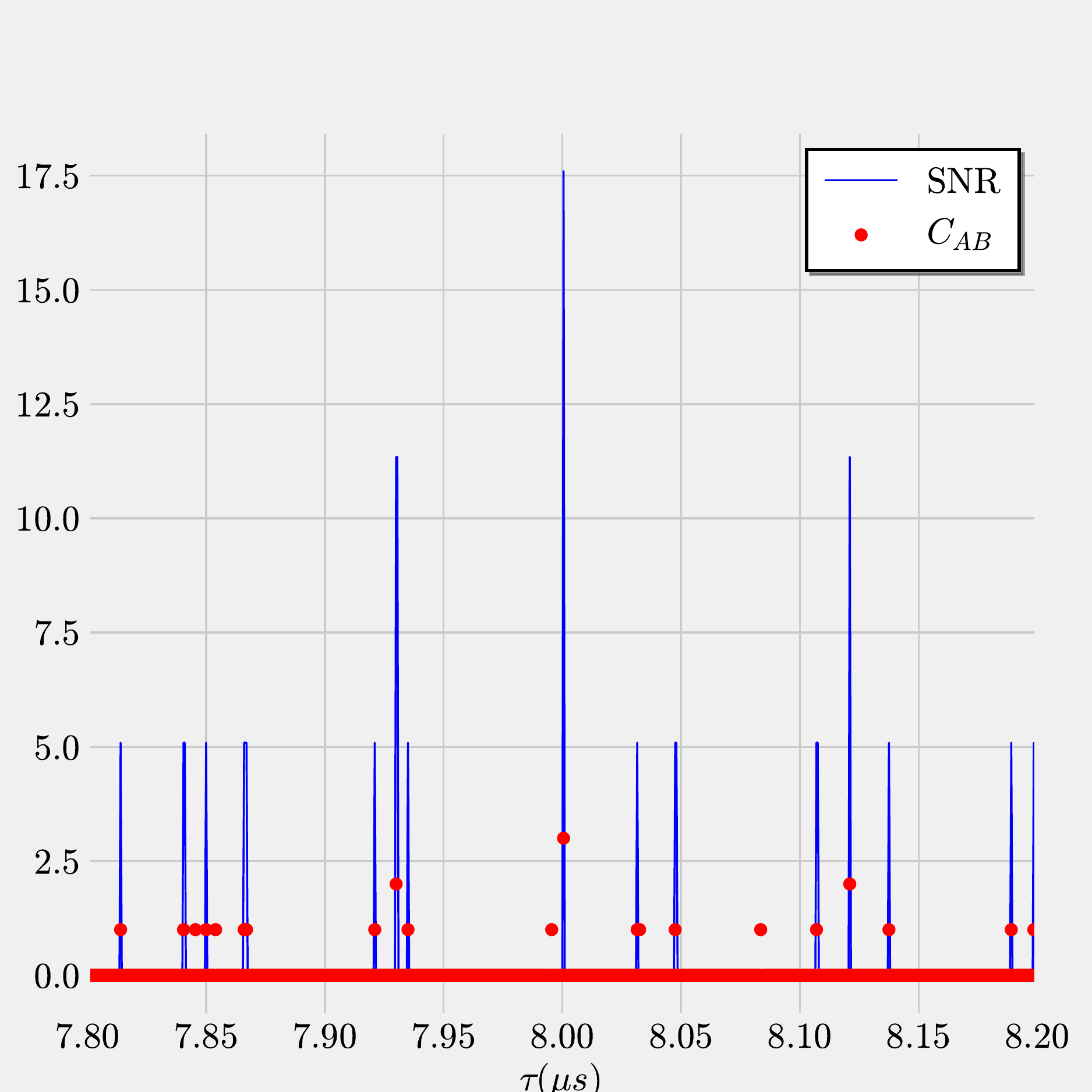}
    \includegraphics[width=0.3\linewidth]{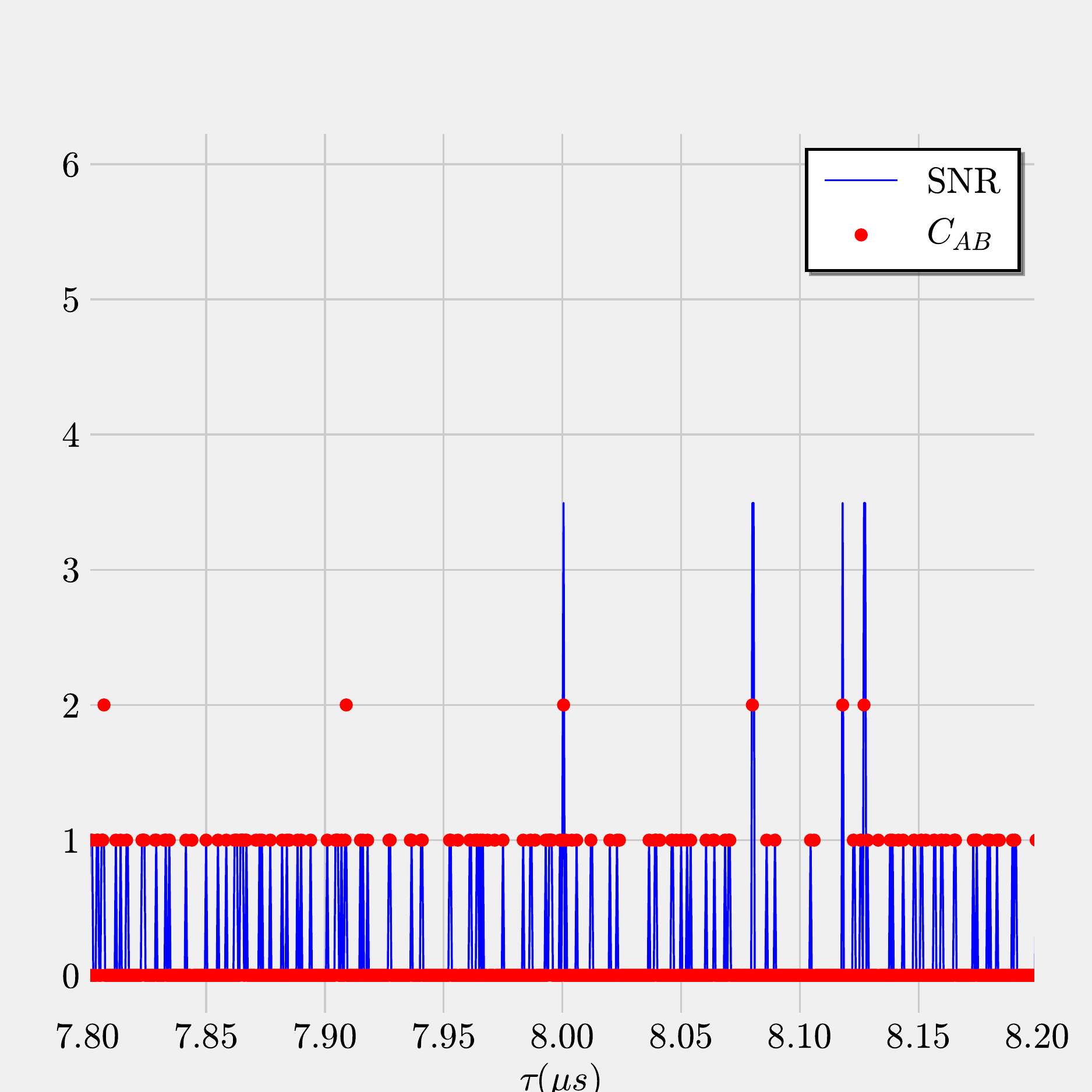}
    \caption{These plots show the effect of background noise in the correlation functions. Correlation functions and SNR are shown for $\theta_0 = 2^{\circ}$. This value is below the critical angle given by Equation (\ref{eqn:38}). The background noise rates are $10^6$ photons/s (left), $10^7$ photons/s (center) and $10^8$ photons/s (right). It is clear that, as the background rate grows past $R\eta = 10^7 \rm /s$, the SNR falls below the threshold value ${\rm SNR}_{th}=5$, making it less probable to detect a unique peak.}
    \label{fig:figure6}
\end{figure*}

Let us now move on to show the effect of background noise on the synchronisation outcomes. In order to study the effect of background noise, we run the simulations at different noise rates.  
It can be seen from Figure \ref{fig:figure6} that the SNR shows significant decrease with increasing $R_{\rm bkg}$ as it grows past the effective rate of detection through the free space channels, i.e., $R\eta$ (as expected from Equation (\ref{eqn:30})). For simplicity, we choose same background rates at the satellite and at the ground station. In a realistic scenario, this is obviously not true---background rates are much smaller at the ground station. Since, the aim here is to just illustrate effect of background photons on the signal to noise ratio, we consider this worst case scenario. 

\subsection{Simulation results for a generic satellite orbit}
\label{subsec:4.2}
\begin{figure*}
    \centering
    \includegraphics[width = 0.45\linewidth]{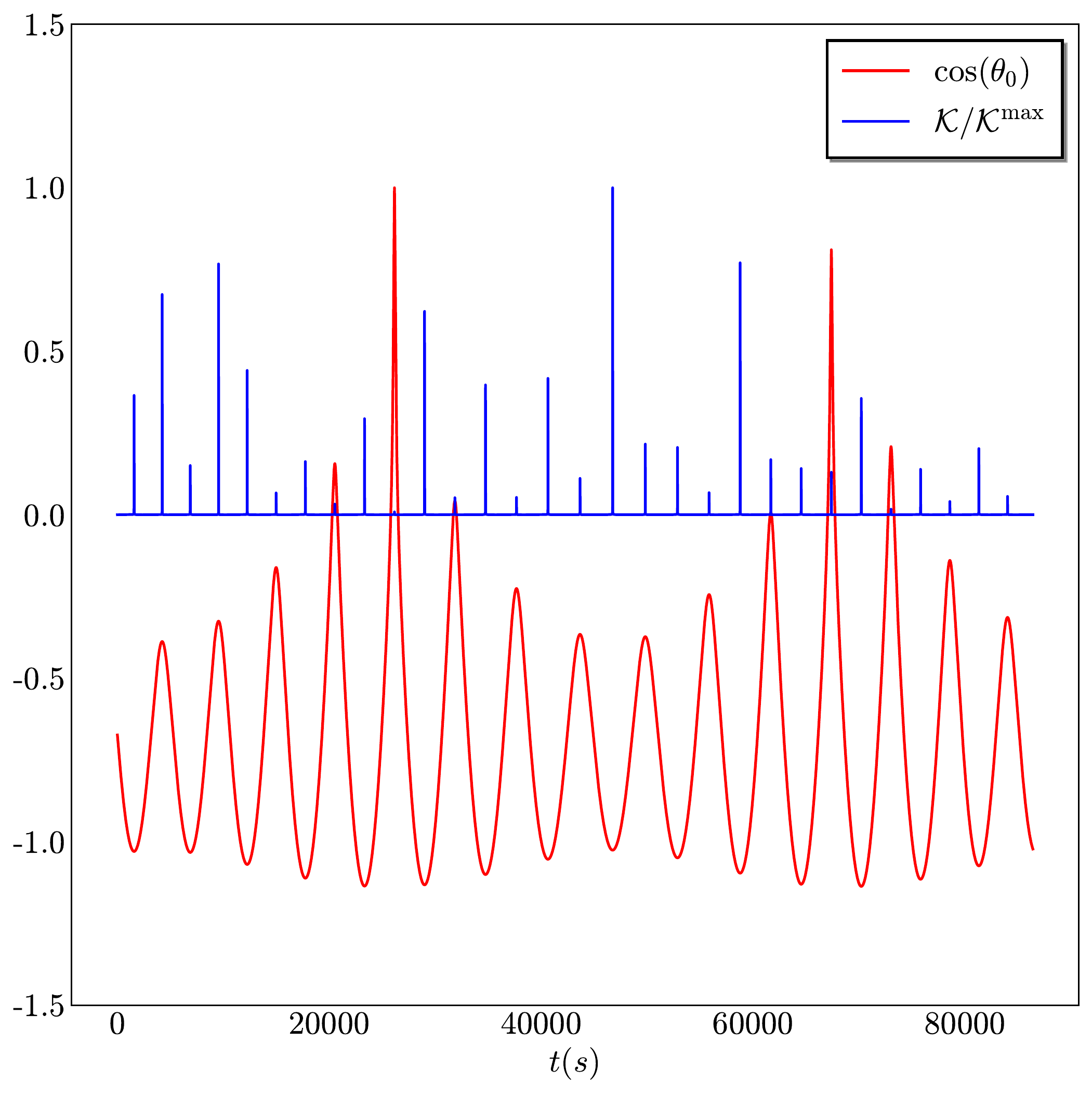}
    \includegraphics[width = 0.45\linewidth]{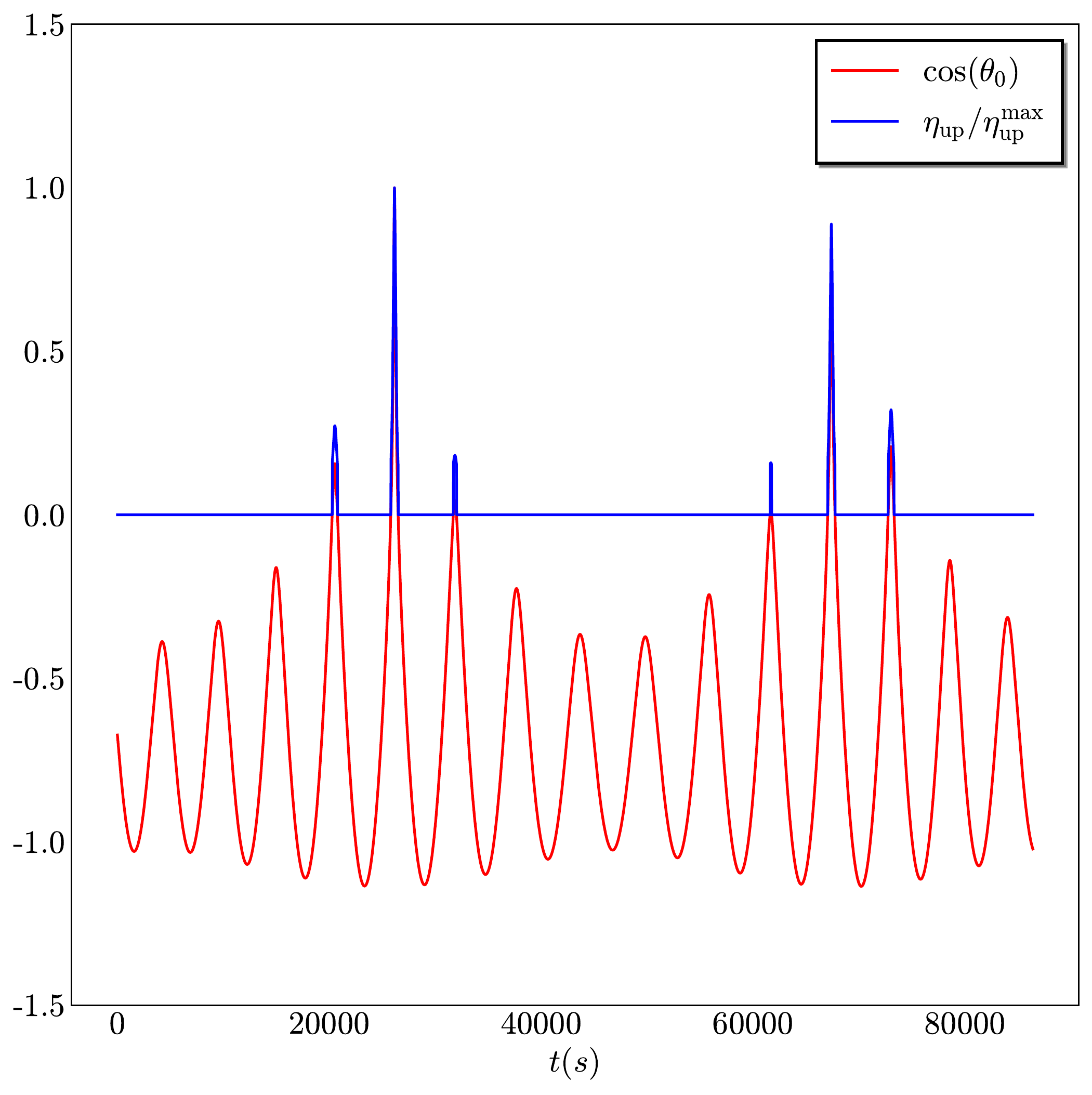}
    \includegraphics[scale = 0.5]{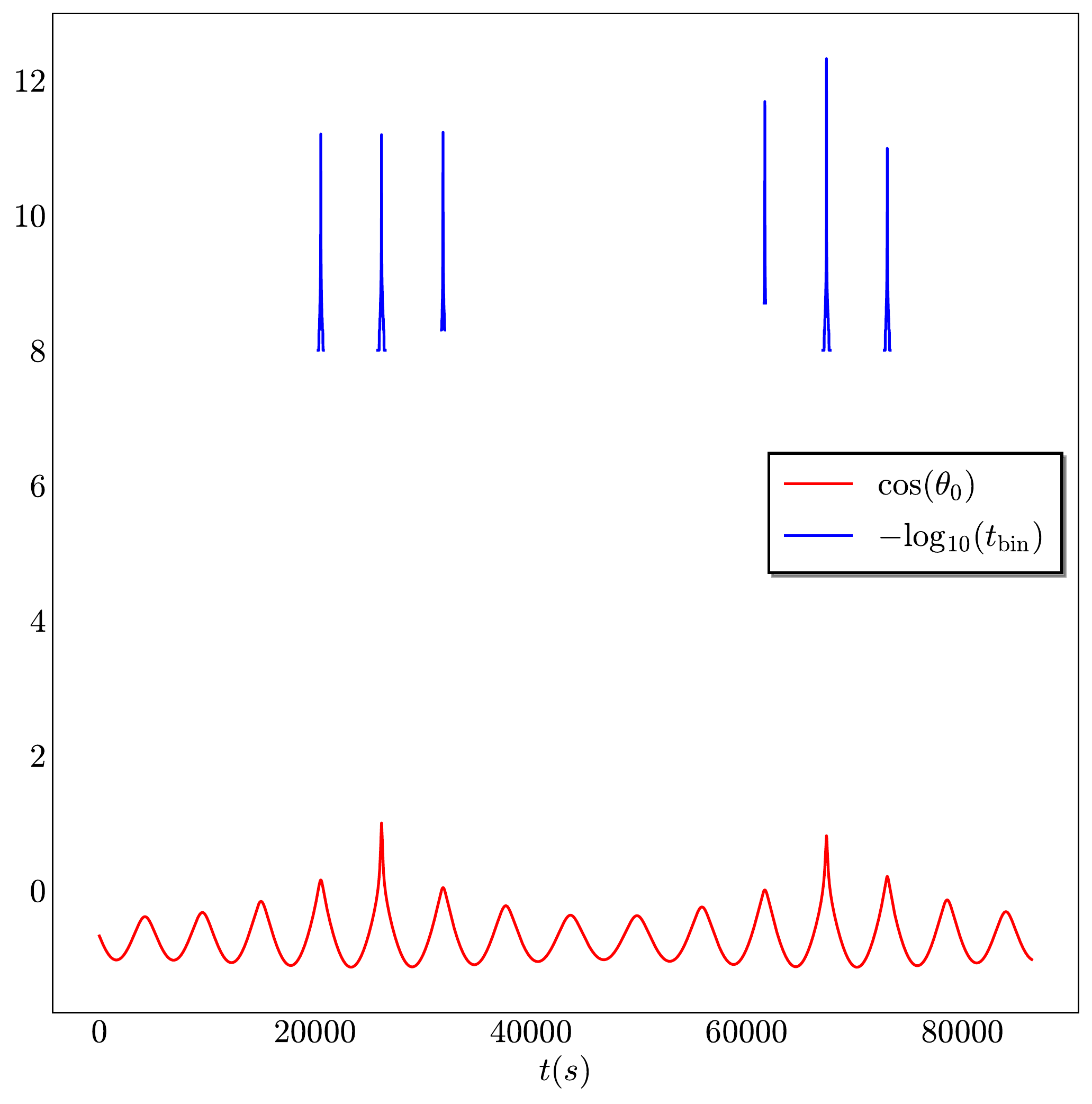}
    \caption{(Top) These plots show values for the two parameters that determine the achievable precision: $\theta_0$ (the angle that the satellite makes with the ground station zenith line) and $\mathcal{K}$ (the geometric factor defined in \eqref{eqn:18}). (Left) The geometrical factor $\mathcal{K}$ is maximized whenever the derivative of the $\theta_0$ vanishes. The optimal acquisition time $t_{\rm Acq}^{\rm opt}$ is proportional to $\mathcal{K}$. (Right) The link transmittance $\eta_{up}$ (recall that the uplink, being the weakest link, determines the limits of precision), as expected is maximized whenever the link distance $L$ is minimized and hence is maximum for overhead passes.
    (Bottom) The synchronization outcome (blue line) determines the sync precision (it is the negative log of the bin size) and it clearly follows the trends set by the two parameters $\mathcal{K}$ and $\eta_{up}$ (we choose $N_{\rm min} = 10$, see Equation (\ref{eqn:35})). A few times during the course of a day, when the link distances are minimized (overhead passes), the precision levels surpass the nanosecond level and even approach the picosecond level. These sync outcomes are currently not achievable by classical techniques for space-based communication like the GPS. The discontinuity in the precision curve (also the transmittance curve) is a consequence the horizon condition (Equation (\ref{eqn:43})).}
    \label{fig:figure7}
\end{figure*}

Next, we run the simulation by allowing the ground station to be located in a general location, not necessarily in the plane of satellite's orbit, which is chosen to be a polar orbit. These simulations are run with the  understanding that, once the acquisition time is chosen optimally ($t_{\rm Acq} = t_{\rm Acq}^{\rm opt}$), the optimal  precision can be achieved. Hence, from here on we use the term precision to mean maximum achievable precision.\footnote{This does not mean that the satellite and ground station trajectories need to be known to perform the QCS successfully. Of course, for any practical application some information of the trajectories is available. For example, in order to successfully send photons to the satellite, its trajectory must be known at least approximately. Given this knowledge, we can find bounds for the relative velocity, calculate a minimum value for $t_{\rm Acq}^{\rm opt}$ and then choose our acquisition time close to this value to run the protocol at the near optimal precision.} 
In order to understand the effect of such a generic orbit on the sync outcomes, let us first look at the two parameters  $\mathcal{K}$ and $\eta_{up/dwn}$ discussed in Section \ref{sec:3}. The radial velocity of the satellite w.r.t. the ground station and the link loss are both minimal whenever the trajectory is at a local minimum, i.e. whenever the link distance $L$ or the zenith angle $\theta_0$ are at a local minimum. We call such situations overhead passes of the satellite. For the simulation results shown in Figure \ref{fig:figure7}, we consider a setting with a polar satellite in orbit at 500 km altitude along the prime meridian,  and a ground station in New York City (40.7128° N, 74.0060° W). The results clearly show that overhead passes give a high number of double detection events (a peak in $C_{AB/BA}$), and consequently a high level of sync precision. For clarity, we define precision as $-\log_{10}(t_{\rm bin})$, such than a lower value of bin size means a higher precision. Since the precision falls down sharply as the satellite moves past the overhead configuration, we see that an analogue of critical angle $\theta_0^{crit}$ (see Sections \ref{3}, \ref{subsec:4.1}) emerges in the generic case as well. The difference being that now it has to be replaced by a solid angle. Such a solid angle can also be described as a shadow the satellite casts over the Earth during its motion in its orbit. The shadow describes a region on Earth at a given instant of time, inside which a ground station can synchronize with the satellite at a certain maximum achievable precision. We now try to estimate the shape and size of such a shadow for different levels of maximum achievable precision.

\begin{figure}
    \centering
    \includegraphics[width = \linewidth]{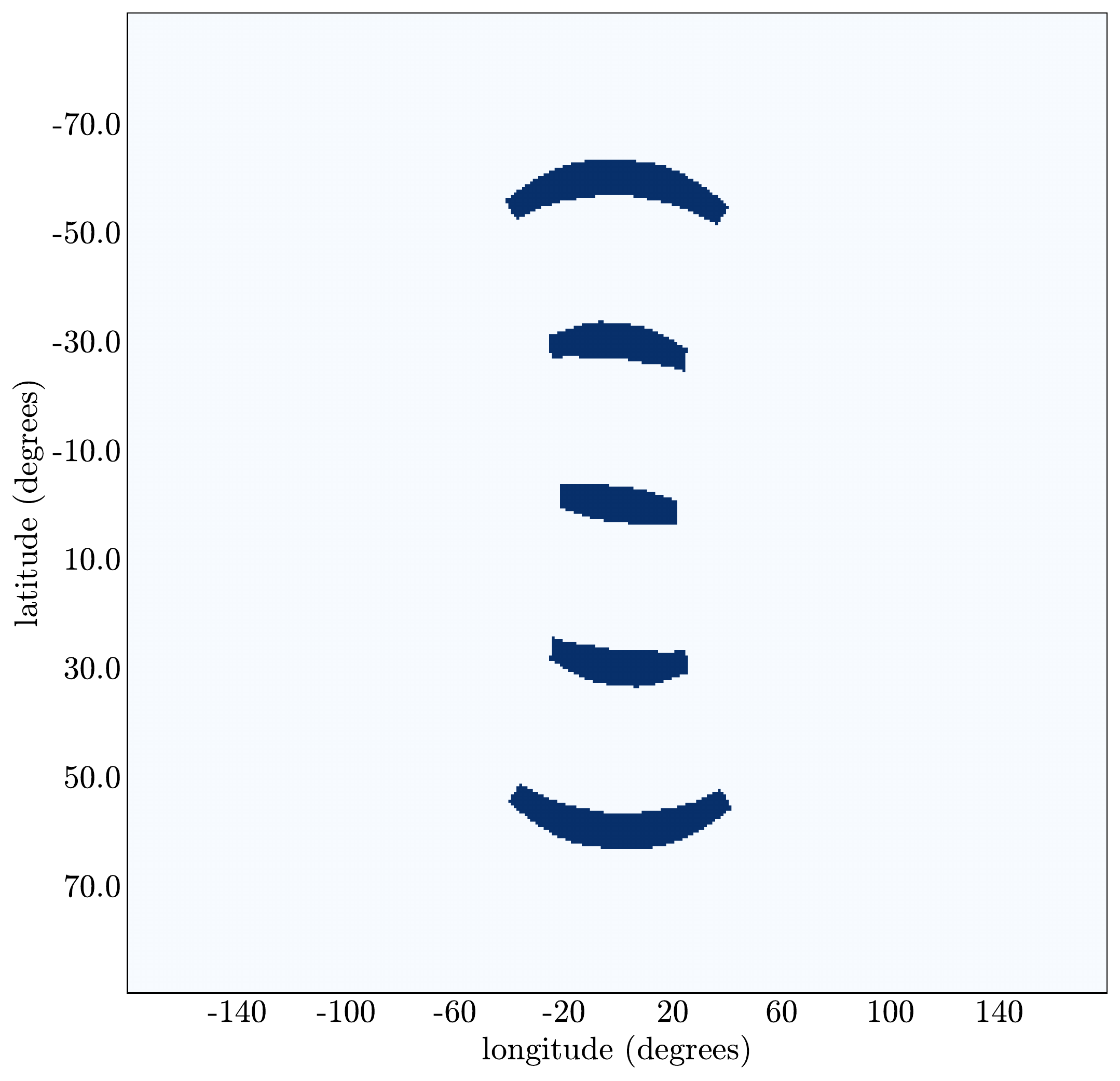}
    \includegraphics[width = \linewidth]{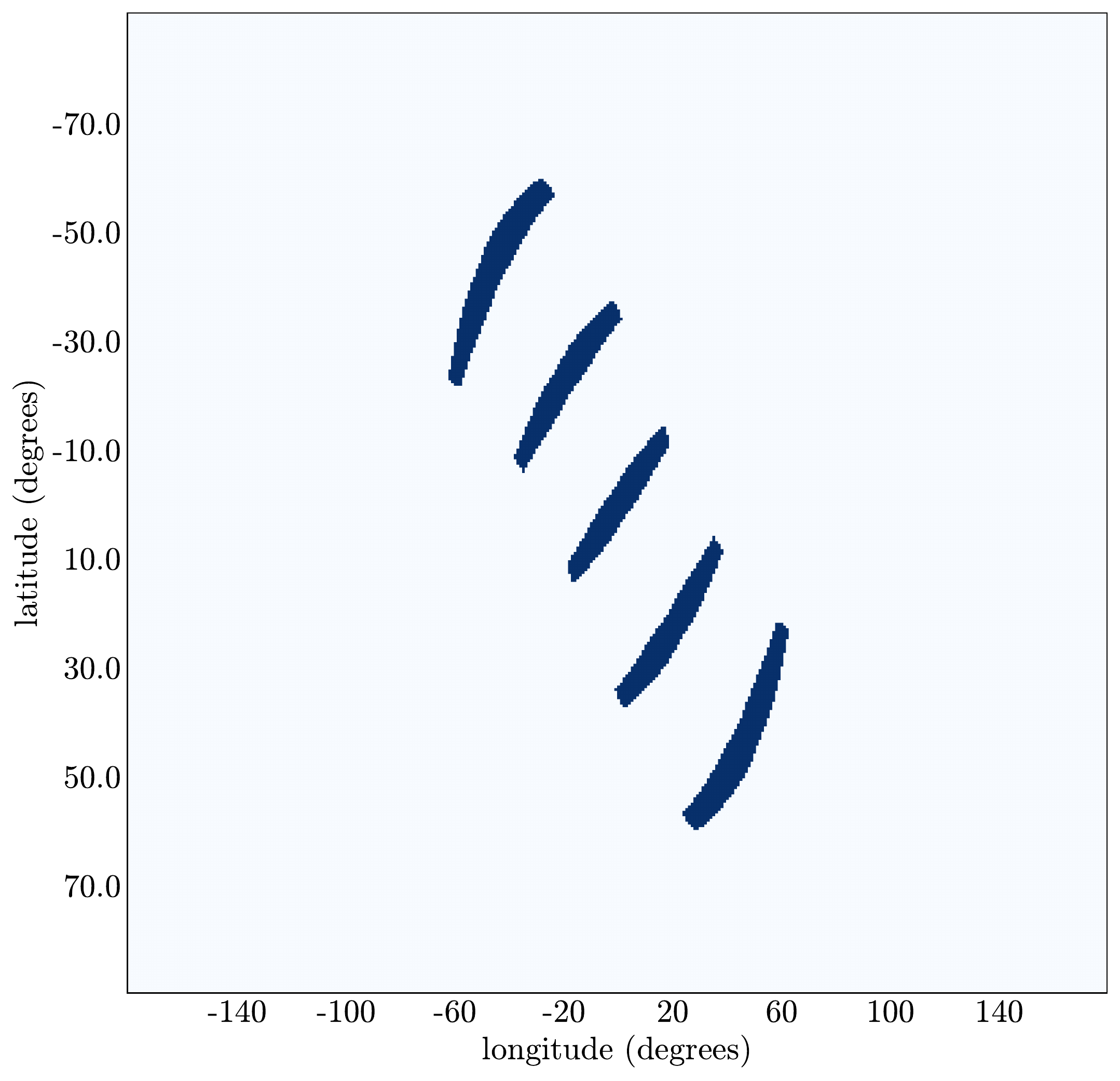}
    \caption{``Precision Shadow'' of a satellite as it passes over different parts of the Earth. A shadow refers to a region within which a certain sync precision can be achieved. Here, within the black shadow a precision of at least 1 nanosecond can be achieved. During real time dynamics the shadow will be a continuous track along the orbit of the satellite. Snippets in time are shown here to illustrate the changing shape and angular size of the shadow as the satellite moves. (Top) Polar orbit along the Prime Meridian. (Bottom) Tilted polar orbit, rotated by $30^\circ$ compared to the case on the left. In both cases the extent of the shadow close to the equator is $\approx 50^\circ$ perpendicular to the satellite trajectory and $\approx 3^\circ$ along the trajectory.}
    \label{fig:figure8}
\end{figure}
\begin{figure*}
    \centering
    \includegraphics[width = 0.4\linewidth]{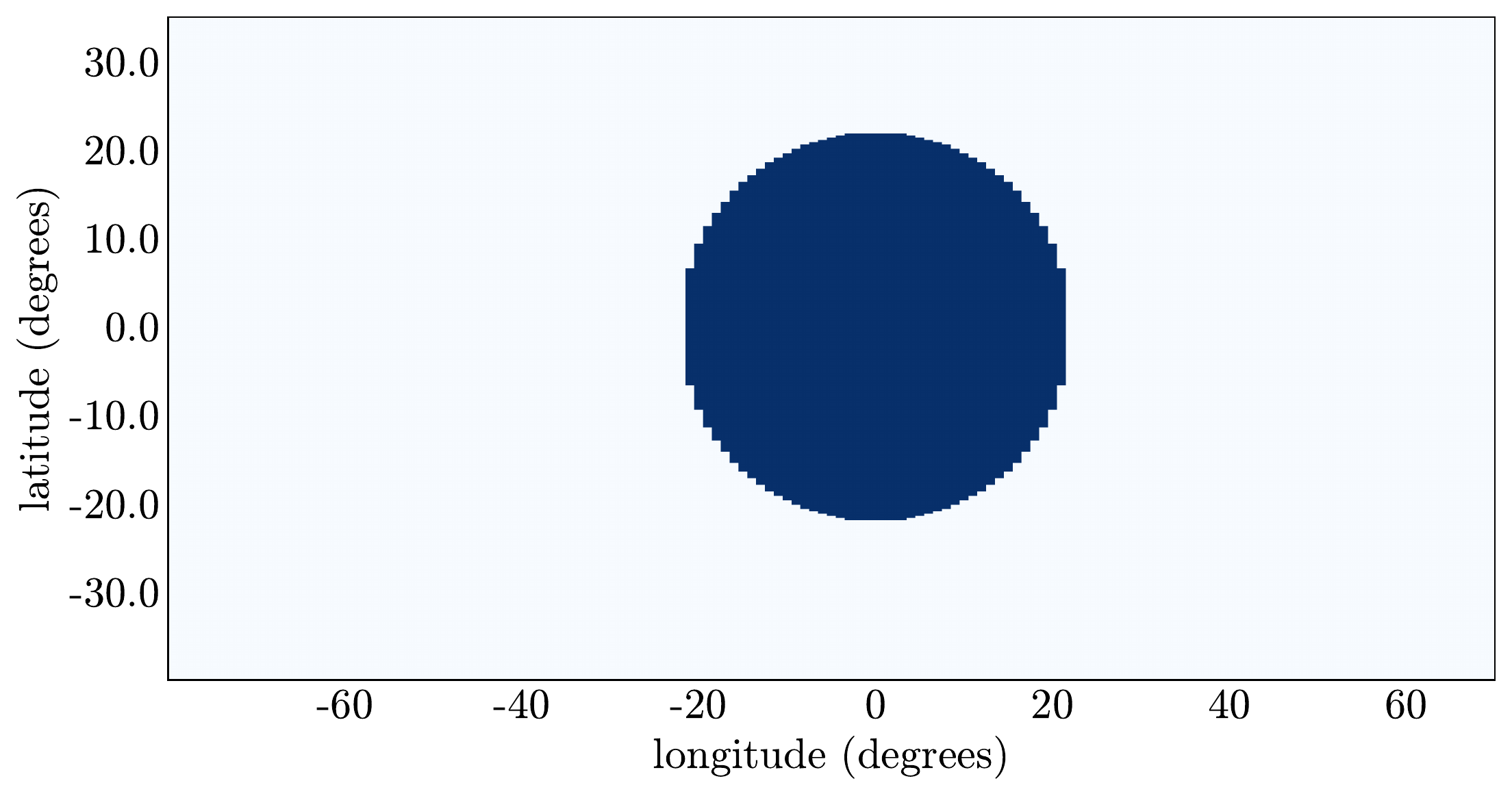}
    \includegraphics[width = 0.5\linewidth]{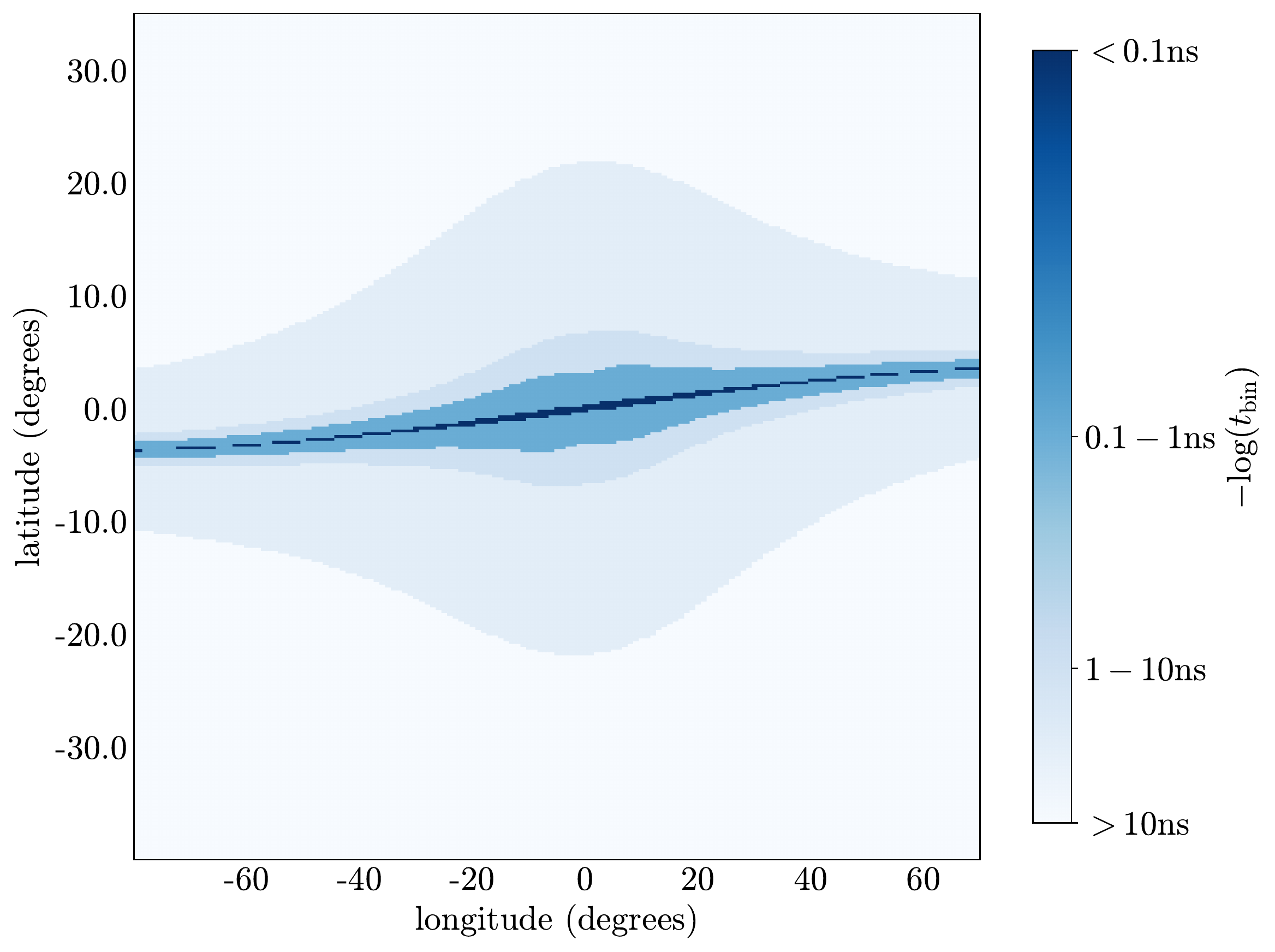}
    \includegraphics[scale = 0.6]{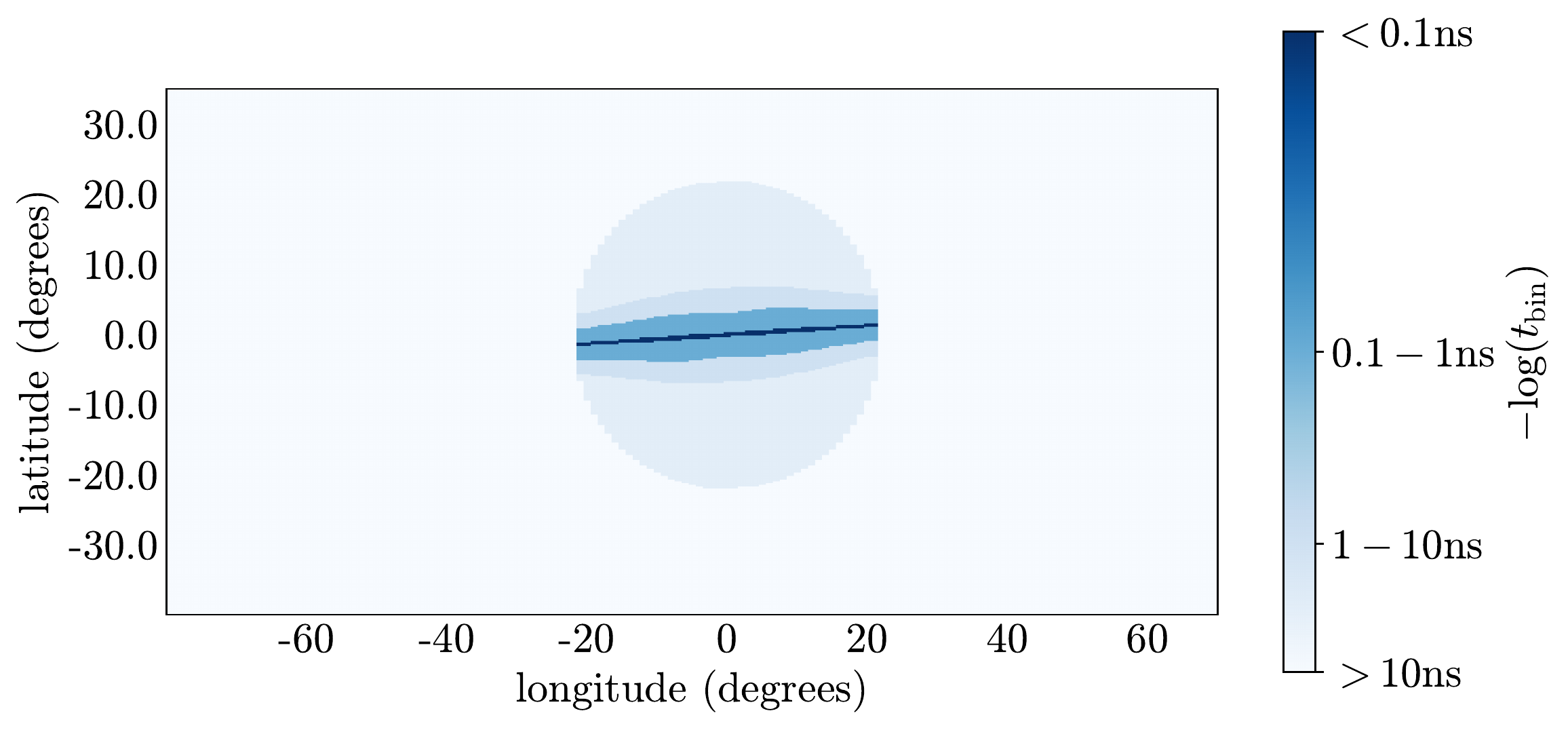}
    \caption{(Top left)
    Region of visibility on Earth of a satellite, which in this example we have placed directly above the intersection of the Prime Meridian and the Equator. (Top right) Regions where a ground station can sync with the satellite at different precisions indicated by different colors. Innermost region can sync at the highest precision since the relative radial velocity is the lowest around overhead pass. This plot does not include the condition of visibility. (Bottom) The Precision Shadow of the satellite, obtained by superposition the top two figures. This Precision Shadow is the main output of our calculation and it contains information about the size and quality of the network.}
    \label{fig:figure9}
\end{figure*}

From the results in Figure \ref{fig:figure7}, it is clear that the shape and size of the shadow of the satellite not only depends on the link distance $L$ through $\eta$ but also on the relative radial velocity of the satellite with respect to the ground station $v_{\rm rel}^{\rm rad}$ through $\mathcal{K}$. In Figure \ref{fig:figure8}, we now show the shadow of the satellite as it passes over different parts of the Earth. Without lack of generality, we choose the longitude along the prime meridian (since an overall rotation along the Earth's axis for both the ground station and satellite have no effect either on $v_{\rm rel}^{\rm rad}$ or on $L$). Clearly, the shadow changes shape as the satellite moves closer to the poles. This is a combined effect of the change in relative velocity and  the fact that arc distances become shorter for a given angular separation as one moves towards the poles. The shadow is closest to an ellipse near the equator, the minor axis being along the direction of satellite motion and the major axis perpendicular to it. This is expected since the satellite velocity is the major contributor to $v_{\rm rel}^{\rm rad}$, and a higher relative velocity means a smaller $t_{\rm Acq}$. In other words, the critical angle $\theta_0^{crit}$ along the direction of satellite motion is the smallest and is the largest perpendicular to the direction of satellite motion. In contrast, if the precision were to just depend on the total number of counts received with no restriction on the acquisition time, the shadow would have been a circle, i.e. same level of precision would have been achieved for points equidistant from the satellite since the loss function is monotonically decreasing with increasing $L$ .
For tilted orbits, the trends remain similar. The shadow still has its minor axis along the orbit of the satellite. We get a tilted ellipse close to the equator which becomes more and more morphed as one moves closer to the poles. These trends are also shown in Figure \ref{fig:figure8}. In both cases the extent of the shadow close to the equator is $\approx 50^\circ$ perpendicular to the satellite trajectory and $\approx 3^\circ$ along the trajectory.
Another important factor in determining the extent of the shadow is the horizon condition. This sets a cutoff for the shadow size, irrespective of the possible acquisition time and the link loss. In Figure \ref{fig:figure9}, we show the effect of the horizon condition in determining the size of the satellite shadow. It is also interesting to look at how the extent of the shadow reduces as one makes the sync requirements more stringent. In Figure \ref{fig:figure9} (Bottom) we show the regions within the shadow where different sync capabilities can be achieved. As expected, the closer one is to the centre of the shadow, i.e., the closer one is to the overhead case, the higher the achievable sync precision.

In summary, in this section we have presented the main results of our numerical analysis. The main output of our simulations are the precision shadows of a satellite, shown in \ref{fig:figure8}. Through these shadows we concretely determine the region on Earth within which two ground stations can be synchronized by means of a satellite at a minimum required level of precision. This information is crucial to determine the size and quality of a QCS network \cite{white_paper}. Our numerical code can be used to obtain these shadows and their time evolution for satellites in arbitrary orbits, and assist in the design of optimal satellite constellations for time distribution.

\section{Synchronising multiple ground stations}
\label{sec:5}
Let us now use the results of the shadow picture to achieve a practical sync requirement. Consider 4 major cities on geographical corners of 
the continental United States, namely: New York, Atlanta, Los Angeles and Seattle. Let us now ask the following question: ``What are the best synchronization outcomes that can be achieved for these 4 cities using a single satellite in a LEO?". We believe this question to be pertinent from two points of view: (1) It describes a proof of principle realization of the proposed QCS protocol using intermediary satellites and therefore paves a way for more ambitious implementations, like augmenting a global quantum internet, building a quantum GPS, etc. And (2) such small-medium scale realizations themselves may have practical benefits, such as time distribution for sensitive scientific experiments, high precision secure time transfer for civil and defense uses, etc.

\begin{figure*}
    \centering
    \includegraphics[scale = 0.55]{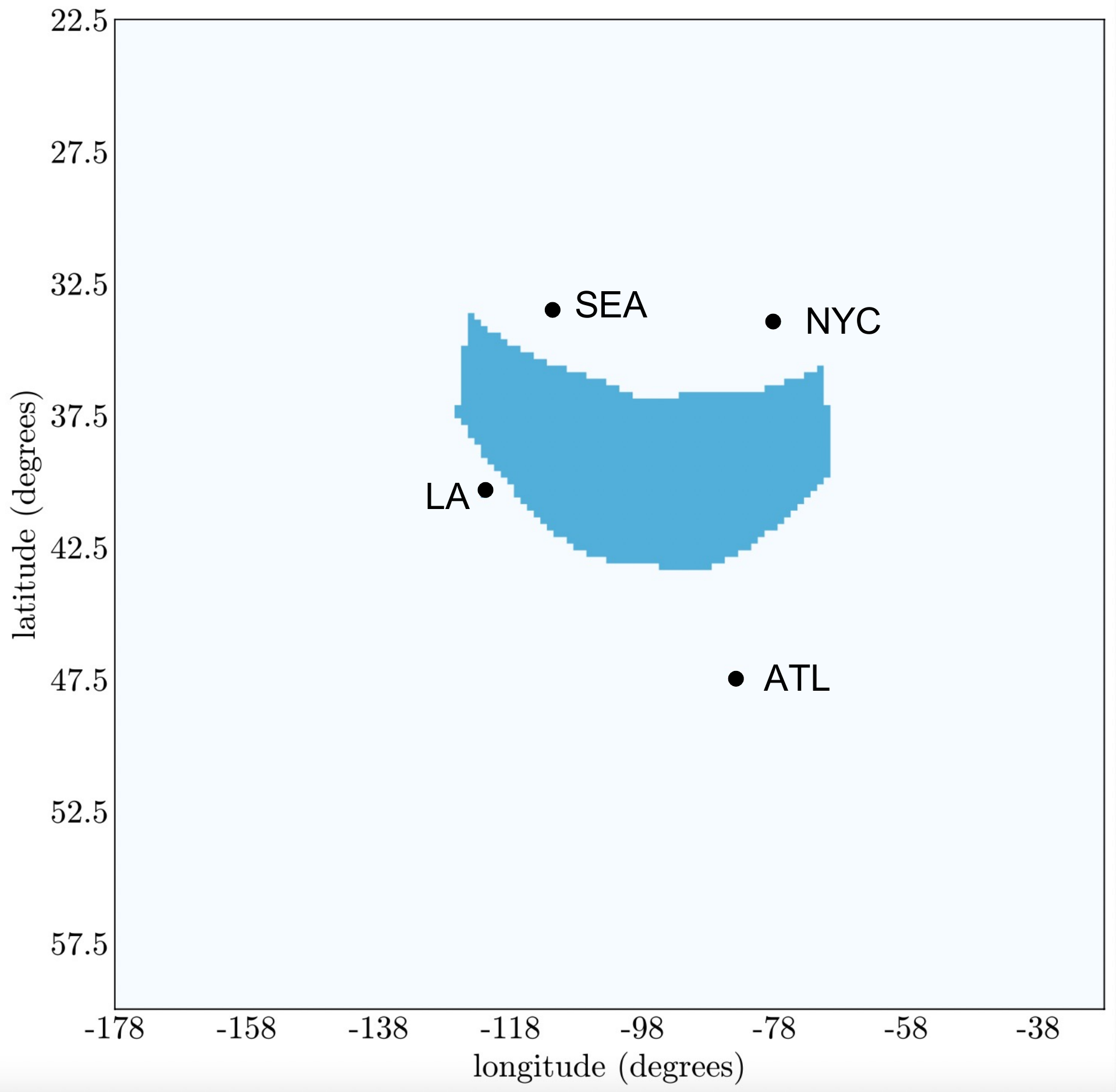}
    \includegraphics[width = 0.45\linewidth]{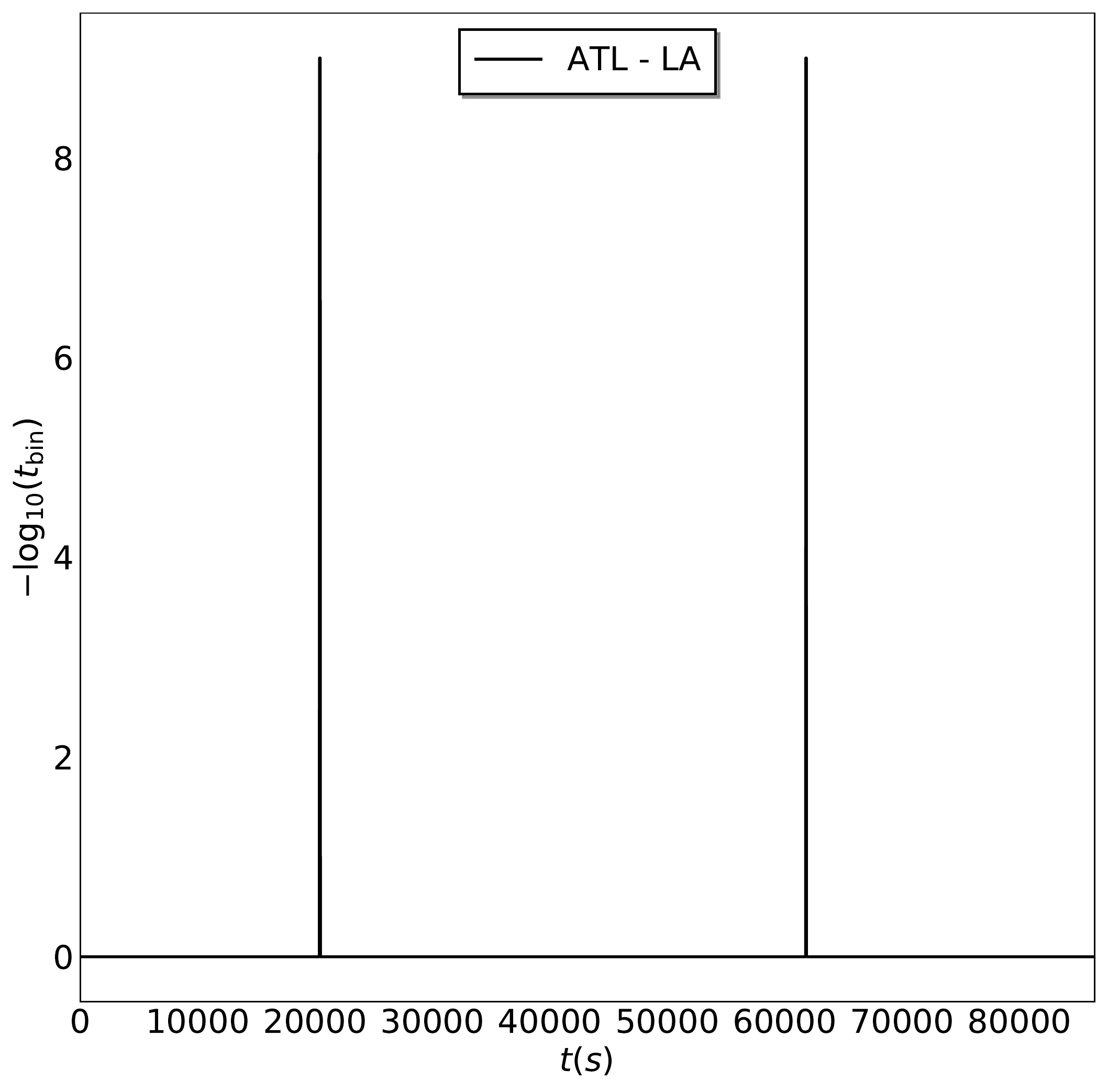}
    \includegraphics[width = 0.45\linewidth]{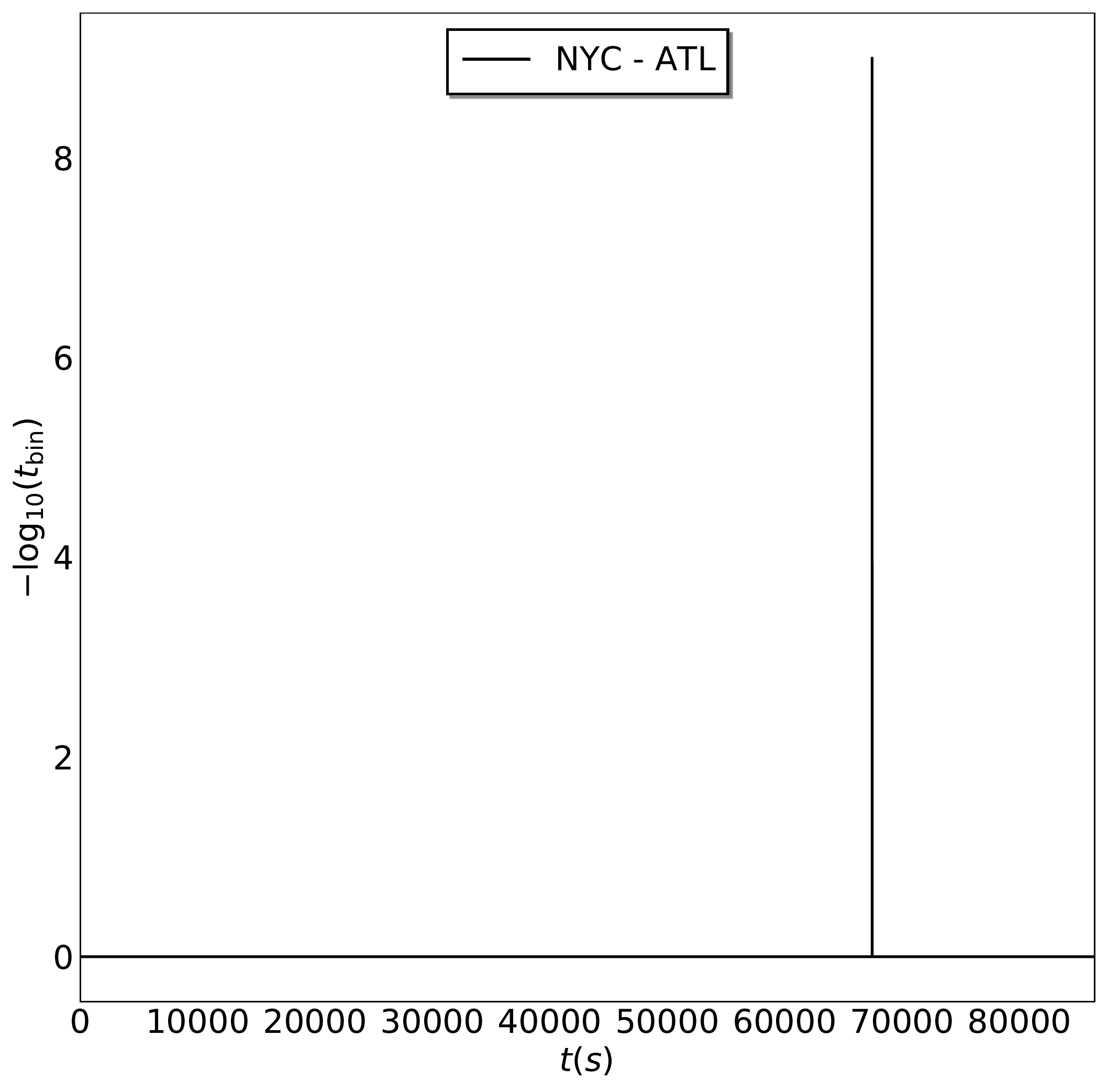}
    \caption{(Top) Satellite shadow when it is just above the center of the ground station configuration NYC-ATL-LA-SEA. All city pairs do not fall within the shadow and hence cannot be simultaneously synchronized. The only two pairs whose latitude difference is less than the width of the shadow are ATL-SEA (Bottom left) and NYC-ATL (Bottom right). Sub-nanosecond precision can be achieved at least once a day in both these cases}
    \label{fig:figure10}
\end{figure*}

Let us first trace out the shadow of the satellite when it is above the center of the ground station configuration mentioned above. This tells us that all four cities cannot be synced simultaneously since the shadow is not wide enough (latitude-wise) to contain all 4 cities at any given instant (See Figure \ref{fig:figure10}). We verify this by looking at the uplink photon counts received simultaneously by each pair of cities. A sync at a given precision is considered successful if both the cities see the satellite (satellite is above the horizon for both) and more than a cut off number of correctly correlated photons are received by each of them in accordance with Equation (\ref{eqn:35}). The sync precision is given by the minimum of the two sync precision values obtained, namely, precision of sync between ground station 1 (GS1) and the satellite and precision of sync between GS2 and the satellite. We find, as expected, that only New York-Atlanta and Atlanta-Los Angeles pairs can be synced at a precision higher than or equal to 1 ns. The latitude difference between the other pair of cities is more than the width of the shadow. 
 
\begin{figure}
    \centering
    \includegraphics[width = \linewidth]{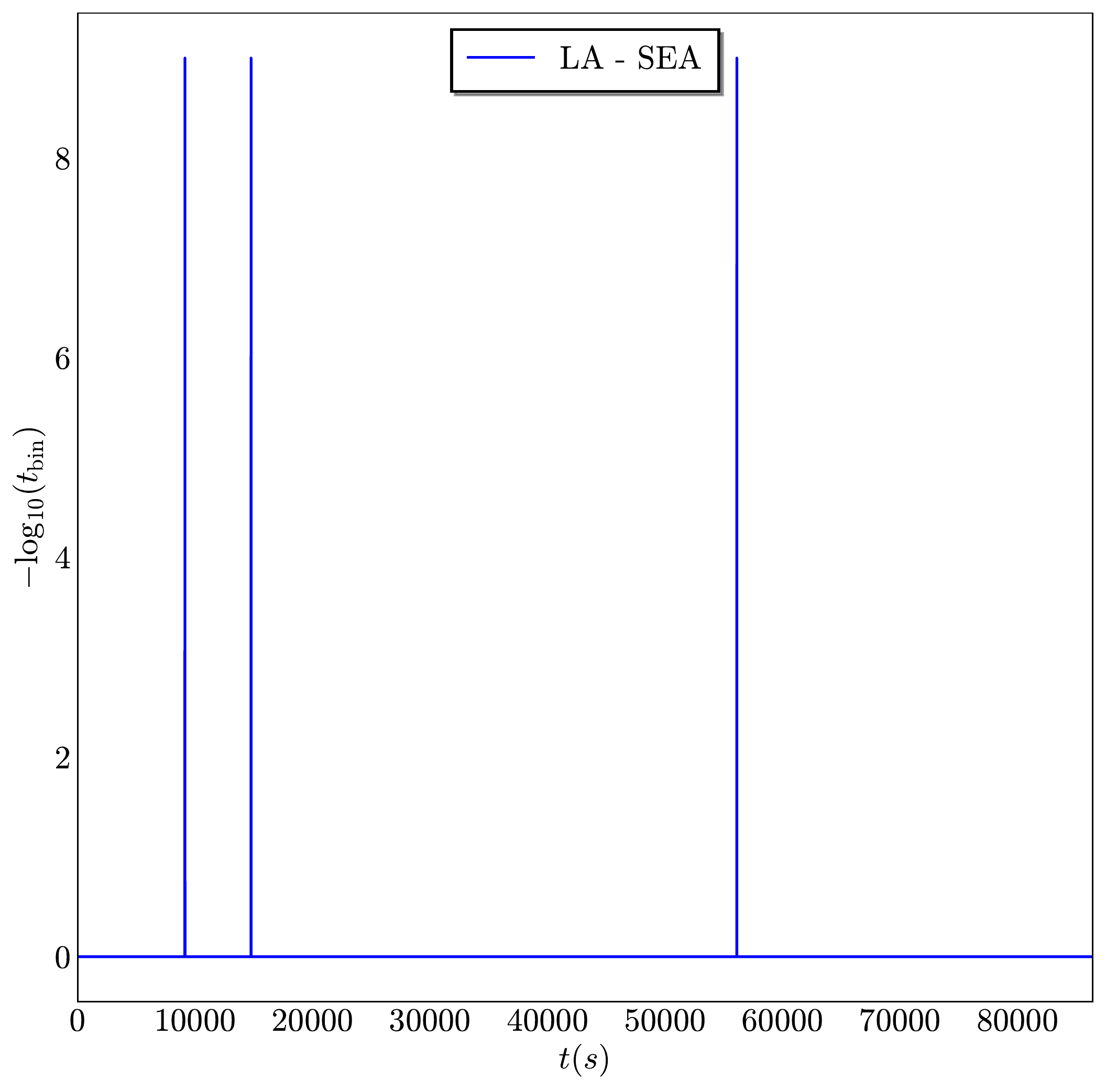}
    \includegraphics[width = \linewidth]{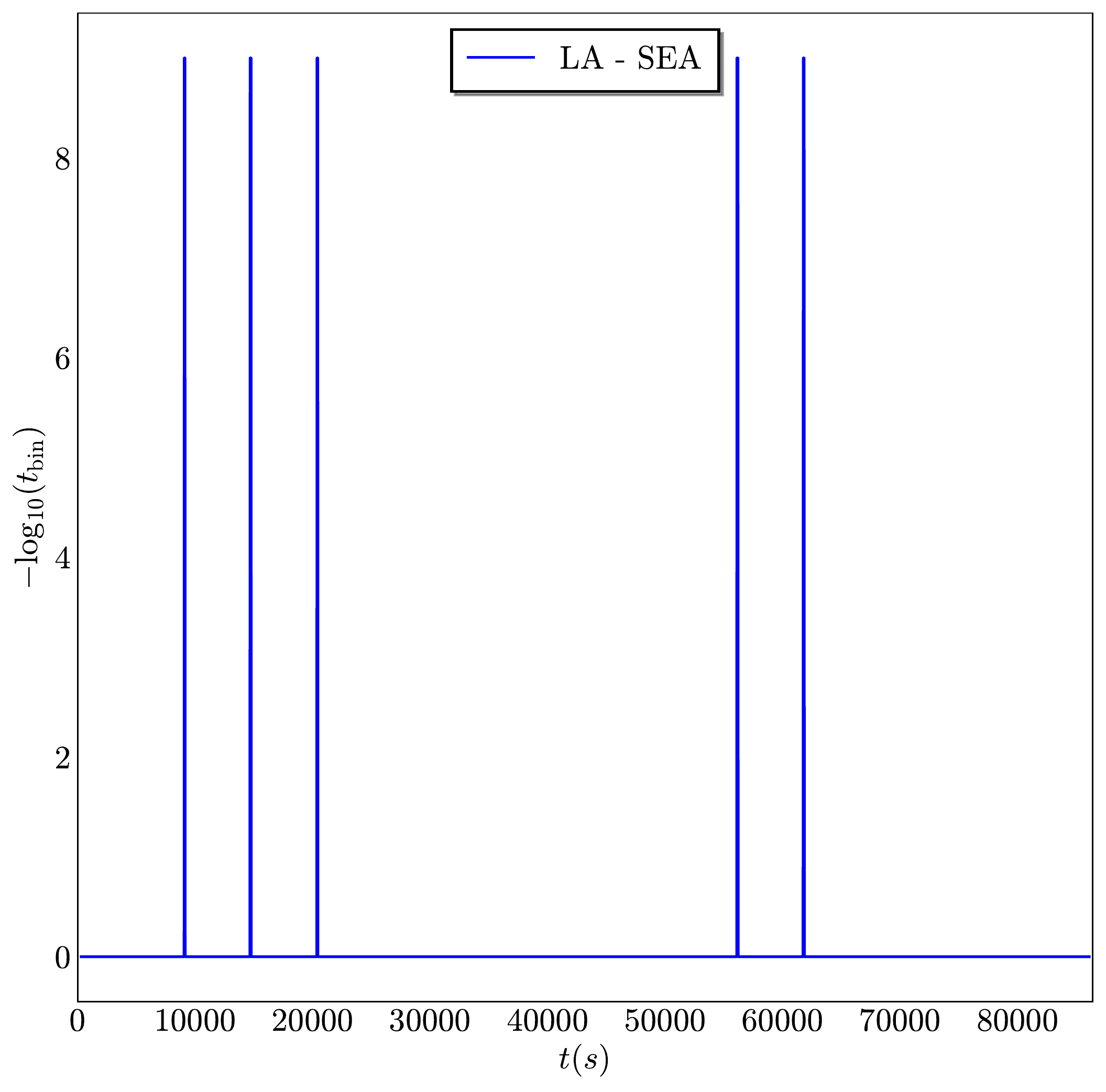}
    
    \caption{Improvement in sync outcomes between Los Angeles and Seattle, both in terms of precision and the number of connections every day, as the holdover time $\tau$ is increased from $\tau = 240 s$ to $\tau = 600 s$.}
    \label{fig:figure11}
\end{figure}

Therefore, now we must relax the sync requirements. If we allow even modestly stable clocks on the satellites (standard Rubidium clocks can hold time up to 1 ns precision for around 10 mins) then we can relax the condition of simultaneous connectivity between the three parties viz. the satellite, and the two ground stations. We can instead define a holdover time $\tau$ within which, if the satellite connects to both cities, the two cities are considered synchronised. 
Larger holdover times lead larger sync coverage areas,  because they allow the satellite to reach stations that are located further away. But of course such an improvement comes at a cost. Since, earlier we were only considering simultaneous connection between GS1, GS2 and the satellite, the precision of the protocol was limited only by the sync precision, assuming that the satellite and ground station clocks can time stamp with much higher precision compared to the sync precision. If we allow for a holdover time, this no longer remains true. The precision of the protocol will be determined not only by the sync precision but also by the precision of the least stable clock. It is reasonable to assume that the ground station clocks will be much more stable than the smaller and lighter clocks onboard a satellite. Therefore, if at a given instant $t$ there is connection between the satellite and ground station 1, then we look for the best precision connection between the satellite and ground station 2 within the interval $(t-\tau/2, t+\tau/2)$. The precision of connection at time $t$ between the two ground stations defined as:
\begin{equation*}
    -\log_{10}(t_{\rm bin}^{GS1-GS2})(t)\, ,
\end{equation*}
is given by the following expression:
\begin{eqnarray}
\max_{t' \in (t-\tau/2, t+\tau/2)}\bigg(\min\big(-\log_{10} (t_{\rm bin}^{GS1-SAT})(t), \\ \nonumber -\log_{10}(t_{\rm bin}^{GS2-SAT})(t'), -\log_{10}(t_{\rm bin}^{SAT})(t')\big)\bigg)\, .
\label{eqn:44}
\end{eqnarray}
In general, the precision of a clock, and hence $-\log_{10}(t_{\rm bin}^{SAT})(t')$, is a complicated function of time. In fact the Allan deviation curve which is the standard measure of a clock's stability, is often obtained empirically. Here, for our simulations we take a simplistic, worst-case-scenario approach. 
\begin{equation}
   -\log_{10}(t_{\rm bin}^{SAT})(t') = 
    \begin{cases} 
      C, & t-\tau/2 < t' < t+\tau/2 \\
      0, & {\rm otherwise}\, ,
   \end{cases}
   \label{eqn:45}
\end{equation}
where $C$ is a constant level of precision.
Figure \ref{fig:figure11} shows the improvement in sync outcomes as the holdover time is increased for SEA-LA. For $\tau = 0$ there was no connection at the 1 ns precision, and as $\tau$ is increases,  we see more frequent connections. 

\begin{figure*}
    \centering
    \includegraphics[width = 0.45\linewidth]{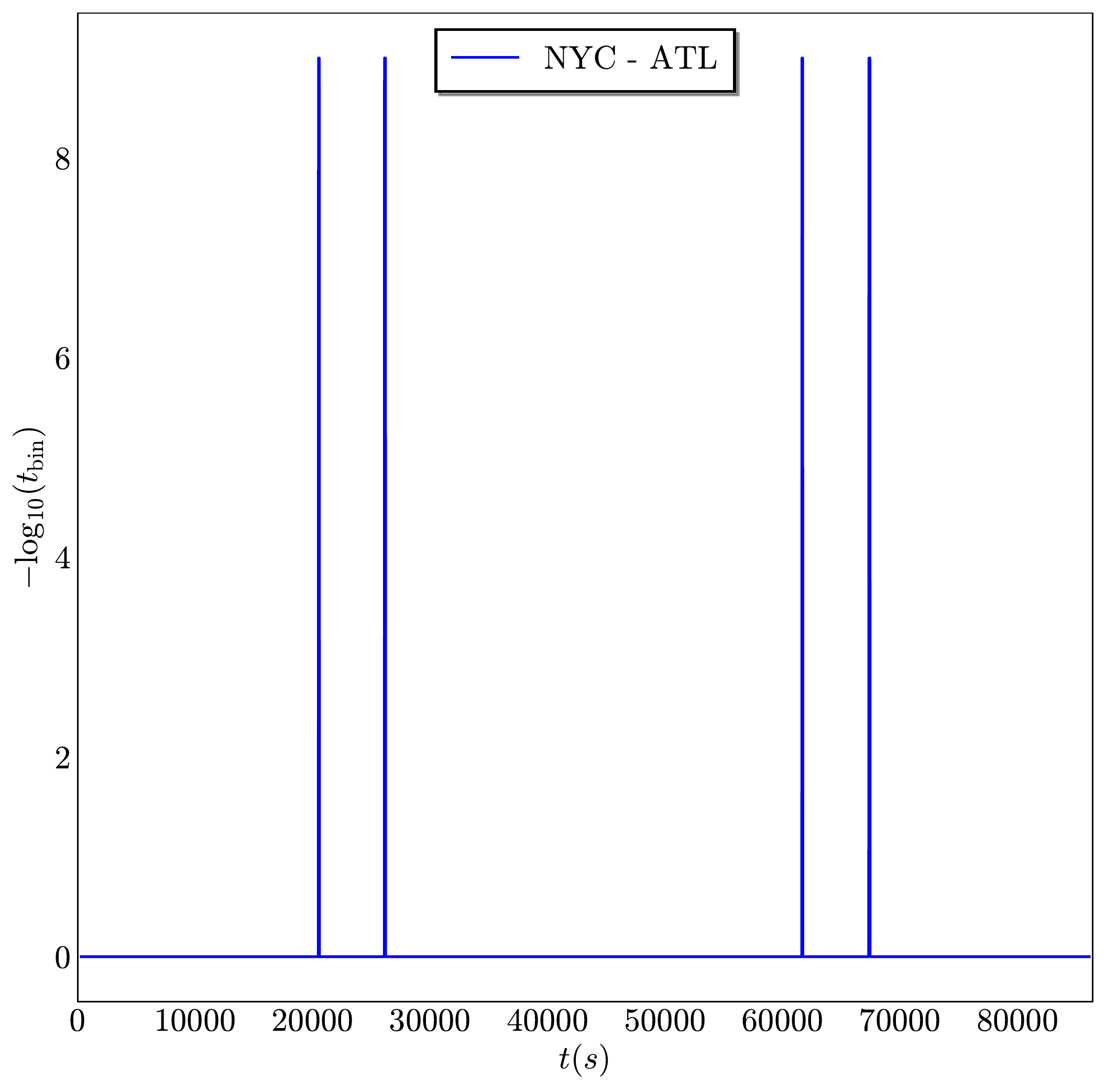}
    \includegraphics[width = 0.45\linewidth]{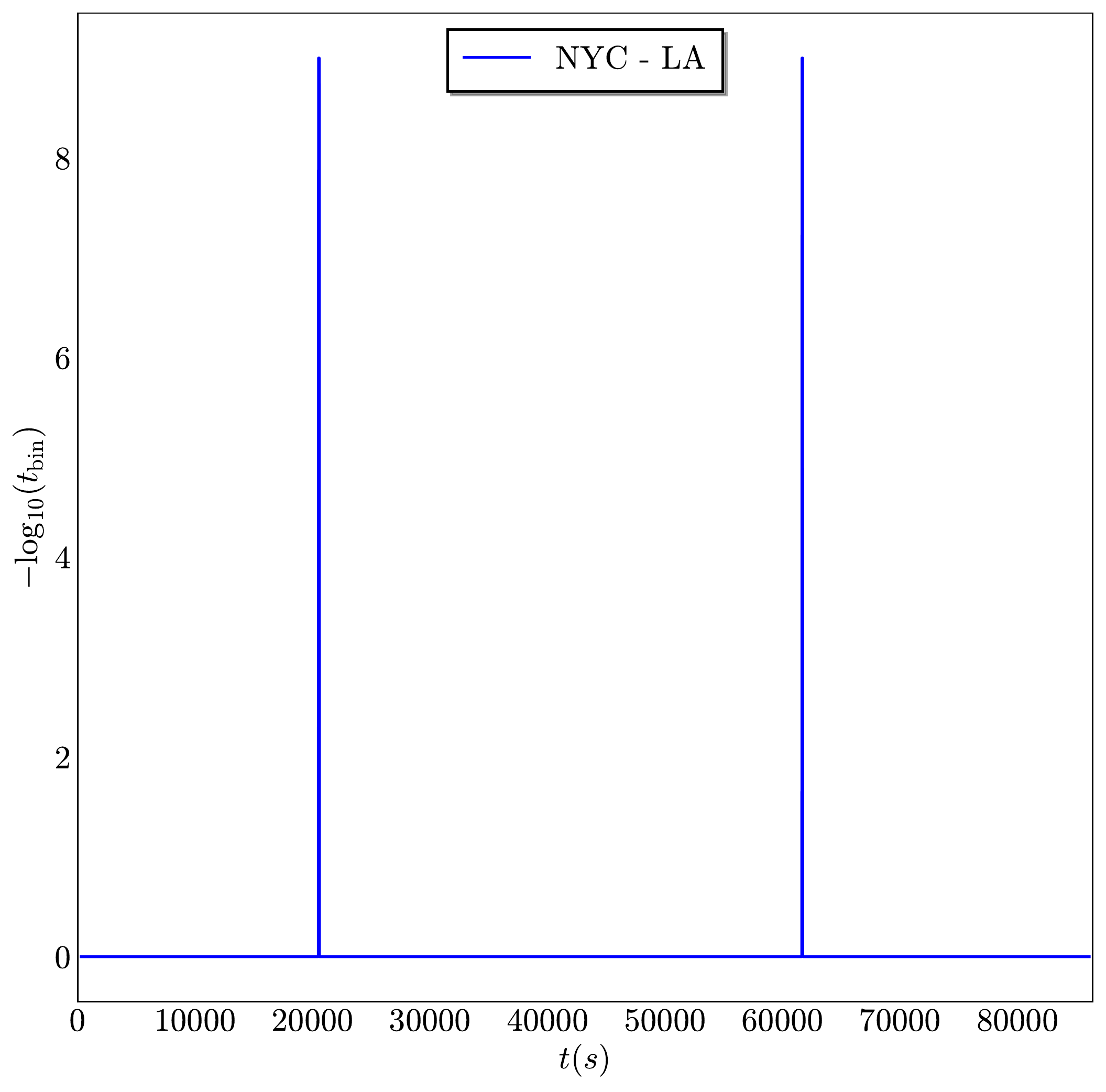}
    
    \includegraphics[width = 0.45\linewidth]{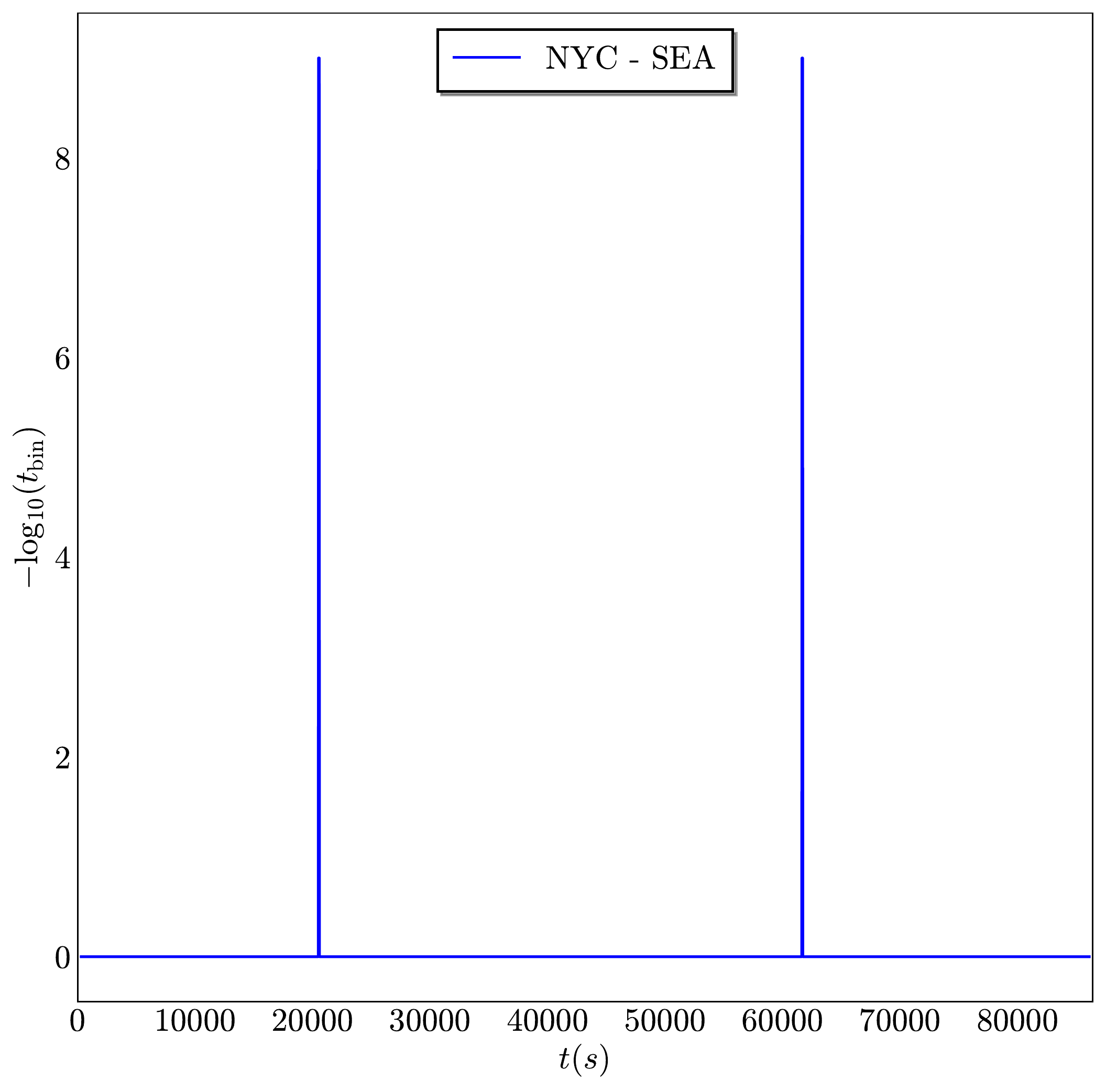}
    \includegraphics[width = 0.45\linewidth]{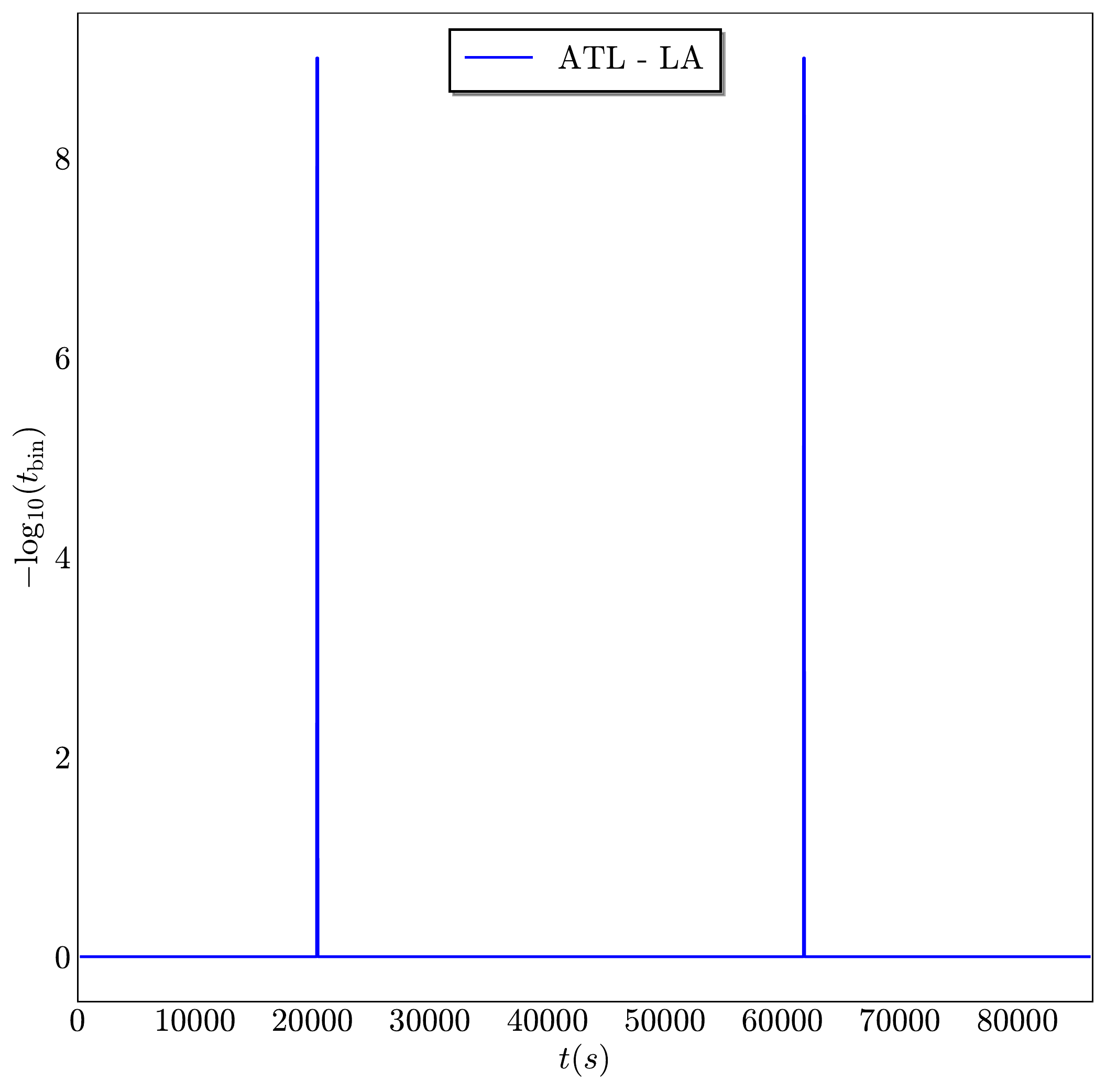}
    
    \includegraphics[width = 0.45\linewidth]{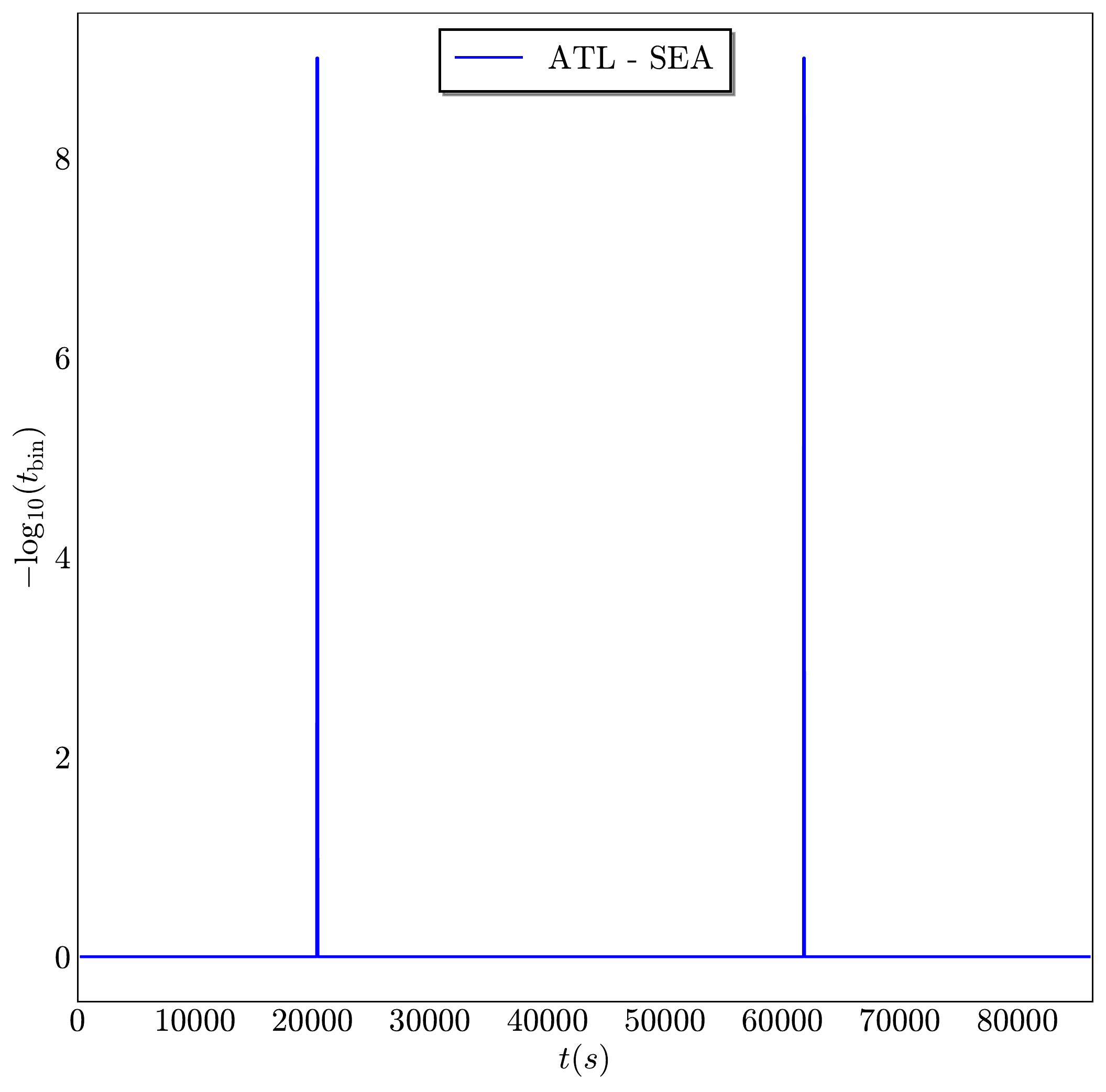}
    \includegraphics[width = 0.45\linewidth]{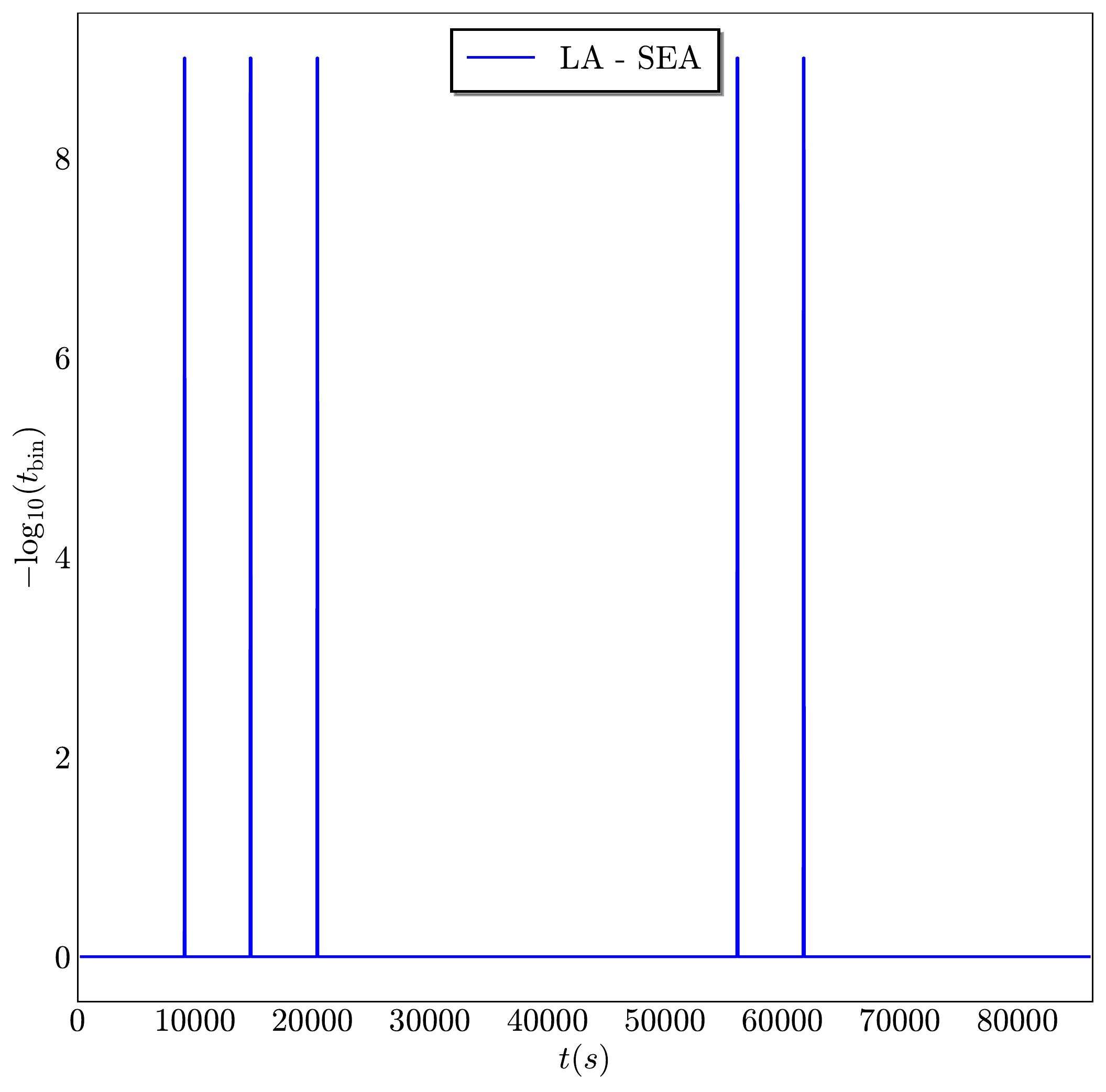}
    \caption{Sync outcomes between all 6 city pairs for a holdover time $\tau = 600 s$ (10 mins). Nanosecond precision can now be achieved in all cases in contrast to the situation with $\tau = 0$ where only two city pairs could achieve the required precision levels.}
    \label{fig:figure12}
\end{figure*}

Figure \ref{fig:figure12} shows that, for $\tau = 600s$ (10 minutes),  6 city pairs in the configuration can be synced at the 1 nanosecond precision ($C = 1$ ns, in Equation (\ref{eqn:45})) more than once a day, in contrast to the $\tau = 0$ case where only 2 city pairs could be synced (Figure \ref{fig:figure10} (Bottom)).

Even though nanosecond level precision can be achieved using a single LEO satellite for the continental US, the connectivity is sparse, with long disconnected intervals between sync events. This situation can be easily alleviated by using a larger constellation of satellites (both adding satellites to same orbit and using multiple orbits). Optimization of satellite resources both in terms of constellation design and onboard system parameters is a complex problem. For a more concrete analysis of figures of merit such connection fractions of the day and longest disconnected intervals (at required levels of precision) which quantify the connectivity of various QCS network designs and system parameters, we refer the reader to a previous work \cite{paper1}.

\section{Conclusions and Future Work} 
\label{sec:6}
In this paper, we have assessed the feasibility of a quantum clock synchronisation network using satellites. We have developed a numerical code to simulate the real time implementation of a QCS protocol, including the effects of relative velocities, photon losses, background noise and detector jitter. Although results presented here are based on some simplifying assumptions, such as constant source and background rates and circular polar orbits, ignoring effects of cloud cover and other weather conditions, etc., the general framework of our code is receptive to loosening of these constraints. Our code is versatile and can be adapted to any given network of ground stations and satellites. Our results are generic in this sense and crucially the trends for network quality and size shown here are supportive of the usefulness of a quantum timing network based on QCS, albeit requiring a more systems and engineering level feasibility study. Further, our code can be also used to analyze quantitatively other entanglement distribution based protocols such as Device Independent QKD, quantum communication networks, etc.

Our detailed analysis has shown that it is possible to provide sub-nanosecond to picosecond level synchronization outcomes for a network spread across the continental US. This is a network size of around 4000 km. It is important to mention that no classical or quantum techniques presently available can provide such precision levels over distances which are this long. We thus propose the QCS network with space-terrestrial optical communication links as the method of choice for high precision, long range time distribution. Since the QCS protocol is yet to be tested between moving clocks, our analytical and simulation results build the ground for future implementations.
For future work we consider the establishment of a master clock in the sky, by interlinking satellites through QCS. By establishing continuous network coverage on global scales, this provides an opportunity to create a high-precision quantum-secure infrastructure for long distance positioning, navigation and timing services.

\appendix

\section{SNR of the correlation functions}
\label{sec:appendixA}
The sharpness of the correlation function peak can be quantified through the signal to noise ratio (SNR). This depends on i) Height of the peak from correctly correlated photons (photons whose time stamps are related by Equations (\ref{eqn:1}) and (\ref{eqn:2}), and ii) the average noise level (spurious peaks) of the correlation function.

Let us look at the threshold due to noise. The peak due to correct correlations should breach this threshold to be detectable. This threshold arises from the following two sources: 
\begin{itemize}
    \item Spurious correlations amongst source photons.  Every source photon detected by A has some time correlation with every photon detected at B, but the correct time correlation given by Equation (\ref{eqn:1}) or (\ref{eqn:2}) with only one photon, which is its entangled pair partner. All the other correlations form a uniformly distributed threshold level.
    Total number of correlations (spurious + correct) $N_{total}$ is given by the area under the $C$ curve. A simple substitution of Equations (\ref{eqn:3}) and (\ref{eqn:4}) into (\ref{eqn:5}) and then integrating w.r.t. time gives:
    \begin{equation}
    \label{eqn:A1}
       N_{total} = \int_0^{t_{\rm Acq}} Cdt = R^2 \eta t_{\rm Acq}^2 \, .
    \end{equation}
    The number of correlated photons detected in time $t_{\rm Acq}$, $N_{corr}$, is given by:
    \begin{equation}
    \label{eqn:A1.5}
       N_{corr} = R\eta t_{\rm Acq}
    \end{equation}
    Therefore, the number of spurious correlations $N_{spur}$ is given by:
    \begin{equation}
    \label{eqn:A2}
        N_{spur} = N_{total} - N_{corr} = R \eta t_{\rm Acq}(Rt_{\rm Acq} - 1) \approx R^2\eta t_{\rm Acq}^2
    \end{equation}
    The last approximation holds since the total number of photons generated within the acquisition time $R\, t_{\rm Acq}$ $\gg$ 1, for reasonably high source rates ($\approx 10^6 - 10^7$ ebits/s). From Equation (\ref{eqn:A2}) we get the mean height of such spurious correlations to be:
    \begin{equation}
    \label{eqn:A3}
        \bar{C}_{spur} = \frac{N_{spur}}{N_{bins}} = \frac{R^2\eta t_{\rm Acq}^2}{t_{\rm Acq}/t_{\rm bin}} = R^2\eta t_{\rm Acq}t_{\rm bin}
    \end{equation}
    \item Noise from background. Thermal photons, photons from the sun, radiation in space, dark counts, etc., might lead to further photon time stamps which are not correctly correlated. They appear independently at A and B and hence they also form a uniformly distributed threshold in the correlation function $C_{AB}$. The average number of correlations due to background photons, again following an argument similar to above, is given by:
    \begin{equation}
    \label{eqn:A4}
        \bar{C}_{bkg} = RR_{\rm bkg} t_{\rm Acq} t_{\rm bin}\, ,
    \end{equation}
    where $R_{\rm bkg} = R_{\rm bkg}^{sat}$  and $R_{\rm bkg} =  R_{\rm bkg}^{gs}$ for $C_{AB}$ and $C_{BA}$, respectively, are the background  photon rates at the ground station and the satellite. 
    Further, if the dark count rate for the detectors is given by $R_{\rm dc}$, then an analogous additional term is added to the average noise level, given by:
    \begin{equation}
    \label{eqn:A5}
        \bar{C}_{\rm dc} = R_{\rm dc} (R\eta + R_{\rm bkg} + R_{\rm dc}) t_{\rm Acq} t_{\rm bin}  \, .
    \end{equation}
    Here we assume that $R \gg R_{\rm dc}$. Also for (LEO) satellite based implementations, since $\eta \approx 0.01$, we have assumed background rates $R_{\rm bkg} \gg R\eta$ and $R_{\rm bkg} \gg R_{\rm dc}$. Therefore, $C_{\rm dc} \ll  C_{\rm bkg}$, and its effect on the SNR has been ignored in this work. Nonetheless, the framework in general allows for its inclusion by introducing a term like Equation (\ref{eqn:A5}).  
\end{itemize}
The total average height of the noise is, therefore, given by:
\begin{equation}
\label{eqn:A7}
   \bar{C}_{noise} = \bar{C}_{spur} + \bar{C}_{bkg} \, .
\end{equation}

Now, we can evaluate  the SNR. The SNR is defined as the ratio of the height of the correlation function peak above the threshold level and the standard deviation of the noise (background + spurious correlations). The noise follows a standard Poisson distribution, and hence the mean is equal to the variance.

The SNR is distinct in the following two time regimes,
\begin{itemize}
    \item $t_{\rm Acq} \leq t_{\rm Acq}^{\rm opt}$
    \begin{equation}
    \label{eqn:A6}
    \rm SNR = \frac{C(\tau_{\rm max}) - \bar{C}_{noise}}{\sqrt{\bar{C}_{noise}}} \, .
    \end{equation}
    using the Equations (\ref{eqn:A2})-(\ref{eqn:A5}) we get,
    \begin{equation}
    \label{eqn:A6.5}
    \rm SNR = \sqrt{\frac{\eta}{t_{\rm bin}(1 + R_{\rm bkg}/R\eta)}} (1 - Rt_{\rm bin}) \sqrt{t_{\rm Acq}}  
    \end{equation}
    In this regime, there exists only one peak due to correctly correlated photons and the SNR increases with $t_{\rm Acq}$. Also, $Rt_{\rm bin} < 1$ for the SNR to make sense. This translates to the fact that each photon should have a unique time stamp within the precision of the clocks. In fact for reasonable choices of $R$ and $t_{\rm bin}$ (e.g. see Table \ref{tab:table1}), $Rt_{\rm bin} \ll 1$. Therefore,
    \begin{equation}
    \label{eqn:A7}
    \rm SNR \approx \sqrt{\frac{\eta}{t_{\rm bin}(1 + R_{\rm bkg}/R\eta)}} \sqrt{t_{\rm Acq}}  
    \end{equation}
    
    \item $t_{\rm Acq} > t_{\rm Acq}^{\rm opt}$.

    Multiple peaks now start appearing. Spurious correlations will now also add to the older peaks, hence not all peaks are of equal height, the average height can be approximated as:
    \begin{equation}
    \label{eqn:A8}
        C(\bar{\tau}_{peak}) \approx R\eta t_{\rm Acq}^{\rm opt} + RR_{\rm bkg}t_{\rm Acq}t_{\rm bin} + R^2\eta t_{\rm Acq}t_{\rm bin}
    \end{equation}
    Therefore, using Equation (\ref{eqn:A6}), the SNR in this case is given by:
    \begin{equation}
    \label{eqn:A9}
        \rm SNR \approx \mathcal{K} \sqrt{\frac{\eta t_{\rm bin}}{\sqrt{(1 + R_{\rm bkg}/R\eta)}}} \frac{1}{\sqrt{t_{\rm Acq}}}
    \end{equation}
    Therefore, for $t_{\rm Acq} > t_{\rm Acq}^{\rm opt}$, the SNR decreases as $(t_{\rm Acq})^{-1/2}$. 
\end{itemize}
This analysis clearly indicates that the SNR is maximum for $t_{\rm Acq} = t_{\rm Acq}^{\rm opt}$ and hence,
\begin{equation}
\label{eqn:A10}
    \rm SNR_{max} \approx \sqrt{\frac{\eta\mathcal{K}}{(1 + R_{\rm bkg}/R\eta)}}
\end{equation}

\section{Operational point of view: How to get the best available level of precision?}
\label{sec:appendixB}
\begin{figure}
    \centering
    \includegraphics[width = \linewidth]{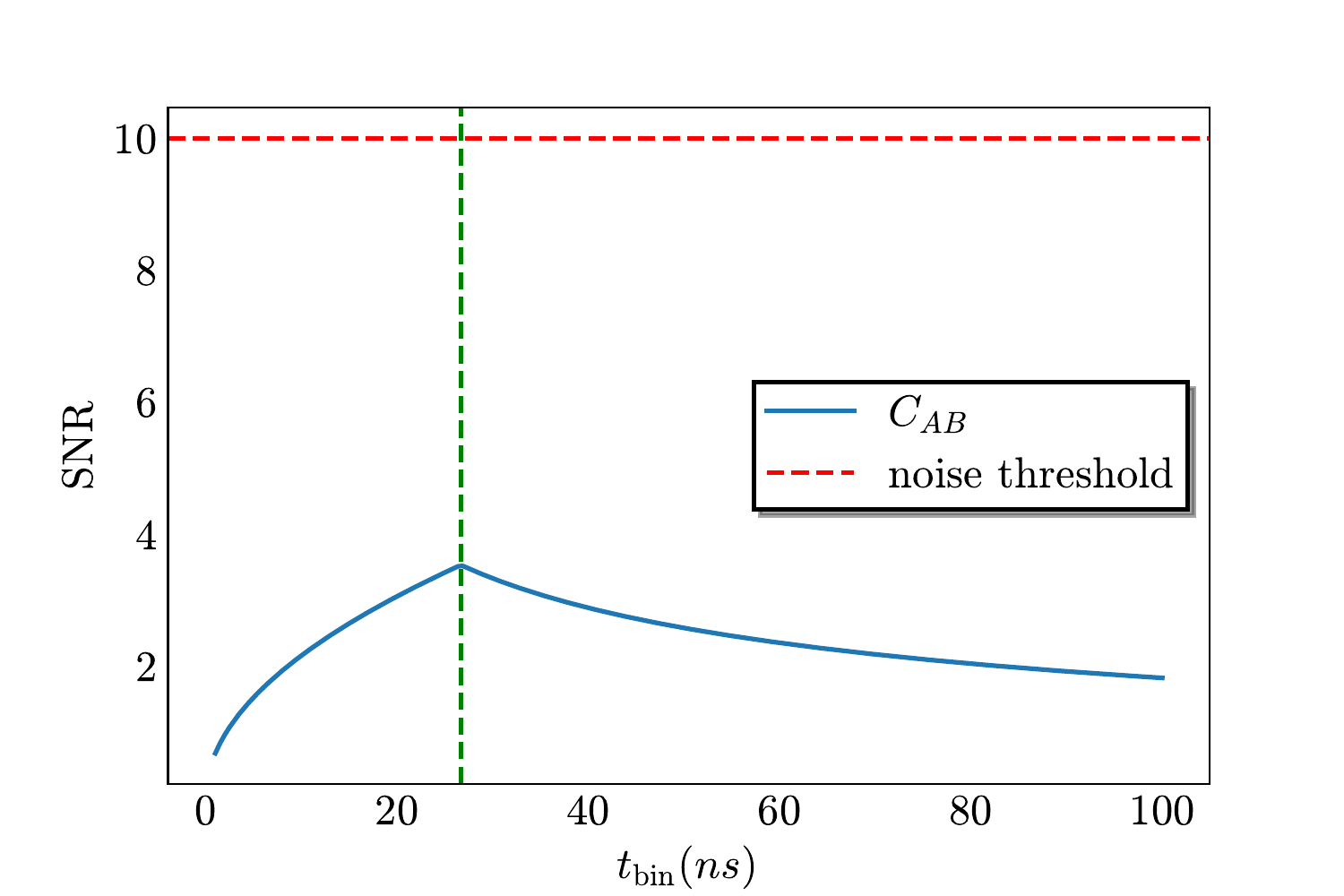}
    \includegraphics[width = \linewidth]{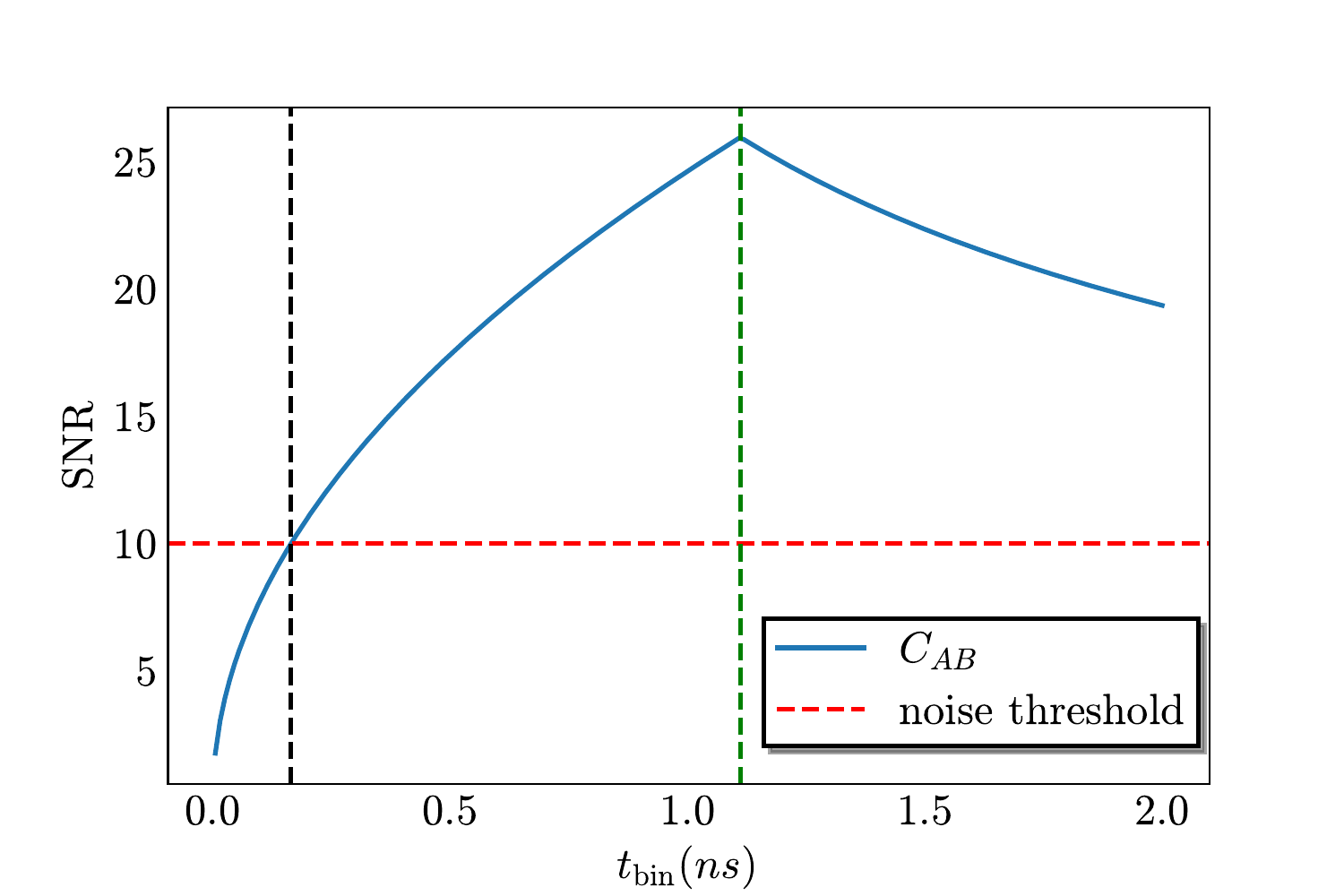}
    \caption{Here we show that precision obtained at the threshold SNR can be substantially higher than what is achievable if we work the max SNR setting. Source rate and background rates are set at  $10^7 \rm ebits/s$ and $10^4 \rm photons/s$ respectively. Loss and relative radial velocity numbers are representative of a 500 km polar orbit passing over New York City. (Top) A very oblique pass of the satellite over the ground station leads to a high loss (35 dB) and high relative velocity (4 km/s) setting. In such cases the max SNR is less than the threshold SNR, and hence, even though we have visibility, the protocol fails at all levels of precision. (Bottom) For a slightly less oblique pass (25 dB loss and 1 km/s relative velocity), working in the threshold SNR regime can provide a precision 10 times higher than what can be achieved by working at the maximum SNR.}
    \label{fig:figure14}
\end{figure}

The limiting value of $t_{\rm bin}$ found via Eqn.(\ref{eqn:35}), in conjunction with Eqn.(\ref{eqn:A10}) is calculated using the relative radial velocity $v^{rad}_{rel}$ of the satellites w.r.t. the ground station. From an operational point of view, the QCS protocol does not require the knowledge of $v^{rad}_{rad}$ and a high degree of precision can be achieved without using any estimate of relative velocity \footnote{This does not eliminate the need for tracking of the satellite for the purposes of alignment etc., the algorithm to find the offset does not need the knowledge of the ground station and satellite positions/velocities.}.
When a satellite is visible from a certain ground station, the satellite holdover time $\tau$ may allow several rounds of the QCS protocol to be conducted (where the true unknown offset remains the same up to a maximum required precision).
Once the time stamp data are collected the cross-correlation functions must be calculated, this requires choosing a bin size (working precision of the protocol $t_{\rm bin}$). We must at the same time also choose an acquisition time window $t_{\rm Acq}$, since the over the total visibility/holdover period large amounts of data are collected which if used in its entirety to plot the correlation functions will lead to substantial peak spreading and loss of SNR. Therefore the data must be divided into smaller acquisition windows. 
Let us begin with the smallest possible $t_{\rm bin}$ (could be the time stamp resolution) and the largest possible $t_{\rm Acq}$ (full time stamp data set available). We would then be working in the regime where the SNR follows Eqn.(\ref{eqn:A9}) and is proportional to $\sqrt{\frac{t_{\rm bin}}{t_{\rm Acq}}}$. The SNR thus depends on the ratio $\frac{t_{\rm bin}}{t_{\rm Acq}}$, as $t_{\rm bin}$ is increased and $t_{\rm Acq}$ is decreased this ratio increases, increasing the SNR up to the point that it reaches $\rm{SNR}_{max}$. We know that $\frac{t_{\rm bin}}{t_{\rm Acq}} = \kappa$ for this optimal setting (see Eqn.(\ref{eqn:A10})). In practice though the following approach could be taken, $t_{\rm bin}$ is increased and $t_{\rm Acq}$ decreased up to the point $\rm SNR = \rm SNR_{th}$, at the same time ensuring $R\eta t_{\rm Acq} \geq N_{\rm min}$. The latter condition is needed to ensure the success of protocol with high probability. For instance even if the $\rm SNR > \rm SNR_{th}$, $R\eta t_{\rm Acq} < 1$ would mean that for some acquisition windows no ebit will be shared between the ground station and satellite, thus the offset would not be obtainable from that round of the protocol. Larger the value of $N_{\rm min}$, larger is the probability that each run of the protocol gets some shared ebits. This gives an operational meaning to $N_{\rm min}$. 
In summary in order to find the best available precision given a set of timestamps (collected over a single holdover window), we vary $t_{\rm bin}$ and $t_{\rm Acq}$ until both the following conditions are satisfied:
\begin{equation}
    \rm SNR \geq \rm SNR_{th},
\end{equation}
\begin{equation}
    C(\tau^{max}) \geq N_{\rm min}
\end{equation}
Working at a sub-optimal $\rm SNR$ still allows us to find the offset because even though we see multiple peaks ($t_{\rm Acq} > \kappa t_{\rm bin})$, all such peaks are above the average noise level. Since the range rate change can be calculated looking at the width of the peak and knowledge of acquisition time, the true offset can then be determined. The precision achievable if we work in this sub-optimal $\rm SNR$ regime (and require $N_{\rm min}$ number of ebits to be collected on average within $t_{\rm Acq}$) is given by:
\begin{equation}
    t_{\rm bin} = \rm SNR_{th}^2 \frac{N_{\rm min}}{R \eta^2} \big( 1+\frac{R_{\rm bkg}}{R\eta} \big) \, .
\end{equation}
In figure \ref{fig:figure14} we show that this precision is higher than what is achievable if we work only in the max SNR regime. Working in this regime can be especially useful for low source rate/high loss or slightly oblique pass (high relative velocity) scenarios. Also for comparison we show a scenario where the max SNR is less than the threshold SNR and thus no working precision can be obtained. 
\section{Effect of detector jitter on SNR and achievable precision}
\label{sec:appendixC}
Let us say that the detector jitter accounts for a uncertainty $\sigma_j$ (standard deviation) in the time stamps of the detected photons. This leads to a peak broadening effect adding onto the effect of range-rate change. Therefore, a smaller maximum precision is now achievable, other factors remaining the same. Appending Eqn.(\ref{eqn:35}) for the effects of jitter, we find that this reduced precision is given by:
\begin{equation}
\label{eqn:C1}
    t_{\rm bin} \geq \frac{N_{\rm min}}{R\eta \mathcal{K}} + \sigma_j\, .
\end{equation}
The above equation can also be understood by realising that since the jitter already leads to a peak broadening by the amount $\sigma_j$ the peak broadening effect due to range-rate change tolerable at a given working precision $t_{\rm bin}$ is smaller. This leads to a smaller $t^{opt}_{Acq}$, and is given by $\kappa(t_{\rm bin} - \sigma_j)$. This has implications for the $\rm SNR$ of the correlation function peaks as well. Since the optimal acquisition time is now reduced to $\kappa(t_{\rm bin} - \sigma_j)$, the max $\rm SNR$ achievable becomes:
\begin{equation}
    \label{eqn:C2}
    \rm SNR_{max} \approx \sqrt{\frac{\eta\kappa}{t_{\rm bin}(1 + R_{\rm bkg}/R\eta)}}(1 - \frac{\sigma_j}{t_{\rm bin}}),
\end{equation}
which is meaningful only when $t_{\rm bin} > \sigma_j$. This reduces to the jitter-less case for $\sigma_j = 0$. In the limiting case of $\sigma_j = t_{\rm bin}$ shows than $\rm SNR = 0$, i.e., no peak can be seen at the working precision of $t_{\rm bin}$. Further, contrary to the jitter-less case, here the $\rm SNR$ is not independent of the precision $t_{\rm bin}$. Thus, imposing a threshold condition on the required $\rm SNR$ (analogous to Eqn.(\ref{eqn:31})), and then solving the equation for $t_{\rm bin}$, gives a relation for the maximum achievable precision in terms of the imposed threshold ${\rm SMR}_{th}$. This is given by:
\begin{equation}
\label{eqn:C3}
    t_{\rm bin} \geq \frac{\sigma_j}{1 - \frac{S_{th}^2}{\eta \kappa}(1+\frac{R_{\rm bkg}}{R\eta})}.
\end{equation}

\begin{figure}
    \centering
    \includegraphics[width = \linewidth]{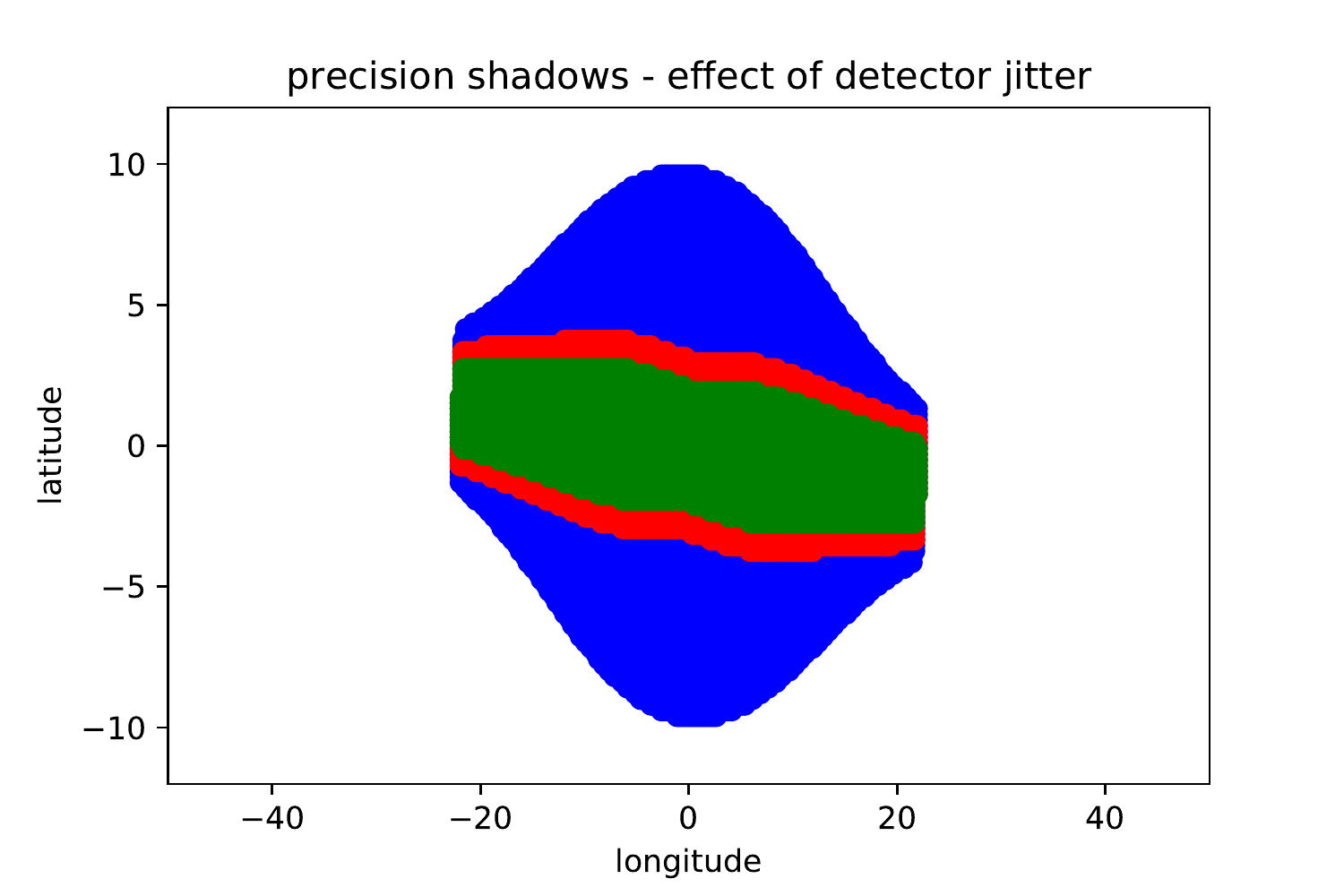}
    \includegraphics[width = \linewidth]{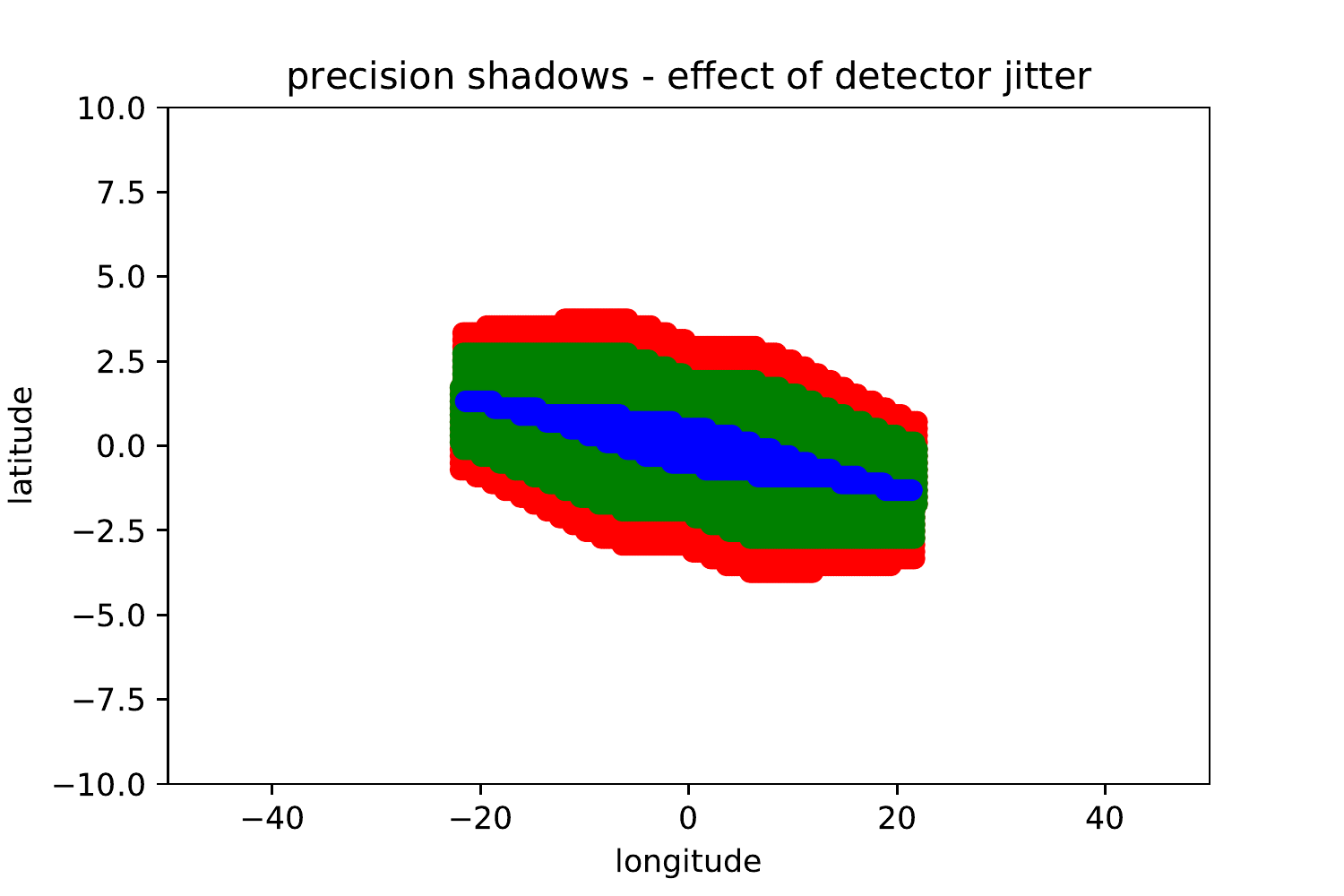}
    \caption{Effect of detector jitter on the precision shadows for 1 ns precision -- Red shadow represents the jitterless case, Green shadow represents the shadow for 300 ps of jitter if the precision is determined using Eqn.(\ref{eqn:C1}) ($N_{\rm min}$ condition) whereas the Blue shadow represents the shadow if Eqn.(\ref{eqn:C3}) (SNR condition) determines the precision. (Top) Precision shadow for low background rate $R_{\rm bkg}/R = 10^{-2}$, Eqn.(\ref{eqn:C1}) dominates over Eqn.(\ref{eqn:C3}) in determining the maximum achievable precision. (Bottom) Precision shadow for high background rate $R_{\rm bkg}/R = 0.5$, Eqn.(\ref{eqn:C3}), i.e., the SNR condition dominates over Eqn.(\ref{eqn:C1}) in determining the maximum achievable precision.}
    \label{fig:figure15}
\end{figure}
\bibliography{main.bib} 
\end{document}